\documentclass[prd,nofootinbib,showpacs,superscriptaddress, preprint]{revtex4-1}
\pdfoutput=1
\usepackage{amsmath,amsfonts,amssymb,graphicx,natbib,bm}
\usepackage{color}
\usepackage{slashed}    
\usepackage[colorlinks,citecolor=blue]{hyperref}
\usepackage{color}
\usepackage{url}
\usepackage{cancel}
\usepackage{slashed}
\usepackage{multirow}
\usepackage{rotating}
\usepackage{braket}
\usepackage{float}
\usepackage{subfigure}

\newcommand{\be}{\begin{eqnarray*}}
\newcommand{\ee}{\end{eqnarray*}}

\newcommand{\bee}{\begin{eqnarray}}
\newcommand{\eee}{\end{eqnarray}}
\newcommand{\beeq}{\begin{equation}}
\newcommand{\eeq}{\end{equation}}

\usepackage{xspace}

\newcommand{\ba}{\begin{array}}
\newcommand{\ea}{\end{array}}
\newcommand{\bd}{\begin{displaymath}}
\newcommand{\ed}{\end{displaymath}}
\newcommand{\besub}{\begin{subequations}}
\newcommand{\eesub}{\end{subequations}}
\newcommand{\bea}{\begin{eqnarray}}
\newcommand{\eea}{\end{eqnarray}}

\def\a{\alpha}

\def\b{\beta}
\def\g{\gamma}

\def\l{\lambda}
\def\m{\mu}

\def\G{\Gamma}

\def\q2 {q^2}

\def\bt{\begin{table}}
\def\et{\end{table}}

\begin{document}
\title{Two Component  Singlet-Triplet Scalar Dark Matter and Electroweak Vacuum Stability}

\author{Amit Dutta Banik}
\email{amitdbanik@mail.ccnu.edu.cn}
\affiliation{Key Laboratory of Quark and Lepton Physics (MoE) and Institute of Particle Physics, Central China
Normal University, Wuhan 430079, China }
\author{Rishav Roshan}
\email{rishav.roshan@iitg.ac.in}
\author{Arunansu Sil}
\email{asil@iitg.ac.in}
\affiliation{Department of Physics, Indian Institute of Technology Guwahati, Assam-781039, India }

\begin{abstract}  
We propose a two component dark matter setup by extending the Standard Model with a singlet and a hypercharge-less 
triplet scalars, each of them being odd under different $Z_2$ symmetries. We observe that the inter-conversion between the 
two dark matter components allow a viable parameter space where masses of both the dark matter candidates can be below 
TeV, even though their individual contribution to single component dark matter rules out any such sub-TeV dark matter. We 
find that a lighter mass of the neutral component of the scalar triplet, playing the role of one dark matter component, compared to 
the singlet one is favored. In addition, the setup is shown to make the electroweak vacuum absolutely stable till the Planck scale, 
thanks to Higgs portal coupling with the scalar dark matter components. 
\end{abstract}

\maketitle
\section{Introduction}
The Standard Model (SM) of particle physics undoubtedly emerges as the fundamental theory of interactions 
after discovery of the Higgs boson at the Large Hadron Collider (LHC) \cite{Chatrchyan:2012xdj,Aad:2012tfa}.  
However there still remains some issues, confirmed by experimental observations and can't be resolved within SM. 
For example, observation of cosmic microwave background radiation by Planck \cite{Aghanim:2018eyx} reveals that 
about 26.5\% of Universe is made up of mysterious dark matter (DM). The SM of particle physics, however, can not 
account for a dark matter candidate. Dark matter direct search experiments such 
as LUX \cite{Akerib:2016vxi}, XENON-1T \cite{Aprile:2018dbl}, PandaX-II \cite{Tan:2016zwf,Cui:2017nnn}  search 
for the evidences of DM-nucleon 
interaction. Till date no such direct detection signal of DM has been detected which limits the DM-nucelon 
scattering cross-section. Apart from dark matter 
there also exists problem with the stability of electroweak (EW) vacuum within the 
Standard Model as the electroweak vacuum becomes unstable at large scale $\Lambda_I\sim 10^{10}$ GeV 
\cite{Buttazzo:2013uya,Degrassi:2012ry,Tang:2013bz,Ellis:2009tp,EliasMiro:2011aa} for top quark mass $m_t=173.2$ GeV  \cite{Tanabashi:2018oca}. This instability of EW vacuum at 
large scale can be restored in presence of additional scalars. 

In order to address the above mentioned issues, we need to go beyond the SM. In this work, we will include
new scalar particles which can serve as dark matter candidate and also stabilize the EW vacuum simultaneously.
It is also to  be noted that the null detection of DM in direct detection (DD) experiments also triggers the possibility of dark
sector to be multi-component which is explored in many literatures in recent \cite{Biswas:2013nn,Fischer:2011zz,Bhattacharya:2013hva,Bian:2013wna,Esch:2014jpa,Karam:2015jta,Karam:2016rsz,Bhattacharya:2016ysw,DuttaBanik:2016jzv,Ahmed:2017dbb,Herrero-Garcia:2017vrl,Herrero-Garcia:2018qnz,Poulin:2018kap,Aoki:2018gjf,Bhattacharya:2018cgx,Aoki:2017eqn,Barman:2018esi,Chakraborti:2018aae,Elahi:2019jeo,Borah:2019aeq,Bhattacharya:2019fgs,Biswas:2019ygr,Bhattacharya:2019tqq,Nanda:2019nqy,Maity:2019hre,Khalil:2020syr,Belanger:2020hyh,Nam:2020twn}. In multi-component dark 
matter scenario DM-DM conversion plays a significant role to determine the observables such as direct 
detection and relic density and also helps to stabilize EW vacuum with increased number of scalars. 
In this work, we consider a multi-component dark matter with scalar singlet and 
scalar triplet with zero hypercharge.

Study of scalar singlet dark matter and its effects on electroweak vacuum is 
done extensively in earlier works \cite{Haba:2013lga,Khan:2014kba,Khoze:2014xha,Gonderinger:2009jp,Gonderinger:2012rd,Chao:2012mx,Gabrielli:2013hma,Ghosh:2017fmr,Bhattacharya:2017fid,Garg:2017iva,DuttaBanik:2018emv,Borah:2020nsz}.  In a pure scalar singlet scenario, due to the presence of
the quartic coupling between Higgs and dark matter can help Higgs quartic coupling become 
positive making the EW vacuum stable till Planck scale, $M_{Pl}$ . It is found that singlet scalar with mass 
heavier than 900 GeV can satisfy the constrains coming from relic density, 
direct detection and vacuum stability \cite{Bhattacharya:2019fgs}. Introducing an inert doublet as a possible 
dark matter component attracts a great amount of attention in recent days. It is found that there exists an intermediate 
region (80 - 500) GeV, beyond which the neutral component of the inert Higgs satisfies the relic and DD constraints \cite{LopezHonorez:2006gr, Honorez:2010re, Belyaev:2016lok, Choubey:2017hsq, LopezHonorez:2010tb, Ilnicka:2015jba, Arhrib:2013ela, Cao:2007rm, Lundstrom:2008ai, Gustafsson:2012aj, Kalinowski:2018ylg, Bhardwaj:2019mts}. Recently 
it has been shown that in multi-component DM scenarios involving inert Higgs doublet(s) and/or singlet scalar, the region can be 
revived \cite{Borah:2019aeq,Bhattacharya:2019fgs}.

Moving toward a further higher multiplet, it is found that an inert triplet can also be a possible dark matter candidate. 
A hypercharge-less ($Y$=0)  inert triplet scalar can serve as  a feasible dark matter candidate similar to inert Higgs 
doublet \cite{LopezHonorez:2006gr, Honorez:2010re, Belyaev:2016lok, Choubey:2017hsq, LopezHonorez:2010tb, Ilnicka:2015jba, Arhrib:2013ela, Cao:2007rm, Lundstrom:2008ai, Gustafsson:2012aj, Kalinowski:2018ylg, Bhardwaj:2019mts,Bandyopadhyay:2020vfc,Bandyopadhyay:2020qpn}. However, 
the allowed mass ranges of inert triplet dark matter is very much different from that of the inert doublet. Similar to the case of 
inert doublet, annihilation of triplet scalar is mostly gauge dominated  which leaves a larger desert region compared to inert doublet.  
Also, due to small mass splitting between charged and neutral triplet scalar, co-annihilation channels into SM particles becomes 
relevant. Earlier studies \cite{Araki:2011hm, Fischer:2011zz, Fischer:2013hwa, Khan:2016sxm,Jangid:2020qgo} reported that a pure inert triplet (with $Y$=0) dark matter, consistent with relic density, direct detection and vacuum stability constrains  can be achieved with triplet mass $\sim 1.9$ TeV and 
triplet Higgs quartic coupling $\sim 0.3$. In addition, the presence of the charge component in the scalar triplet also provides interesting discovery prospect in the collider searches \cite{Chiang:2020rcv}. The small mass splitting among the neutral and the charge component  of the scalar triplet dictates the decay of the charged component only to the neutral component and  to the soft pion or the soft lepton pairs and once  produced these soft pions can lead to the disappering charge track in the detector. 

The other possibility is to have $Y=2$ triplet scalar which is also investigated. It was shown in \cite{Araki:2011hm} that with $Y=2$, dark matter mass $M_{\rm DM}\geq$ 2.8 TeV is allowed. For $Y=2$ possibility, things are further restricted, mostly from the direct detection bounds. It is to be noted that unlike $Y=2$ inert triplet scalar,  neutral particles of $Y=0$ inert triplet scalar does not have direct  interaction with the $Z$ boson which arise from the kinetic term in case of $Y=2$ triplet.  As a result, additional quark nucleon scattering via $Z$ boson exchange occurs for $Y=2$ inert triplet. This interaction term contributes to dark matter direct detection significantly and because of large scattering cross-section, most of the available parameter space is ruled out \cite{Araki:2011hm}. In this work we concentrate on $Y=0$ triplet scalar.

As mentioned above, due to large gauge dominated annihilation, the relic density of $Y=0$ inert triplet dark matter remains 
under-abundant up to $\sim$1.8 TeV. Therefore, this leaves a great opportunity to explore the phenomenology of multi-component dark 
matter setup involving the inert triplet and a singlet scalars. Similar to the case of scalar singlet, the triplet Higgs quartic coupling also 
helps to stabilize the EW vacuum. In this work however, we want to explore the below-TeV regime of both the dark matters in the two-component framework as this sub-TeV region is of great importance from the collider and dark matter experiments. 
In a work \cite{Fischer:2011zz}, although the authors explored a multi-component DM scenario with an inert triplet and a singlet scalars, the detailed effects of inter-conversion of DMs were not appropriately addressed in view of coupled Boltzmann equations. Furthermore, the work of \cite{Fischer:2011zz} considered the results of DM direct detection experiments ($e.g.$ XENON 100 data) which were not so stringent at the time of their analysis compared to the recent XENON 1T results. 
In our study however, we aim to show the pivotal importance of the conversion coupling in realizing the correct DM relic density by solving the coupled Boltzmann equations while taking into account the most recent DD experimental constraints into account. At the same time we also emphasise on the Higgs portal coupling of both the dark matters as they play a significant role in dark matter phenomenology and also in making the EW vacuum absolutely stable till $M_{Pl}$. We therefore search for a viable parameter space in this multi-component dark matter scenario that satisfies constraints from dark matter observables as well as electroweak vacuum stability can also be achieved.

The paper is organized as follows. The model is introduced in section~\ref{model} and the various
theoretical and experimental constraints deemed relevant are detailed in section \ref{constraints}. Sections \ref{DMPh} sheds 
light on the DM phenomenology. We then discuss the status of vacuum stability in \ref{VS} in this scenario and finally conclude in section \ref{conclusion}.
   
\section{Model}
\label{model}

In the present setup, we extend the Standard Model particle content by introducing a $SU(2)_L$ triplet 
scalar $T$ having hypercharge $Y=0$ and a $SU(2)_L$  singlet scalar $S$. In addition, we include 
discrete symmetries $Z_{2}\times Z_{2}^{\prime}$ under which all the SM fields are even while additional 
fields transform differently. In Table \ref{t1} we provide the charge assignments of these additional fields 
under the SM gauge symmetry and the additional discrete symmetries imposed on the framework. Both 
the scalar singlet $S$ and the neutral component of $T$ can play the role of the dark matter candidates 
as they are charged odd under different $Z_2$ and hence stable. Therefore the present setup can 
accommodate a two-component dark matter scenario. 
\begin{table}[H]
\begin{center}
\vskip 0.5 cm
\begin{tabular}{|c|c|c|c|c|}
\hline
    Particle & $SU(2)$ &   $U(1)_Y$  & $Z_2$ &  $Z_2^{\prime}$           \\
\hline        
$H$         &  2     &  $\frac{1}{2}$         &  + & +                \\
\hline
$T$         &  3     &  0         &  - & +                \\

\hline
$S$         &  1     &  0         &  + & -                \\
\hline
\end{tabular}
\end{center}  
\caption{Scalar particles and their charges under different symmetries.}
\label{t1}
\end{table}

The most general renormalisable scalar potential of our model, $V(H,T,S)$, consistent with $SU(2)_{L}\times U(1)_Y\times Z_2 
\times Z_{2}^{\prime}$ consists of (i) $V_H$: where sole contribution of the SM Higgs is included, (ii) $V_T$: involving contribution 
from scalar triplet $T$ only, (iii) $V_S$ : contribution of scalar singlet $S$ only and (iv) $V_{int}$: specifying interactions among $H,~T,~S$. This is expressed as below: 
\begin{eqnarray}
V(H,T,S) &=& V_H + V_T + V_S + V_{int} .
\label{p1}
\end{eqnarray}
where

\besub
\bea
V_{H} &=& -\mu_H^2 H^\dag H + \lambda_H (H^\dag H)^2,  \\
V_{T} &=& \frac{M_T^2}{2}~{\rm{Tr}}[T^2]+\frac{\lambda_T}{4!}(~{\rm{Tr}}[T^2])^2,  \\
V_{S} &=& 
\frac{M_S^2}{2} S^2+ \frac{\lambda_S}{4!} S^4, 
\eea
\eesub
and 
\bea
V_{\rm{int}}&=& 
 \frac{\lambda_{HT}}{2}(H^\dag H)~{\rm{Tr}}[T^2]+ \frac{\lambda_{HS}}{2}(H^\dag H)S^2+\frac{\kappa}{4}~{\rm{Tr}}[T^2]S^2.
 \label{p1int}
 \eea
In the above expression of Eq.~(\ref{p1}), $H$ denotes SM Higgs doublet. After the electroweak symmetry breaking (EWSB), the 
SM Higgs doublet obtains a vacuum expectation value (vev) $v=246$ GeV. On the other hand, $T^0$ and $S$ do not acquire any non-zero vacuum expectation value, thereby $Z_2$ and $Z_2^{\prime}$ remains unbroken so as to guarantee the stability of the 
dark matter candidates. 

The scalar fields can be parametrised as
\bea
H = \left( \begin{array}{c}
                          w^+  \\
                          \frac{1}{\sqrt{2}}(v+h+iz)  
                 \end{array}  \right) \, ,                     
&& ~~~~~T =\left( \begin{array}{c}\
                           \frac{1}{\sqrt{2}}T^0~~~~-T^+ \\
        				      -T^- ~~~~-\frac{1}{\sqrt{2}}T^0
                 \end{array}  \right) \, ,
~~~~~S \,\, , 
\eea
and after the EWSB, the masses of the physical scalars are given as 
\bea
m_h^2&=&2 \lambda_H v^2,\nonumber  \\
m_{T^0,T^\pm}^{2}&=& M_T^2+ \frac{\lambda_{HT}}{2}v^2,\nonumber \\
m_{S}^{2}&=& M_S^2+ \frac{\lambda_{HS}}{2}v^2\ \, .
\label{mass}   
\eea 

In Eq.~(\ref{mass}), $m_h=125.09$ GeV \cite{deFlorian:2016spz}, is the mass of SM Higgs. It is to be noted that although mass of neutral and charged triplet scalar are degenerate, a small mass difference of $\Delta m$ is generated via one loop correction \cite{Cirelli:2005uq,Cirelli:2009uv} and therefore $T^0$ can be treated as a stable DM candidate. This mass difference is 
expressed as 
\bea
\Delta m&=& (m_{T^{\pm}}-m_{T^0})_{1-loop} = \frac{\alpha ~m_{T^0}}{4\pi}\bigg{[}f\bigg{(}\frac{M_W}{m_{T^0}}\bigg{)}-c_{W}^2f\bigg{(}\frac{M_Z}{m_{T^0}}\bigg{)}\bigg{]},
\label{loop1}   
\eea 
where $\a$ is the fine structure constant, $M_W,~M_Z$ are the masses of the W and Z bosons, $c_W=\cos\theta_W=M_W/ M_Z$ and $f(x)=-\frac{x}{4}\bigg{[}2x^3\rm{ln}(x)+(x^2-4)^{3/2}\rm{ln}\bigg{(}\frac{x^2-2-x\sqrt{x^2-4}}{2}\bigg{)}\bigg{]}$where $x=\frac{M_{W,Z}}{m_{T^0}}$. It turns out that 
in the limit $x\rightarrow0$ $i.e.$ $m_{T^0} \gg$ $M_W$ or $M_Z$, $f(x)\rightarrow 2\pi x$ and $\Delta m$ can be expressed as \cite{Cirelli:2005uq}
\bea
\Delta m&=&\frac{\a}{2}M_W\sin^2\frac{\theta_W}{2} \simeq166~ \rm{MeV}. 
\label{deltam}   
\eea 

The couplings $\l_{HS}$ and $\l_{HT}$ denote the individual Higgs portal couplings of two DM candidates $S$ and $T^0$ respectively whereas the coupling $\kappa$ provides a portal which helps in converting one dark matter into another 
(depending on their mass hierarchy). For our analysis purpose, we first implement this model in LanHEP \cite{Semenov:2014rea}, choosing the independent parameters in the scalar sector as:
$$(m_{T^0},~m_S,~\lambda_{HS},~\lambda_{HT},\kappa).$$

\section{Theoretical and experimental constraints}
\label{constraints}
\subsection{Theoretical constraints}
The parameter space of this model is constrained by the theoretical consideration like the vacuum stability, perturbativity and unitarity of the scattering matrix.
These constraints are as follows:
\begin{itemize}
\item[(i)]{\bf{Stability:}}
Due to the presence of extra scalars ($T$ and $S$) in our model, the SM scalar potential gets modified which can be seen from Eq (\ref{p1}). In order to ensure that the potential is bounded from below, the quartic couplings in the potential must satisfy the following co-positivity conditions. Following \cite{Kannike:2012pe,Chakrabortty:2013mha} we have derived the copositivity conditions for our present setup: 
\besub
\bea
\lambda_H (\mu),~ \lambda_T (\mu) ,~ \lambda_s (\mu) \geq 0  \\
\lambda_{HT}(\mu)~+~\sqrt{\frac{2}{3}\lambda_{H}(\mu)\lambda_{T}(\mu)}~\geq 0  \\
\lambda_{HS}(\mu)~+~\sqrt{\frac{2}{3}\lambda_{H}(\mu)\lambda_{S}(\mu)}~\geq 0  \\
\kappa(\mu)~+~\sqrt{\frac{1}{9}\lambda_{T}(\mu)\lambda_{S}(\mu)}~\geq 0 
\,\, .  
\eea
\label{copos} 
\eesub
where $\m$ is the running scale. These condition should be satisfied at all the energy scales  till $M_{Pl}$ in order to ensure the stability of the entire scalar potential in any direction.

\item[(ii)]{\bf{Perturbativity:}} A perturbative theory expects that the model parameters should obey: 
\bea
|\l_{i}|,~|\kappa|< 4\pi~ \rm{and}~ |g_i|,|y_{\a \b}|<\sqrt{4\pi}  .
\label{pert} 
\eea
where $\l_i$ and $\kappa$ represents the scalar quartic couplings involved in the present setup whereas $\rm{g}_i$ and $\rm{y}_{\a \b}$ denotes the SM gauge and Yukawa couplings respectively. We will ensure the perturbativity of the couplings present in the model till the $M_{Pl}$ energy scale by employing the renormalisation group equations (RGE).

\item[(iii)]{\bf{Tree level unitarity:}} One should also look for the constraints coming from perturbative unitarity associated with the S matrix corresponding to scattering processes involving all two particle initial and final states \cite{Horejsi:2005da,Bhattacharyya:2015nca}. In the present setup, there are 13 neutral and 8 singly charged combination of two particle initial/final states. All the details are provided in the Appendix \ref{appenA}. 
The constraints imposed by the tree level unitarity of the theory are as follows:
\bea
|\l_H|<4\pi, ~\bigg{|}\frac{\l_{T}}{3}\bigg{|}< 8\pi ,\nonumber \\
|\l_{HT}|< 8\pi, |\l_{HS}|< 8\pi, |\kappa|< 8\pi, \nonumber \\
\rm{and}~ |x_{1,2,3}|< 16 \pi   
\eea
where $|x_{1,2,3}|$ are the roots of the following cubic equation:
\be
x^3+x^2(-36\l_H-3\l_S-5\l_T)+x(-27\kappa^2-36\l_{HS}^2-108\l_{HT}^2+108\l_H\l_S
+180\l_H\l_T\nonumber \\
+15\l_S\l_T)+972\kappa^2\l_H-648\kappa\l_{HS}\l_{HT}+324\l_{HT}^2+180\l_{HS}^2\l_T\l_S-540\l_H\l_T\l_S\nonumber \\
= 0.   
\ee

\end{itemize}

\subsection{Experimental constraints}

Below we provide a brief discussion of all the important experimental constraints applicable on the present set up.
\begin{itemize}
\item[(i)]{\bf{Electroweak precision parameters:}} A common approach to study beyond the SM is considering the electroweak precision test. The presence of an additional scalar triplet in the setup may contribute to the oblique parameters. These extra  contributions to 
the oblique parameters coming from the present setup are given as \cite{Forshaw:2001xq,Khan:2016sxm,Cai:2017wdu}

\besub
\bea
S&\simeq & 0, \\
T&=& \frac{1}{8\pi}\frac{1}{s_W^2c_W^2}\bigg{[}\frac{m_{T^0}^2+m_{T^{\pm}}^2}{M_Z^2}-\frac{2m_{T^0}^2m_{T^{\pm}}^2}{M_Z^2(m_{T^0}^2-m_{T^{\pm}}^2)}\rm{ln}\bigg{(}\frac{m_{\it{T}^{\rm 0}}^2}{m_{\it{T^{\pm}}}^2}\bigg{)}\bigg{]}\\ \nonumber 
&\simeq & \frac{1}{6\pi}\frac{1}{s_W^2c_W^2}\frac{(\Delta m)^2}{M_Z^2}\\
U&=& -\frac{1}{3\pi}\bigg{[}m_{T^0}^4\rm{ln}\bigg{(}\frac{m_{\it{T}^{\rm 0}}^2}{m_{\it{T^{\pm}}}^2}\bigg{)}\frac{(3m_{\it{T^{\pm}}}^2-m_{\it{T}^{\rm 0}}^2)}{(m_{\it{T}^{\rm 0}}^2-m_{\it{T^{\pm}}}^2)^3}+\frac{5(m_{\it{T}^{\rm 0}}^4+m_{\it{T^{\pm}}}^4)-22m_{\it{T}^{\rm 0}}^2m_{\it{T^{\pm}}}^2}{6(m_{\it{T}^{\rm 0}}^2-m_{\it{T^{\pm}}}^2)^2}\bigg{]}\\ \nonumber 
&\simeq & \frac{\Delta m}{3\pi ~m_{T^{\pm}}} .  
\eea 
\label{STU} 
\eesub
The contribution to the $S$ parameter from the triplet scalar fields is negligible. It is clear from Eq.(\ref{STU}b) and Eq.(\ref{STU}c) that the contributions to the $T$ and $U$ parameters are also very much suppressed and hence negligible as $m_{T^0}$ and $m_{T^{\pm}}$ are almost degenerate ($\Delta m=166$ MeV).

\item[(ii)]{\bf{Invisible Higgs decays:}} Invisible Higgs decays provide chance for exploring the possible DM-Higgs boson coupling. If the DM particles are lighter than half of the SM Higgs mass ($m_h$), the Higgs ($h$) can decay to the DM and can contribute to the invisible Higgs decay. Under such circumstances, we need to employ the bound on the invisible Higgs decay width of the SM Higgs boson as \cite{Tanabashi:2018oca}: 
\besub
\bea
Br(h\rightarrow \rm{Invisible})<0.24, \\
\frac{\Gamma(h\rightarrow \rm{Invisible})}{\Gamma(h\rightarrow SM)+\Gamma(h\rightarrow \rm{Invisible})} < 0.24 .
\label{invi}   
\eea 
\eesub
where $\Gamma(h\rightarrow \rm{Invisible})=\Gamma(h\rightarrow \it{T^{\rm 0}}\it{T^{\rm 0}})+\Gamma(h\rightarrow SS)$ when $m_{T^0},~m_S<~\frac{m_h}{2}$ and $\Gamma(h\rightarrow SM)=4.2$ MeV. In the present setup we  focus mostly in the parameter space where $m_{T^0},~m_S>~\frac{m_h}{2}$ so the above constraint is not applicable.

\item[(iii)]{\bf{LHC diphoton signal strength:}} Due to the presence of the interaction between the SM Higgs $h$ and the triplet scalar $T$ in Eq.(\ref{p1int}), the charged triplet scalar $T^{\pm}$ can contribute significantly to $h\rightarrow \gamma\g$ at one loop. The Higgs to diphoton signal strength can be written as
\bea
\mu_{\g\g}&=&\frac{\sigma(gg\rightarrow h\rightarrow \g\g)_{\rm{triplet}}}{\sigma(gg\rightarrow h\rightarrow \g\g)_{\rm{SM}}}\simeq \frac{Br(h\rightarrow \g\g)_{\rm{triplet}}}{Br(h\rightarrow \g\g)_{\rm{SM}}} .
\label{hgg1}   
\eea
\bea
\frac{Br(h\rightarrow \g\g)_{\rm{triplet}}}{Br(h\rightarrow \g\g)_{\rm{SM}}}&=&\frac{\Gamma(h\rightarrow \g\g)_{\rm{triplet}}}{\Gamma(h)_{\rm{triplet}}}\times \frac{ \Gamma(h)_{\rm{SM}}}{\Gamma(h\rightarrow \g\g)_{\rm{SM}}}.
\label{hgg2}   
\eea  
Now when triplet is heavier than $m_h/2$, we can further write 
\bea
\mu_{\g\g}&=&\frac{\Gamma(h\rightarrow \g\g)_{\rm{     triplet}}}{\Gamma(h\rightarrow \g\g)_{\rm{SM}}} .
\label{hgg3}   
\eea 
The analytic expression of $\Gamma(h\rightarrow \g\g)_{\rm{triplet}}$ can be expressed as \cite{Ayazi:2014tha}

\bea
\G(h\rightarrow \g\g)&=&\frac{G_f \a^2 m_h^3}{128\sqrt{2}\pi^3}\bigg{|}\frac{4}{3}\mathcal{A}_{1/2}(x_i)+\mathcal{A}_{1}(x_i)+2v\l_{HT}\frac{gM_W}{c_{W}^2m_{T^{\pm}}^2}\mathcal{A}_{0}(x_i)\bigg{|}^2 .
\label{hgg4}   
\eea 
where $G_f$, is the Fermi constant. The form factors $\mathcal{A}_{1/2}(x_i),\mathcal{A}_{1}(x_i)~\rm{and}~\mathcal{A}_{0}(x_i),$ are induced by top quark, $W$ gauge boson and $T^{\pm}$ loop respectively. The formula for the form factors are listed below.

\besub
\bea
\mathcal{A}_{1/2}(x_i)&=& 2[x_i+(x_i-1)f(x_i)]x_i^{-2}, \\ 
\mathcal{A}_{1}(x_i)&=& -[3x_i+2x_i^2+3(2x_i-1)f(x_i)]x_i^{-2}, \\ 
\mathcal{A}_{0}(x_i)&=& -[x_i-f(x_i)]x_i^{-2}.  
\eea 
\eesub
 where $x_i=\frac{m_h^2}{4m_i^2}$ and $f(x)= (\sin^{-1}\sqrt{x})^2$.\\

In order to ensure that $\mu_{\g\g}$ lies within the experimental uncertainties, the analysis should respect the latest signal 
strength from ATLAS \cite{Aaboud:2018xdt} and CMS \cite{Sirunyan:2018koj}. The measured value of $\mu_{\g\g}$ are given by $\mu_{\g\g}=0.99\pm 0.14$ from ATLAS and $\mu_{\g\g}=1.17\pm 0.10$ from CMS. 
 
 \item[(iv)]{\bf{Disappearing charged track:}}  Despite having an analogous spectrum to the inert Higgs doublet multiplet, 
 the inert Higgs triplet model can not produce similar sort of collider signals as associated with inert Higgs doublet. For example, 
 due to the presence of very small mass splitting among the charged and neutral components, it becomes difficult to look for the collider signals such as di-lepton or multi-leptons plus missing energies in case of inert triplet. Hence, one needs to look for different types of collider signals. The involvement of a charged component ($T^{\pm}$) of the triplet scalar in the present scenario provides an interesting discovery prospect at LHC. When produced in pp collisions, the charged triplet scalar can only 
decays to the neutral component and a soft pion or soft lepton pair (due to the presence of small mass splitting, 166 MeV), after getting produced these soft pions yields a disappearing charged track in the detector. Recently in \cite{Chiang:2020rcv}, it was shown that searches for disappearing
tracks at LHC presently excludes a real triplet scalar lighter than 287 GeV
with $\mathcal{L} = 36~\rm{fb^{-1}}$. The reach can extend to 608 GeV and 761 GeV with the collection
of $\mathcal{L} = 300~\rm{fb^{-1}}$ and $3000~\rm{fb^{-1}}$ respectively. 

\item[(v)]{\bf{Relic density and Direct detection of DM:}} The parameter space of the present model is to be constrained 
by the measured value of the DM relic abundance from the Planck experiment \cite{Aghanim:2018eyx}. 
One can further restrict the parameter space by applying bounds on the DM direct detection cross-section coming from the experiments like LUX \cite{Akerib:2016vxi}, XENON-1T \cite{Aprile:2018dbl}, PandaX-II \cite{Tan:2016zwf,Cui:2017nnn}. Detailed discussions on the dark matter phenomenology are presented in section \ref{DMPh}.

\item[(vi)]{\textbf{LEP constraints:}} LEP \cite{ALEPH:2004aa} has set constraints on the masses of the charged and the neutral scalars as $m \geq 100$ GeV from the non-observance of any such related events. We can therefore use the same  
as a conservative lower bound for the mass of the charged scalar ($T^{\pm}$) involved in our construction, $m_{T^{\pm}}\geq 100$ GeV. 
In case of scalar triplet, since the mass splitting among the charged and neutral component is of the order of $166$ MeV, the same 
lower bound can be equivalently applied on the mass of the neutral component\cite{FileviezPerez:2008bj}, the DM component.
\end{itemize}


\section{Dark Matter Phenomenology}
\label{DMPh}

\begin{figure}[H]
\centering
\subfigure[]{
\includegraphics[scale=0.25]{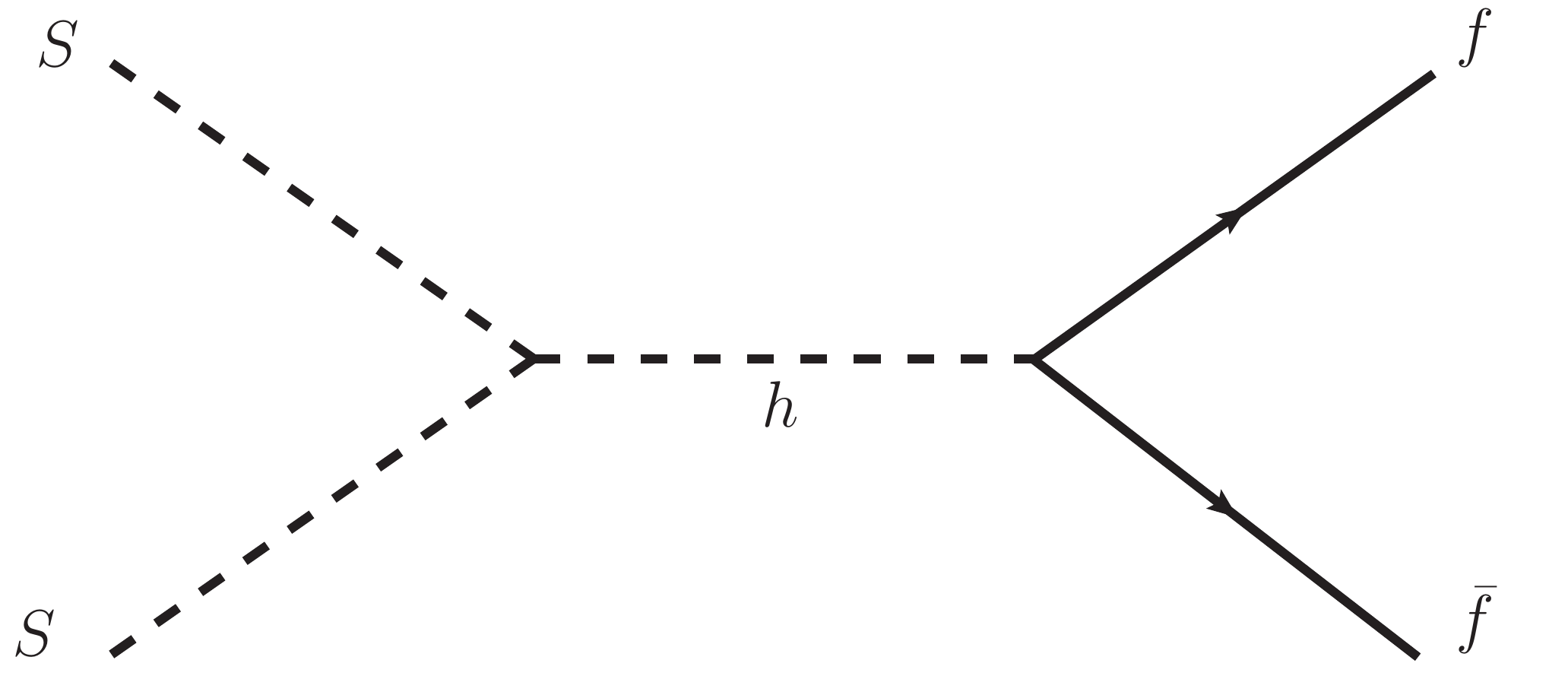}}
\subfigure[]{
\includegraphics[scale=0.25]{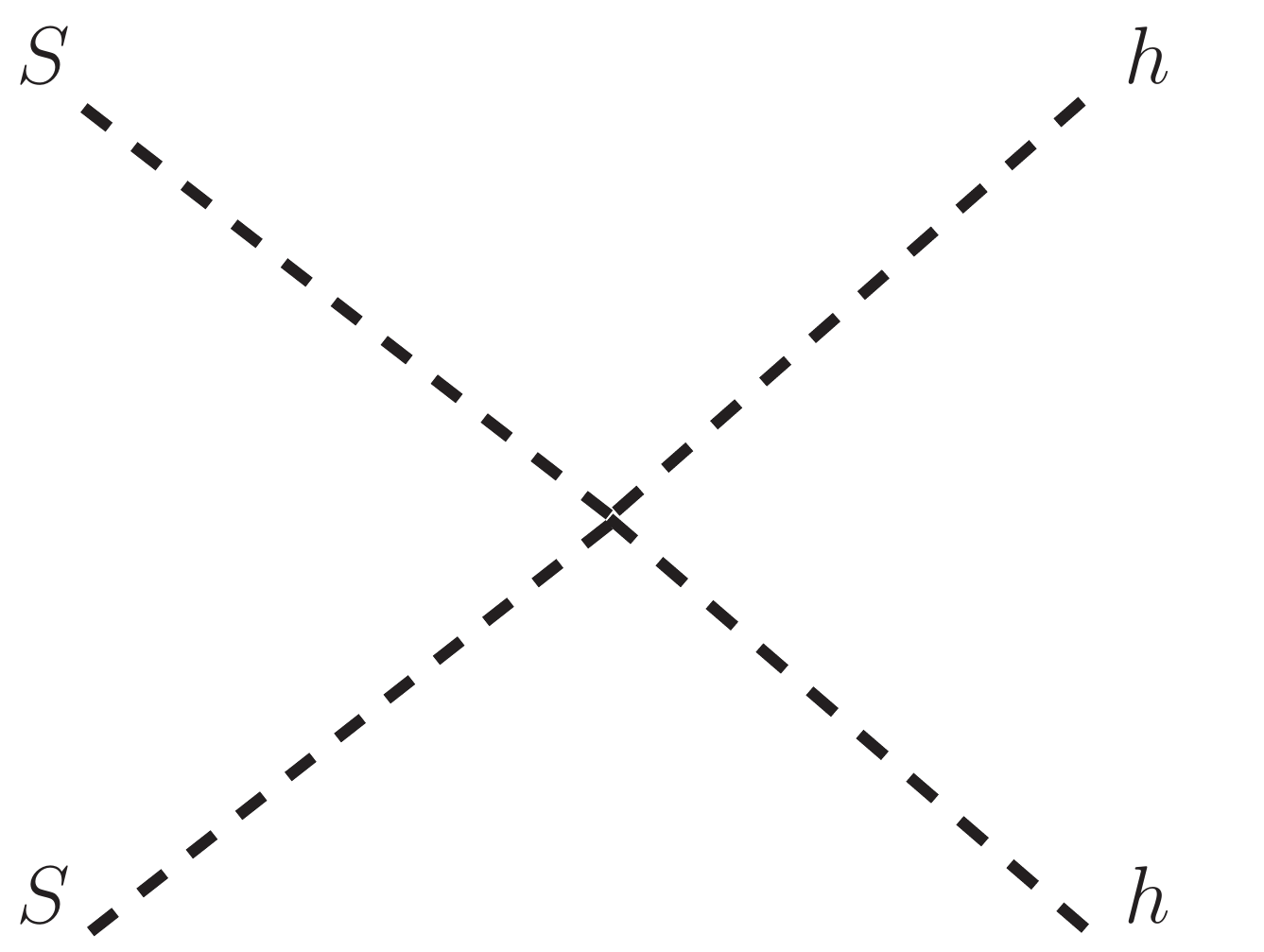}}
\subfigure[]{
\includegraphics[scale=0.25]{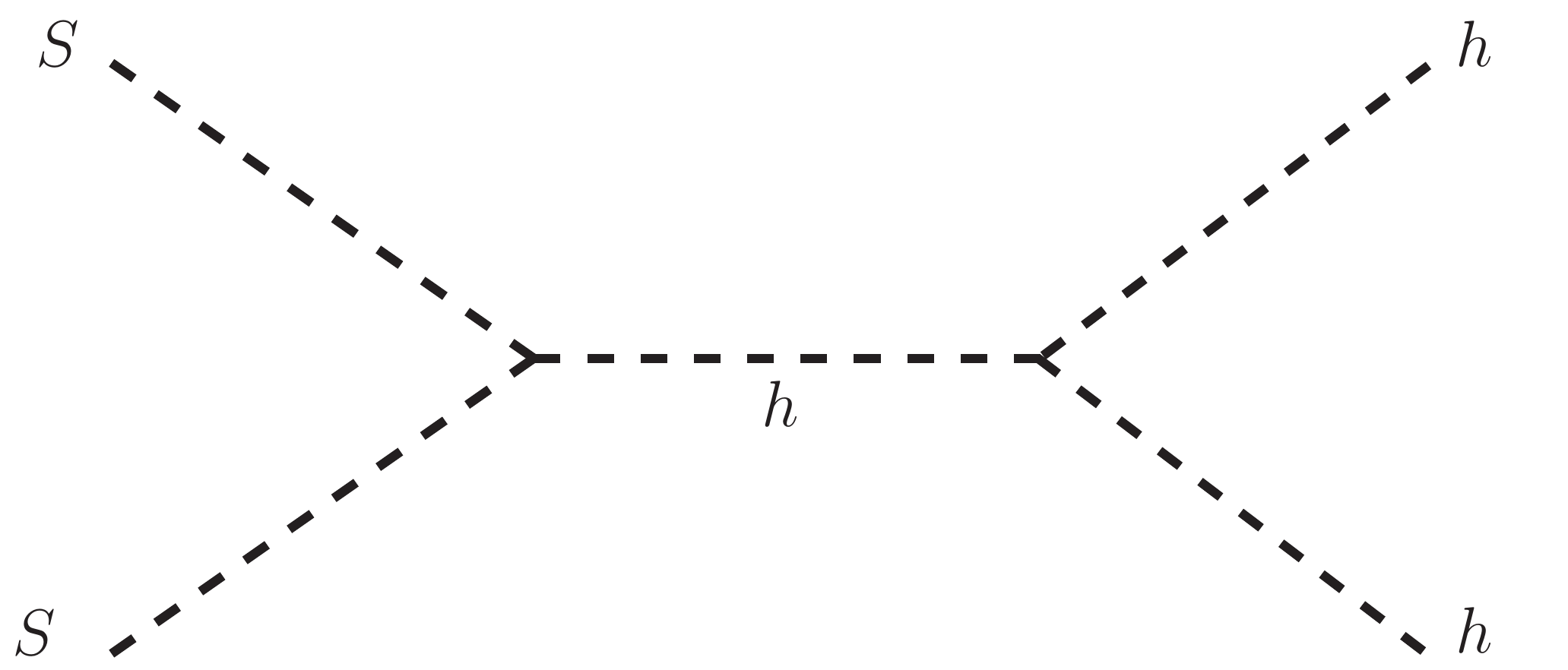}}
\subfigure[]{
\includegraphics[scale=0.25]{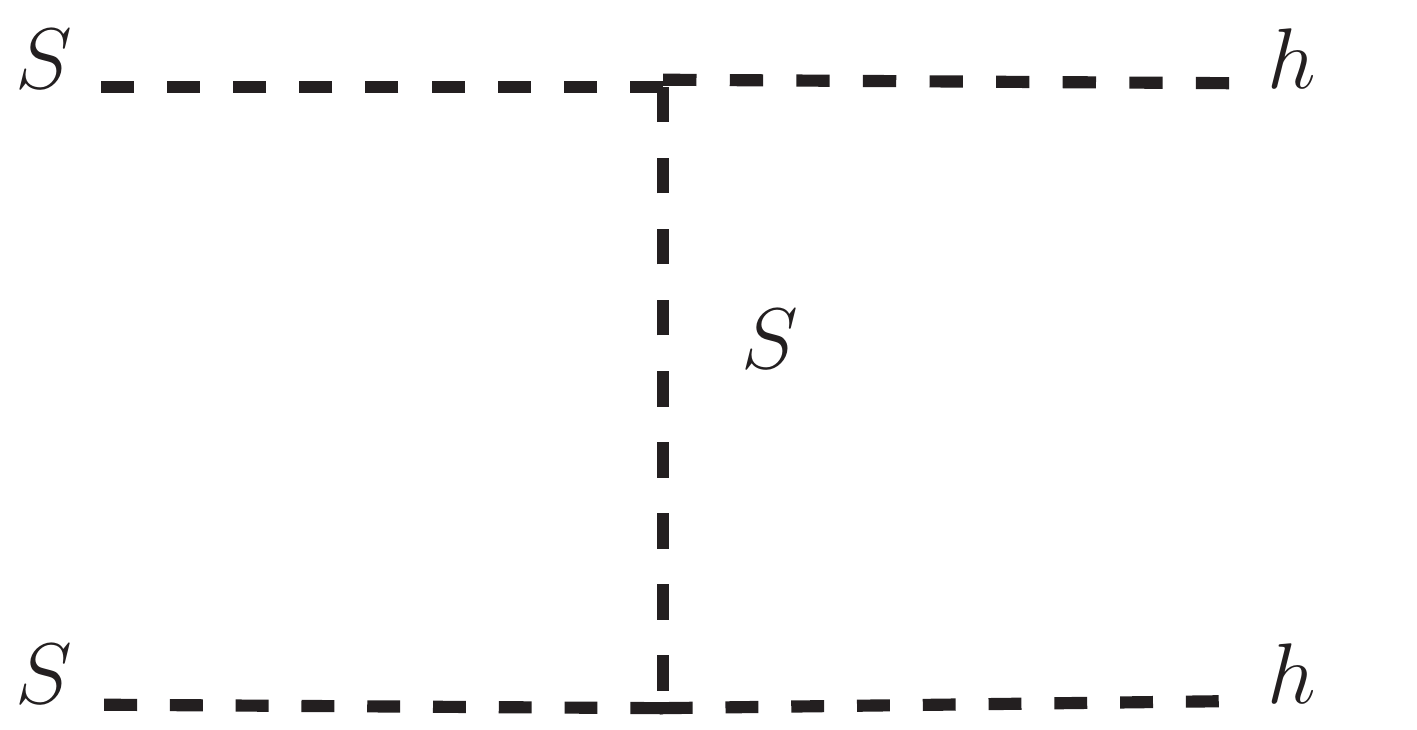}}
\subfigure[]{
\includegraphics[scale=0.25]{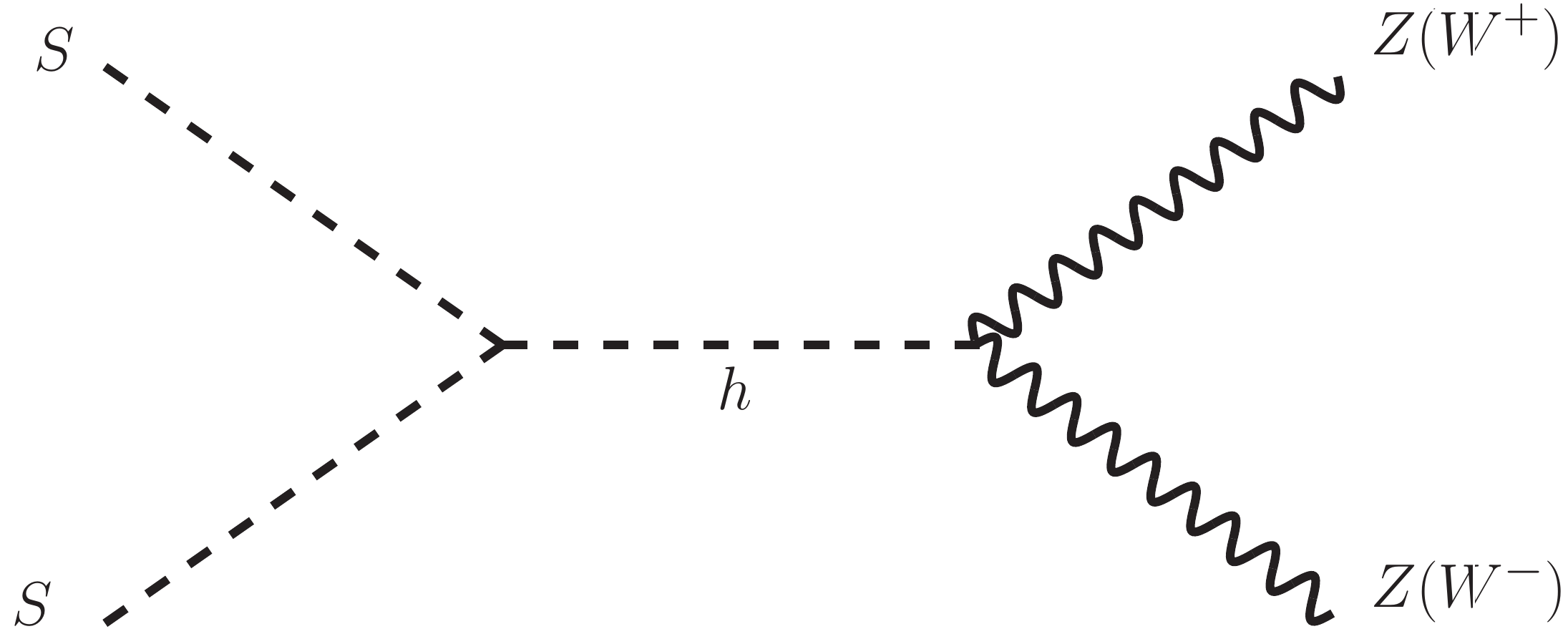}}
\caption{Annihilation channels for scalar singlet dark matter $S$.}
\label{feynS}
\end{figure}

\begin{figure}[H]
\centering
\subfigure[]{
\includegraphics[scale=0.20]{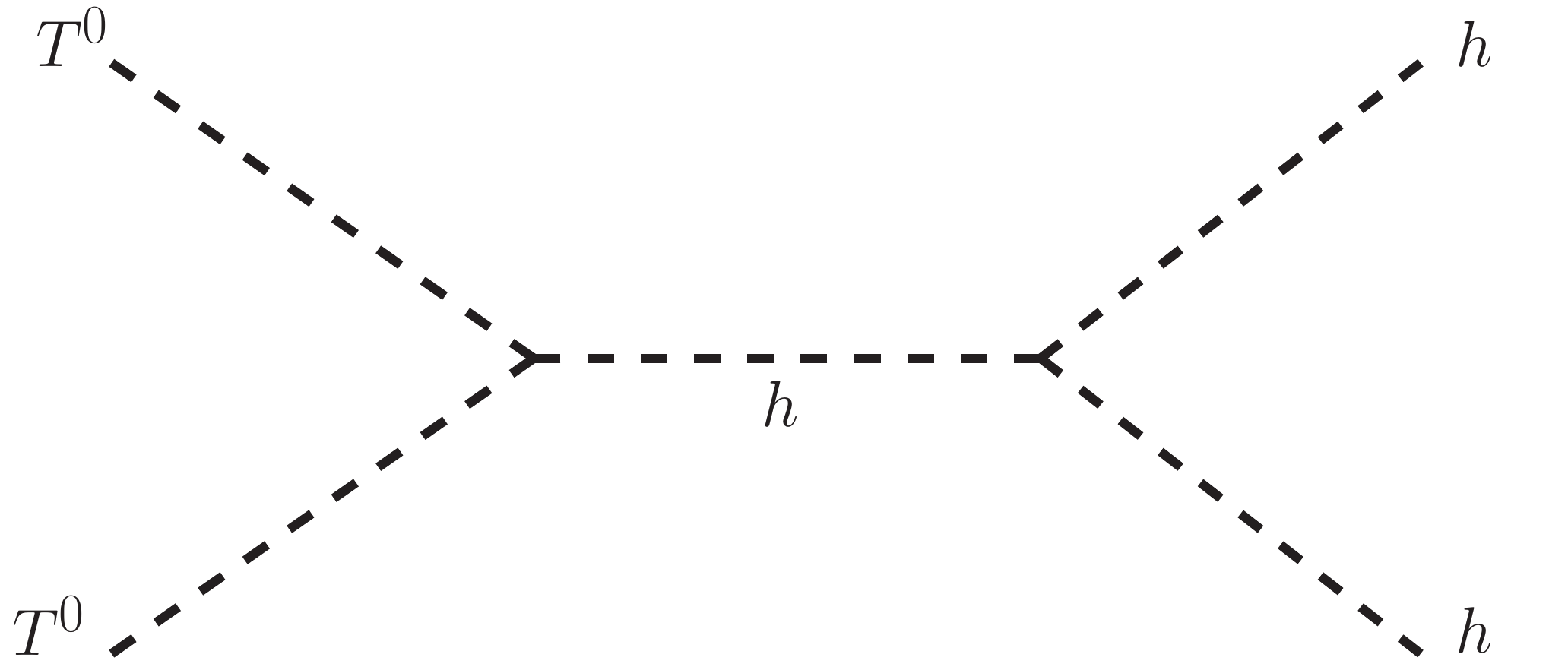}}
\subfigure[]{
\includegraphics[scale=0.20]{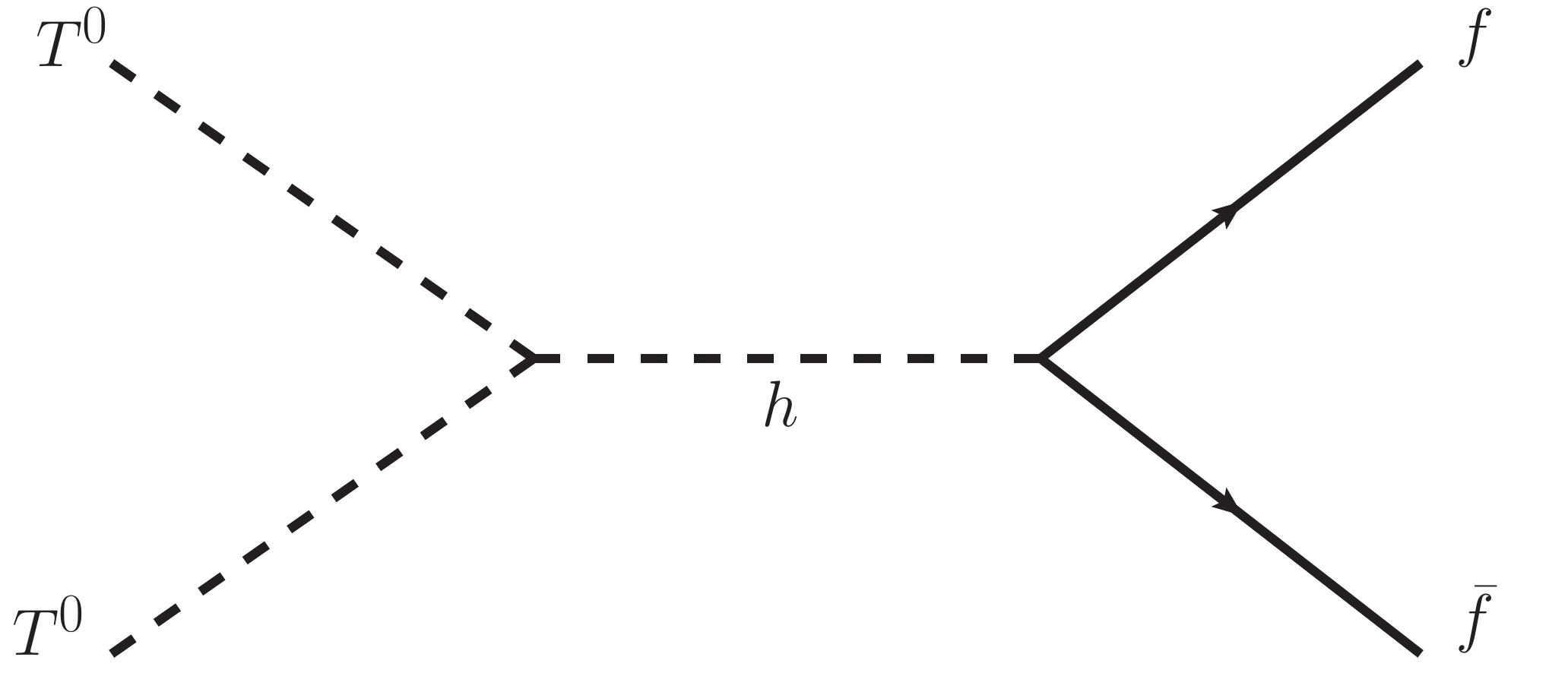}}
\subfigure[]{
\includegraphics[scale=0.20]{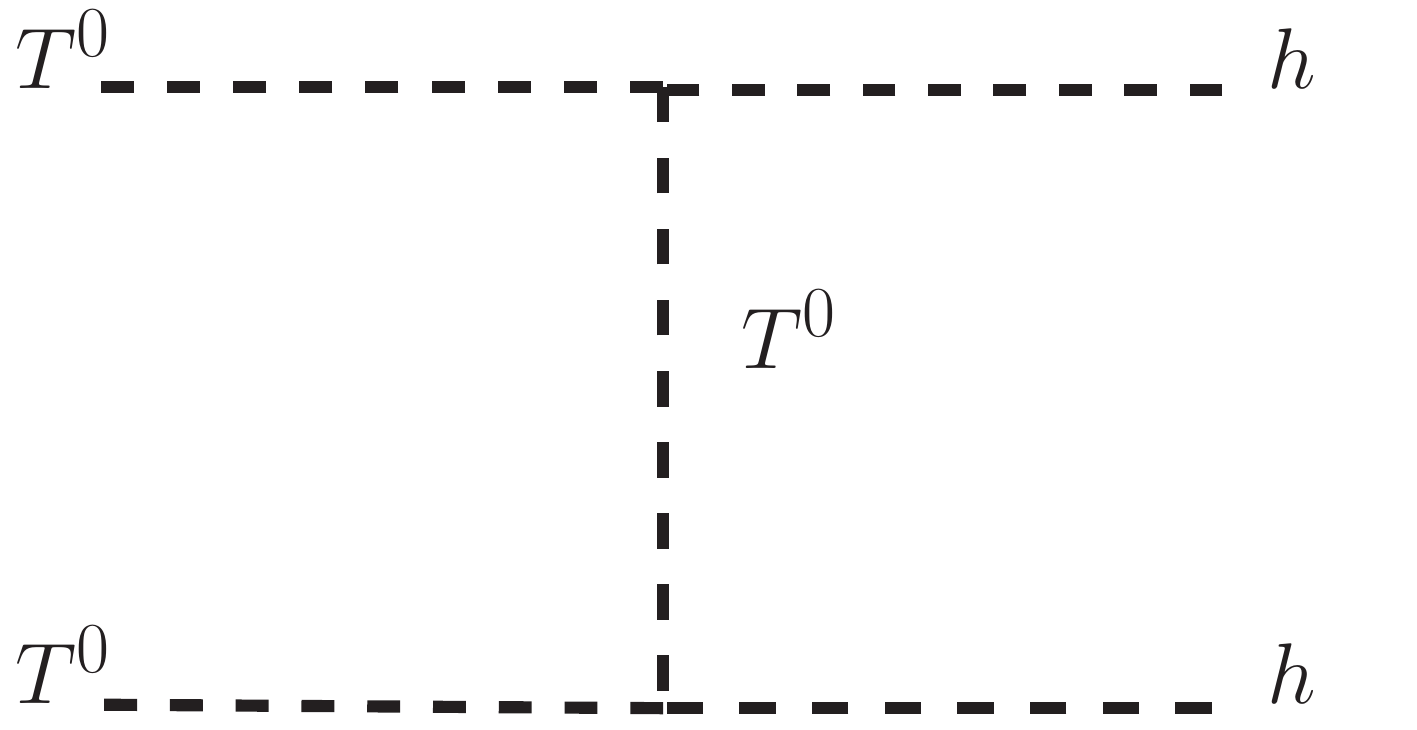}}
\subfigure[]{
\includegraphics[scale=0.20]{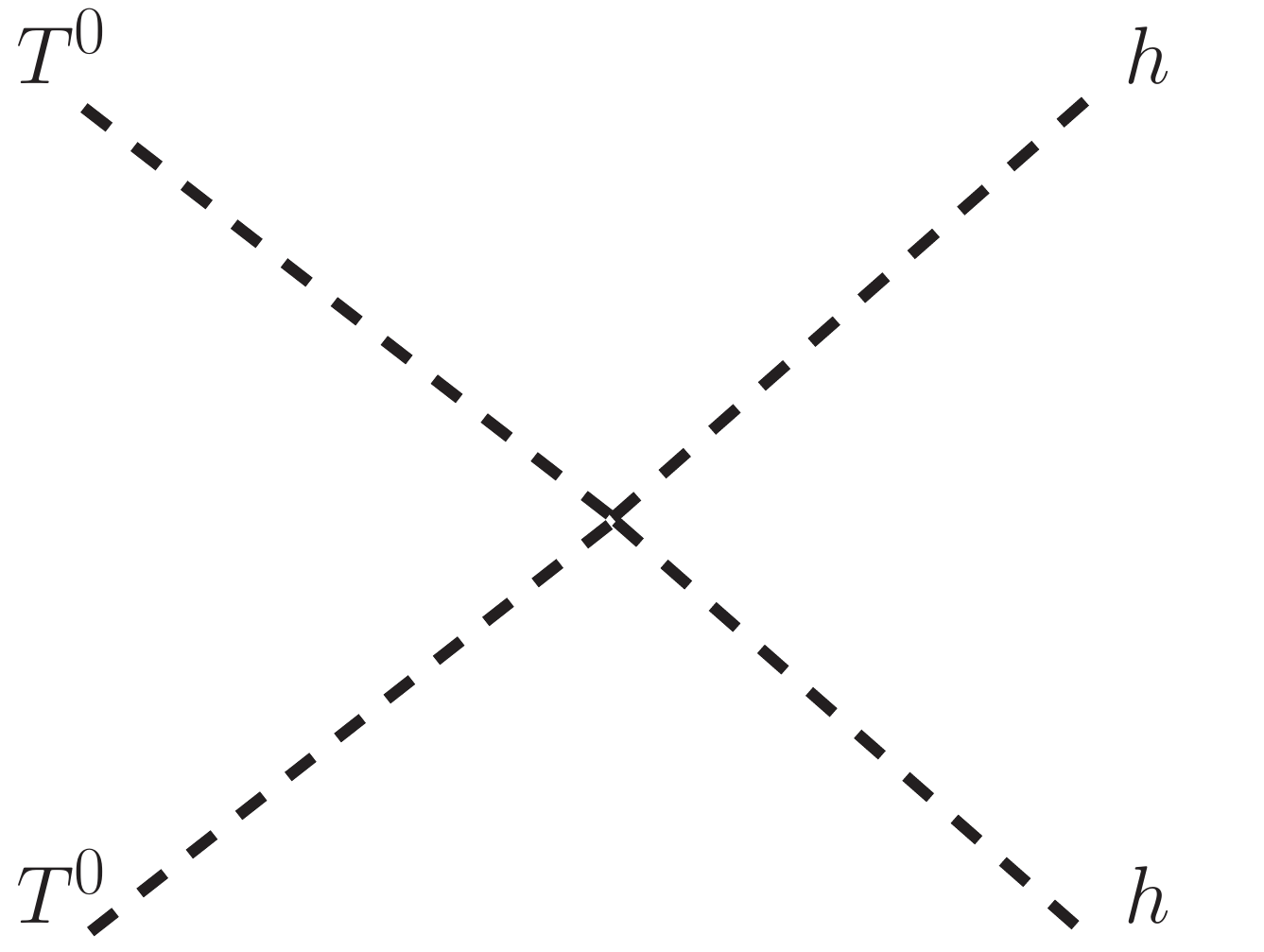}}
\subfigure[]{
\includegraphics[scale=0.20]{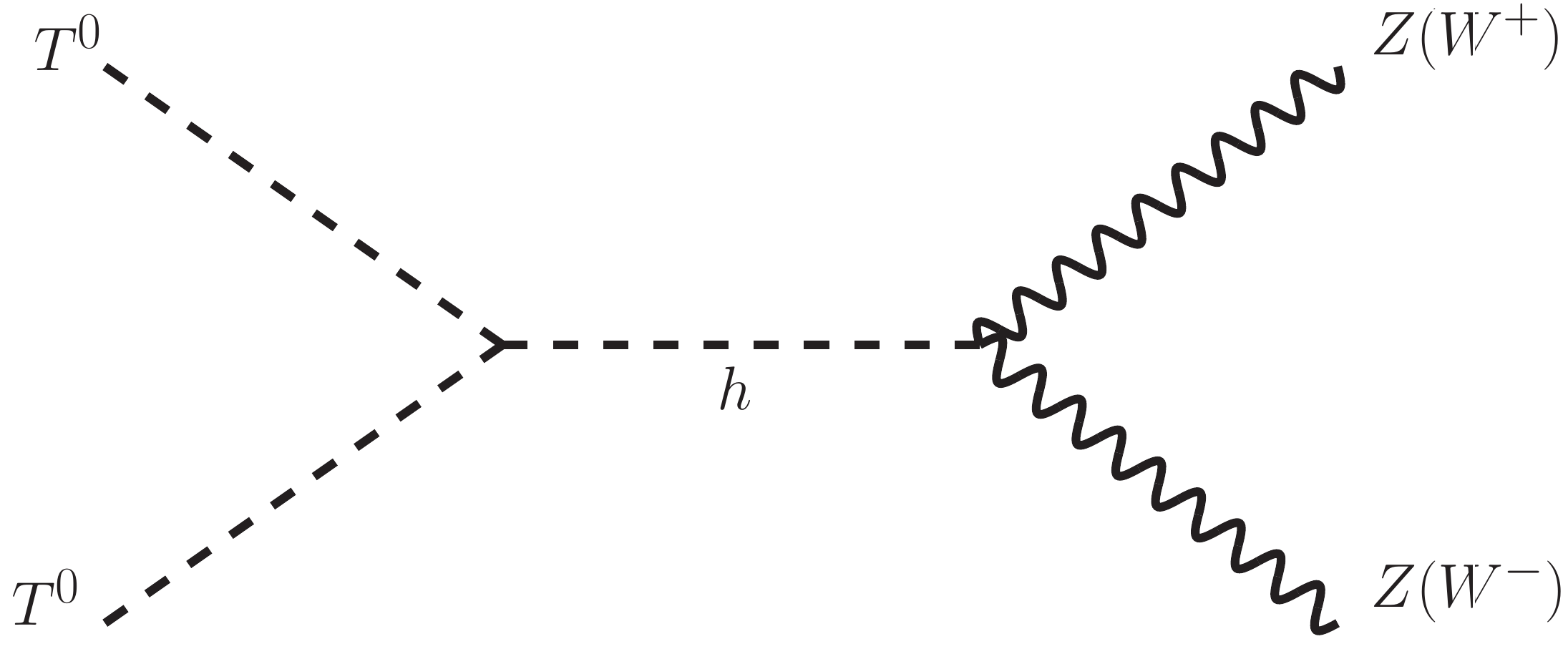}}
\subfigure[]{
\includegraphics[scale=0.20]{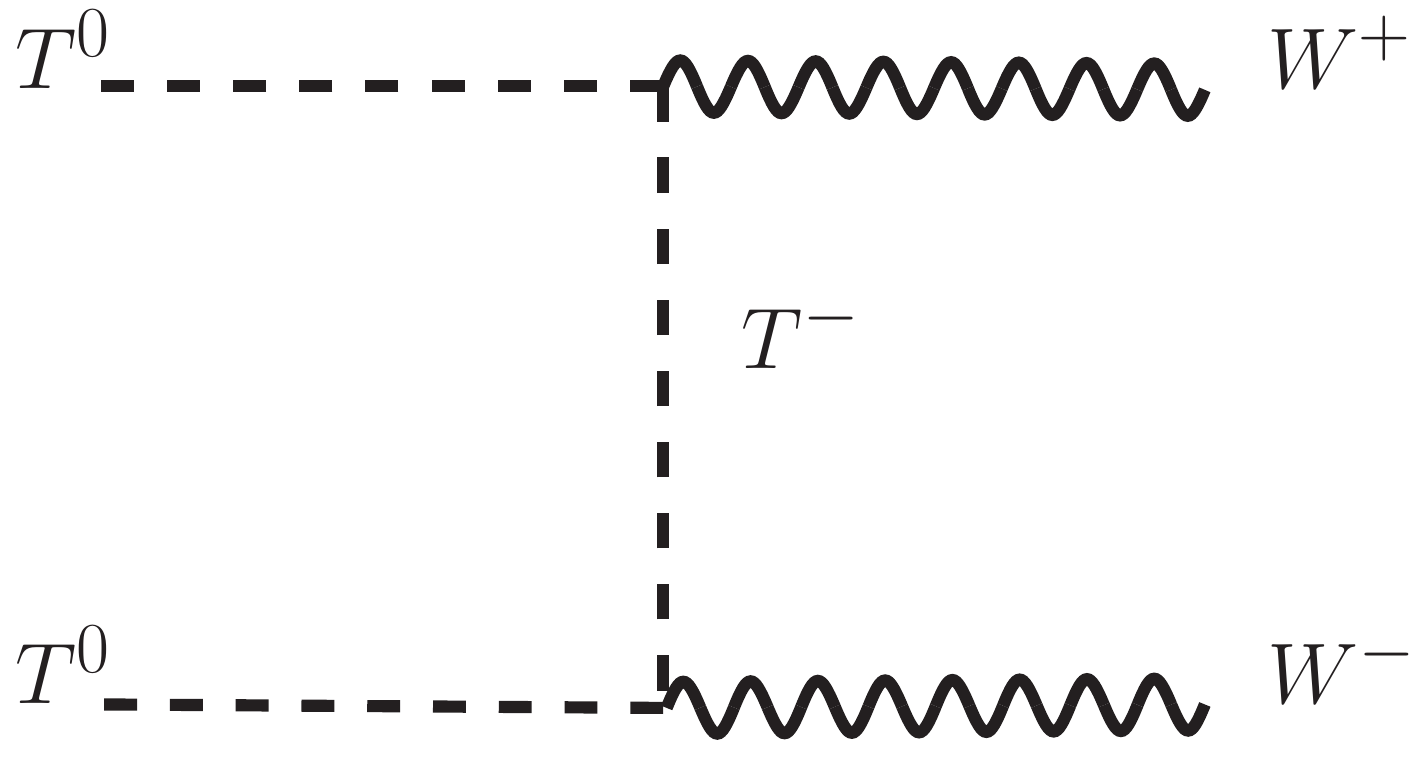}}
\subfigure[]{
\includegraphics[scale=0.20]{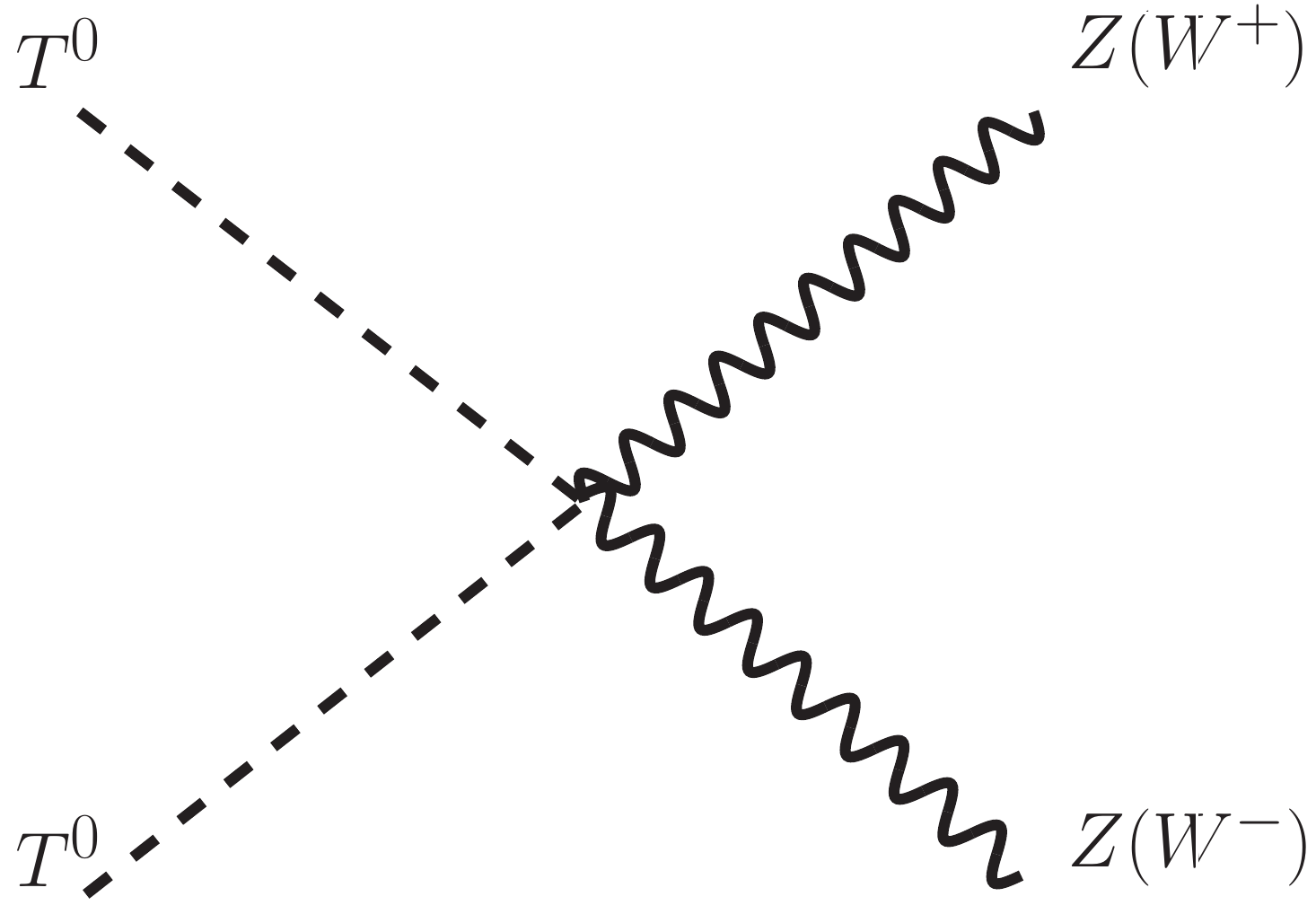}}
\caption{Annihilation channels for triplet scalar dark matter $T^0$.}
\label{feynT}
\end{figure}

\begin{figure}[H]
\centering
\subfigure[]{
\includegraphics[scale=0.22]{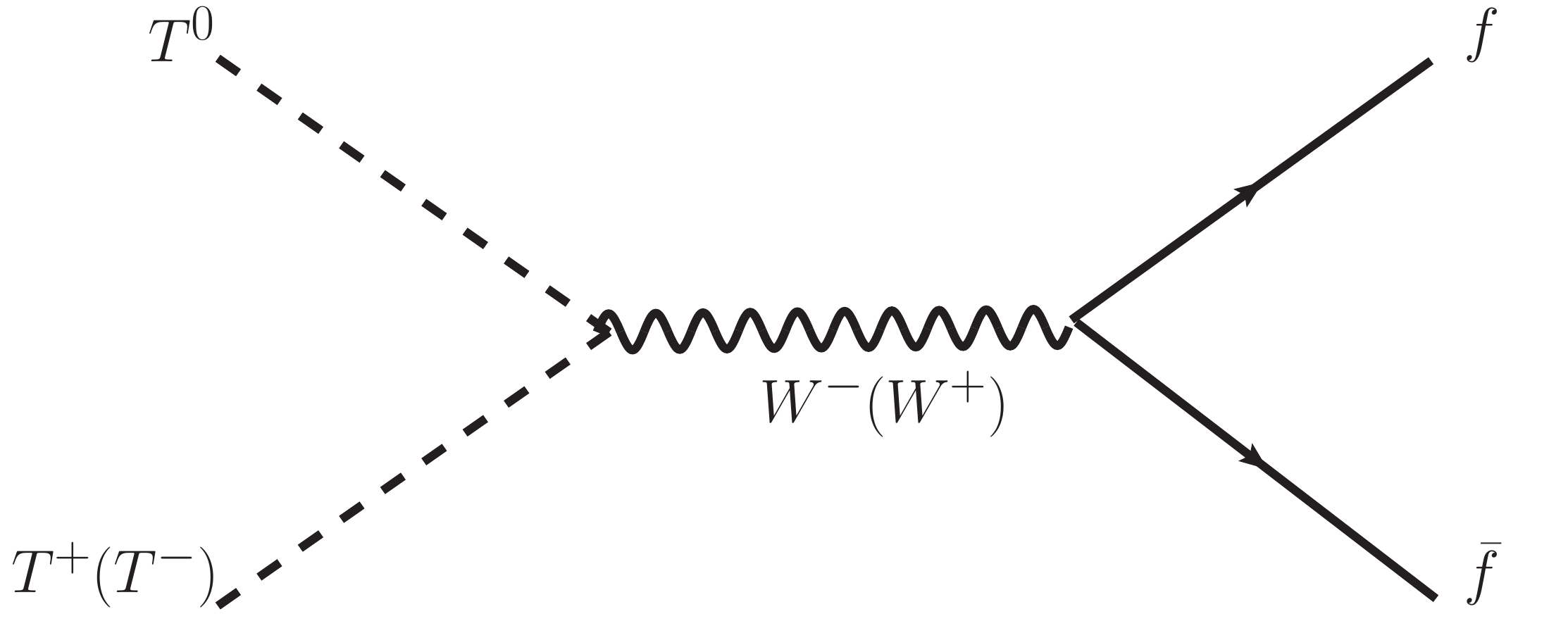}}
\subfigure[]{
\includegraphics[scale=0.22]{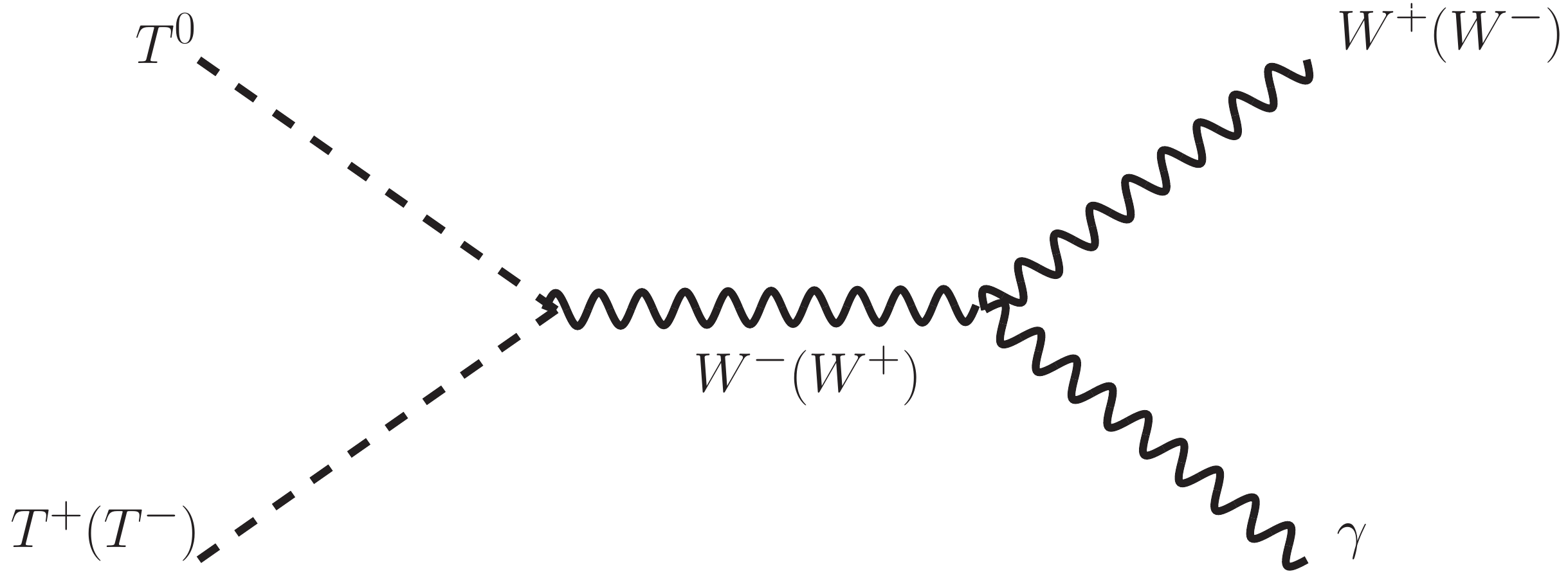}}
\subfigure[]{
\includegraphics[scale=0.22]{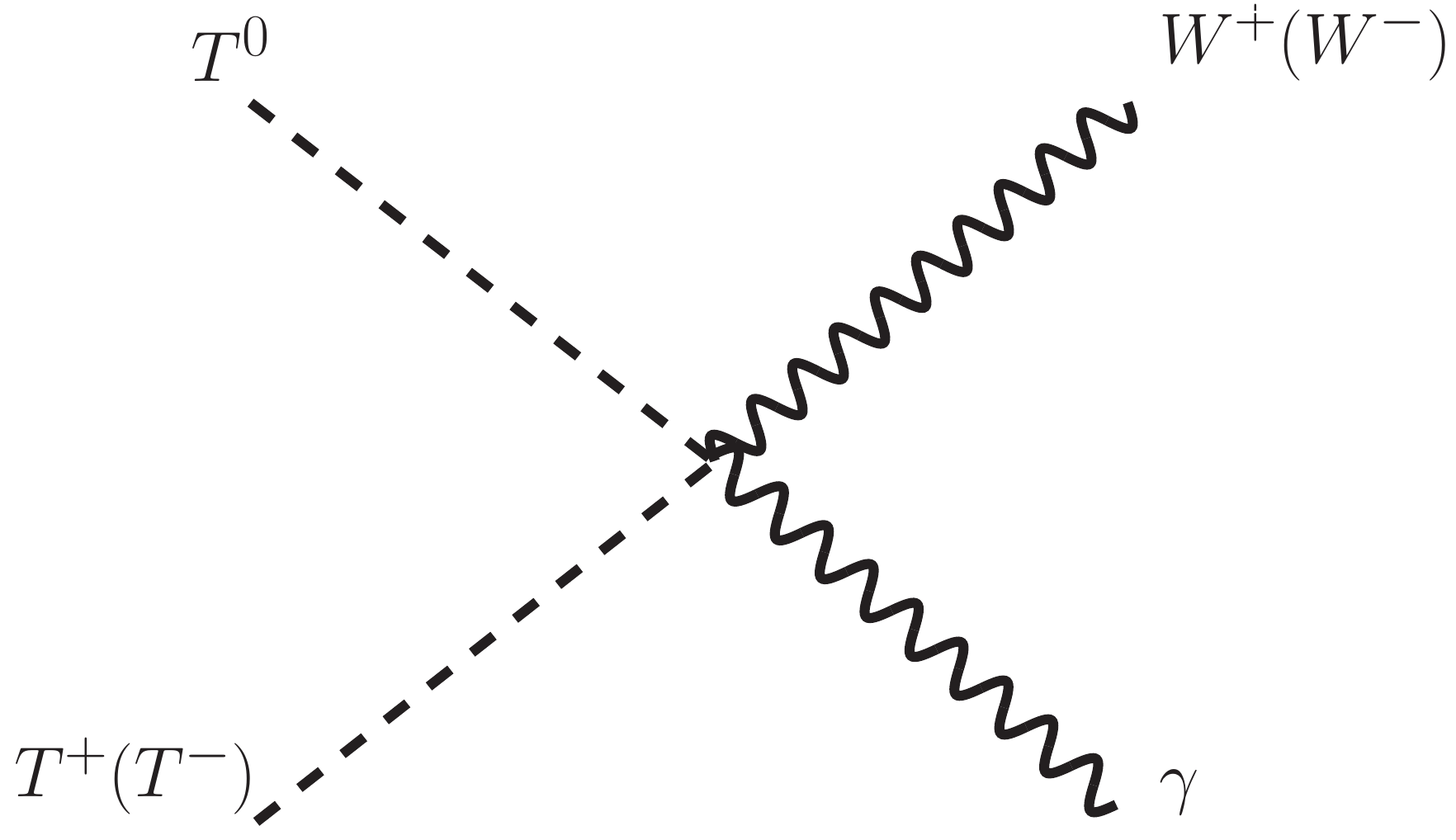}}
\subfigure[]{
\includegraphics[scale=0.22]{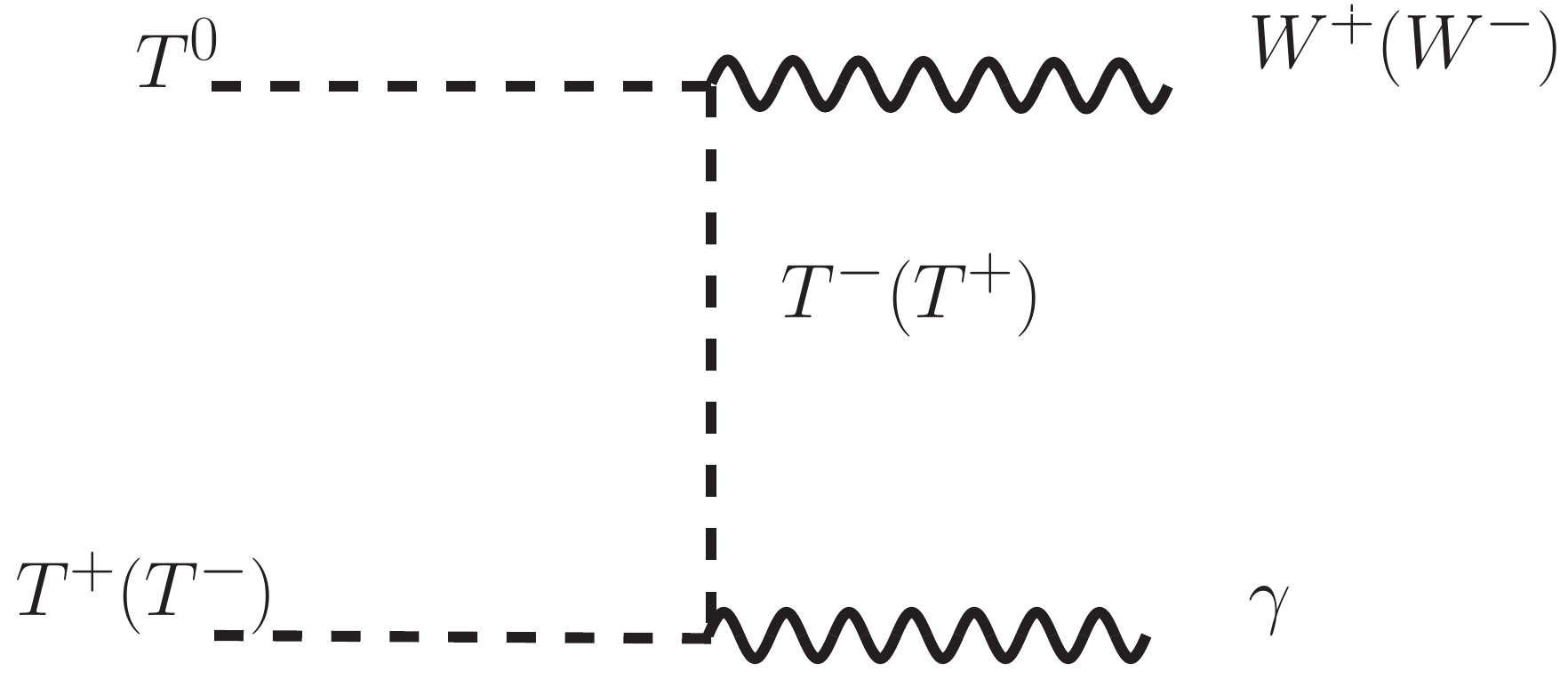}}
\subfigure[]{
\includegraphics[scale=0.22]{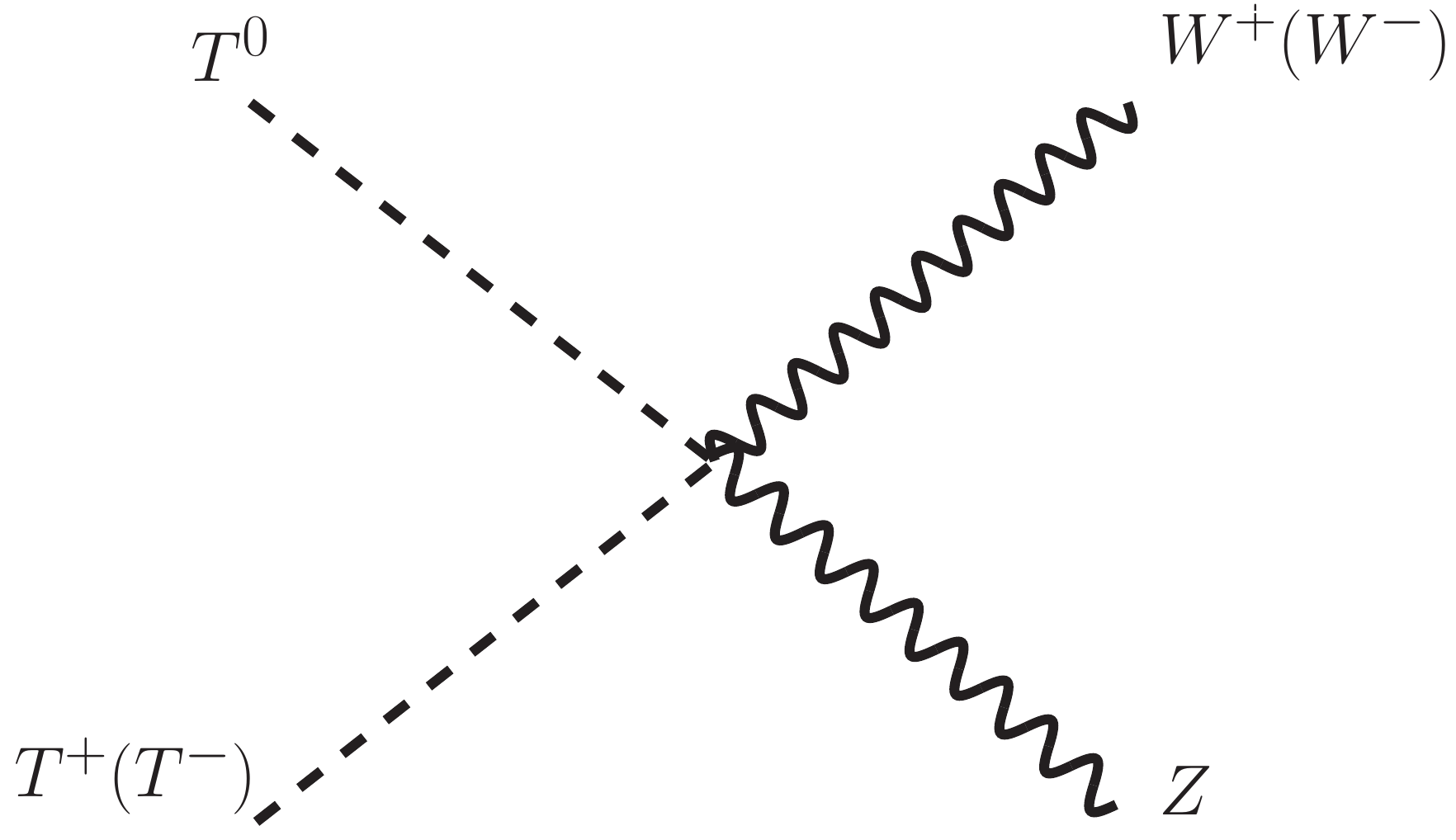}}
\subfigure[]{
\includegraphics[scale=0.22]{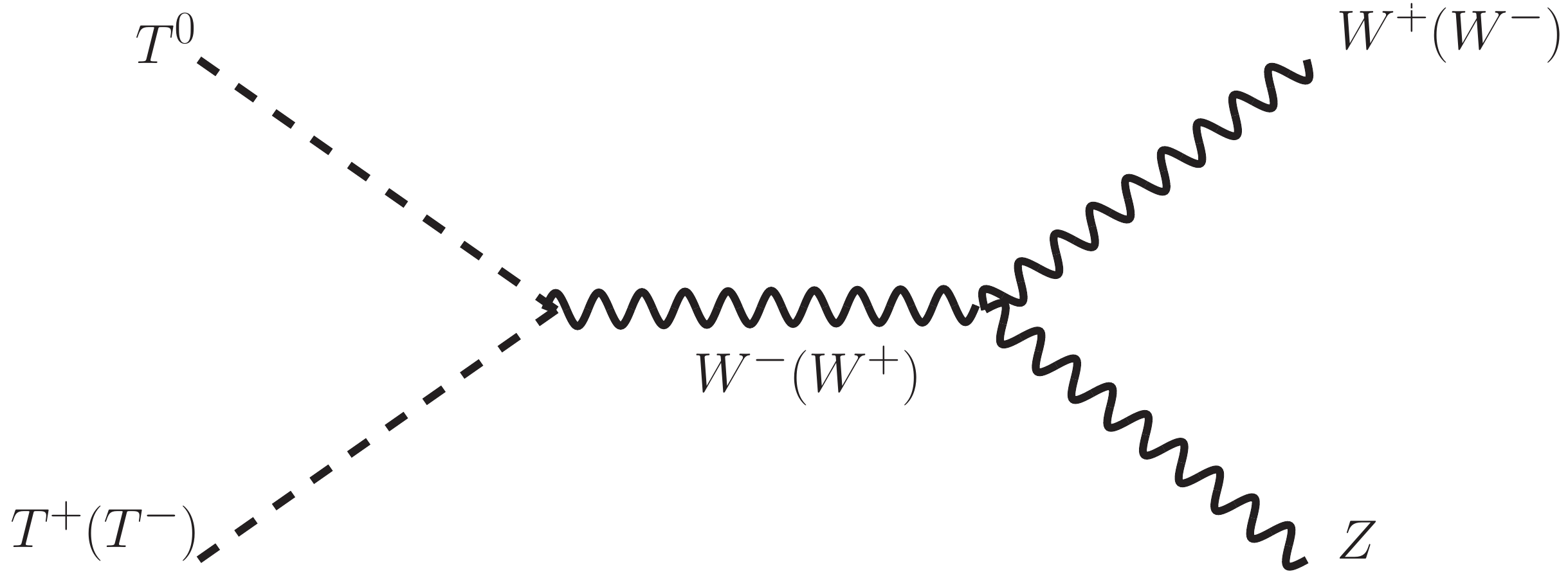}}
\subfigure[]{
\includegraphics[scale=0.22]{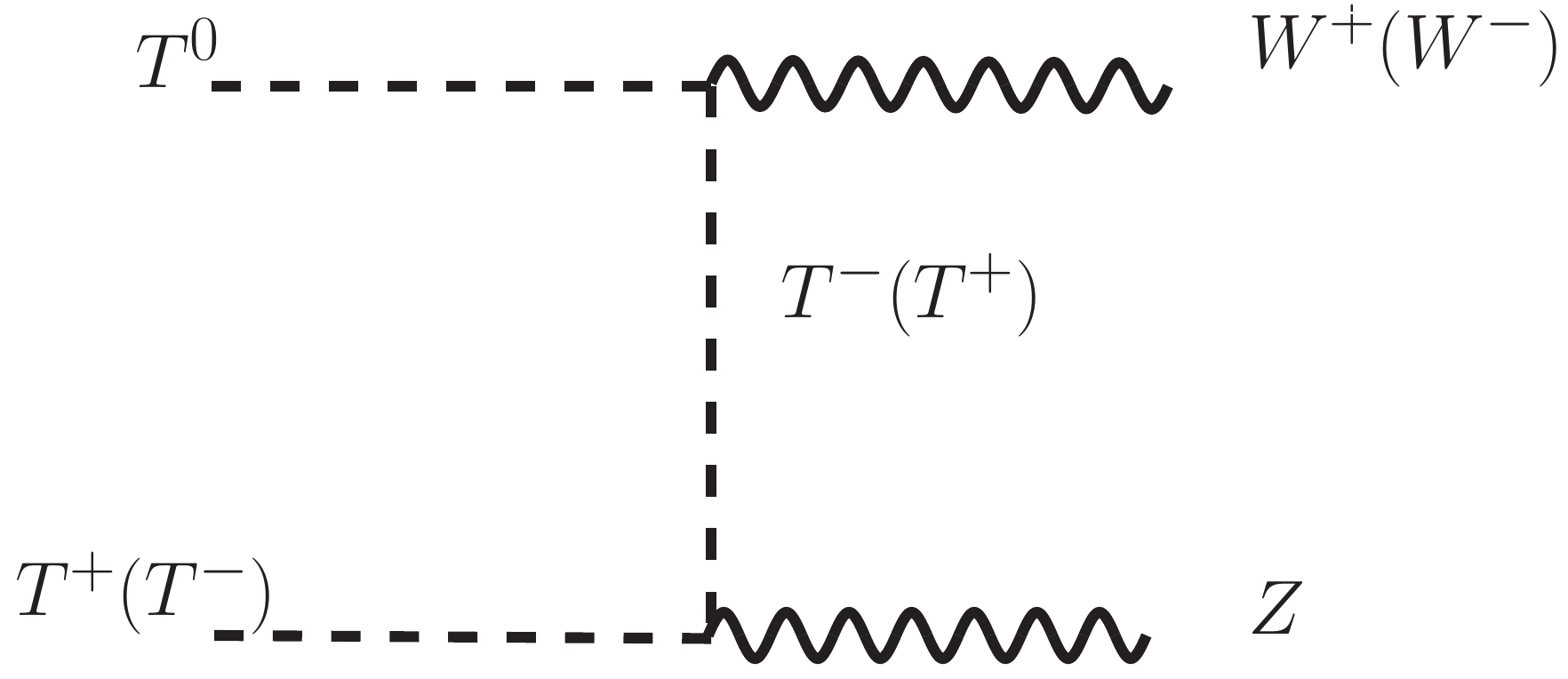}}
\subfigure[]{
\includegraphics[scale=0.22]{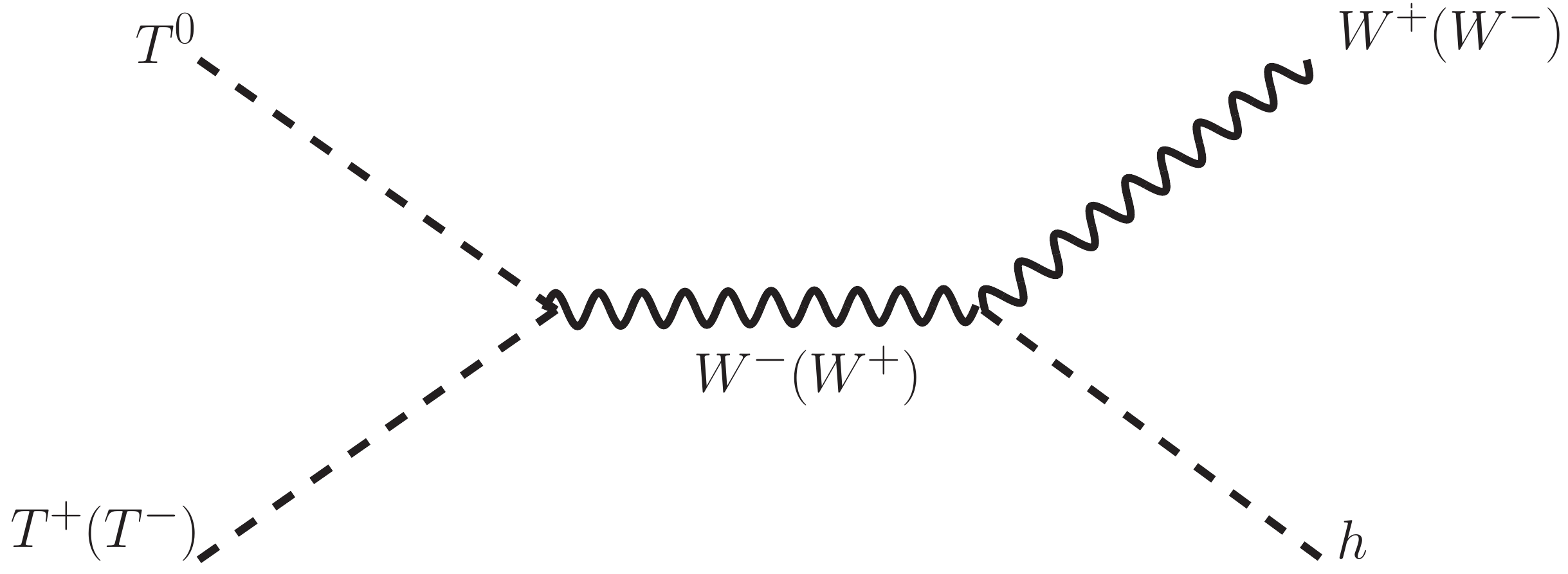}}
\subfigure[]{
\includegraphics[scale=0.22]{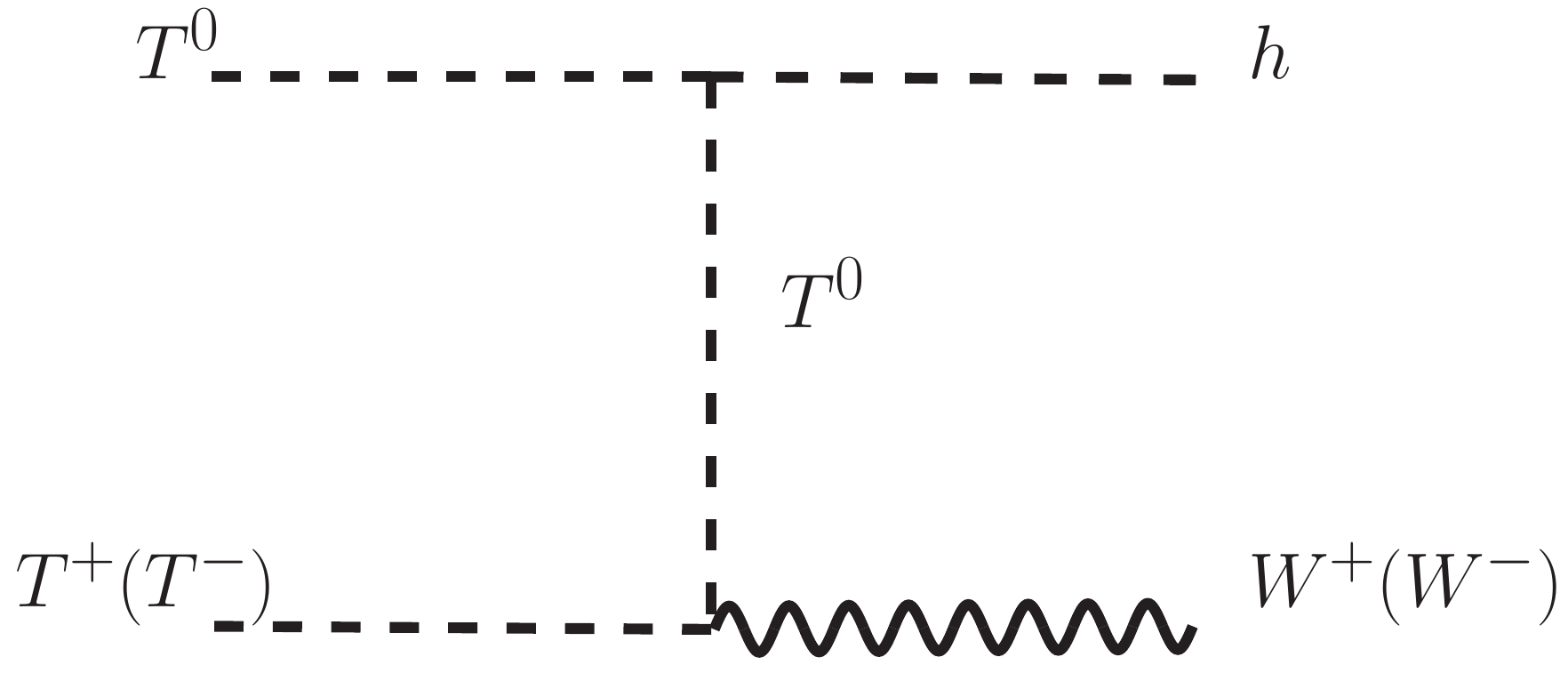}}
\subfigure[]{
\includegraphics[scale=0.2]{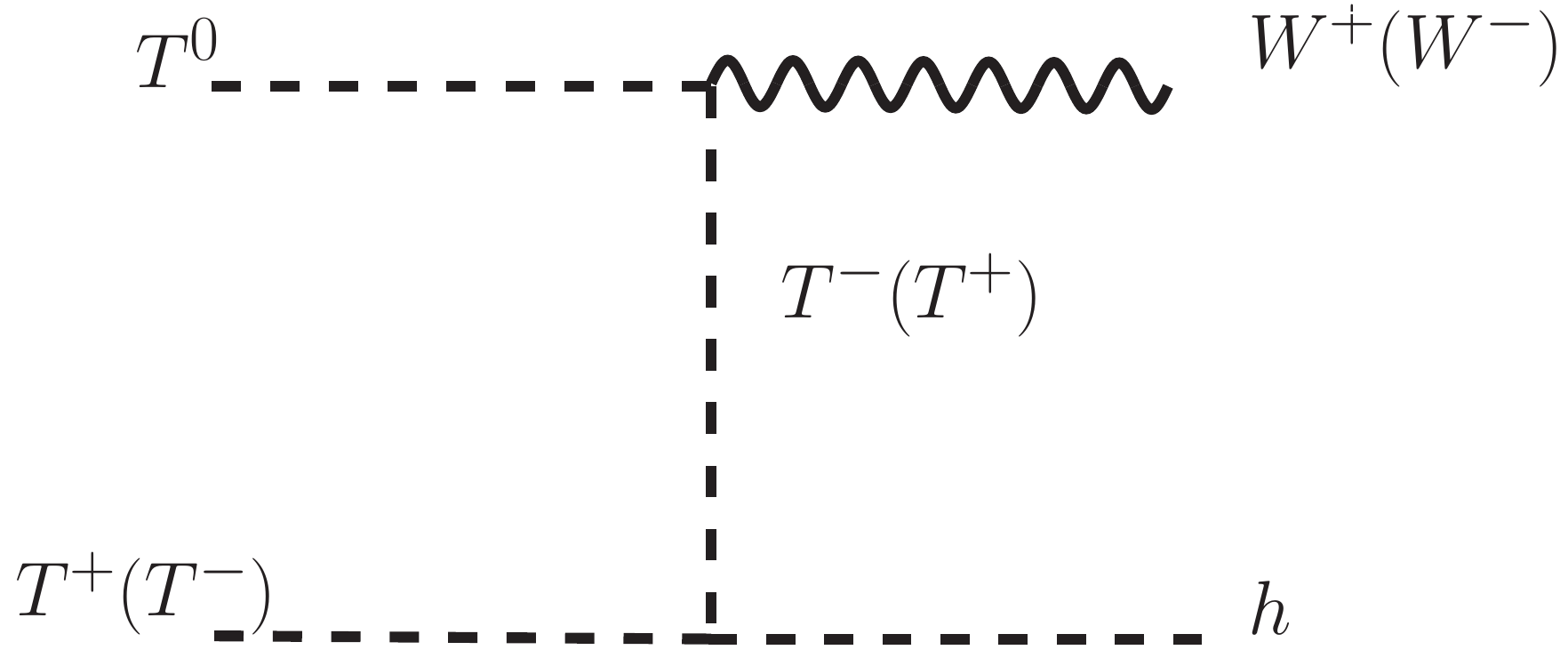}}
\subfigure[]{
\includegraphics[scale=0.22]{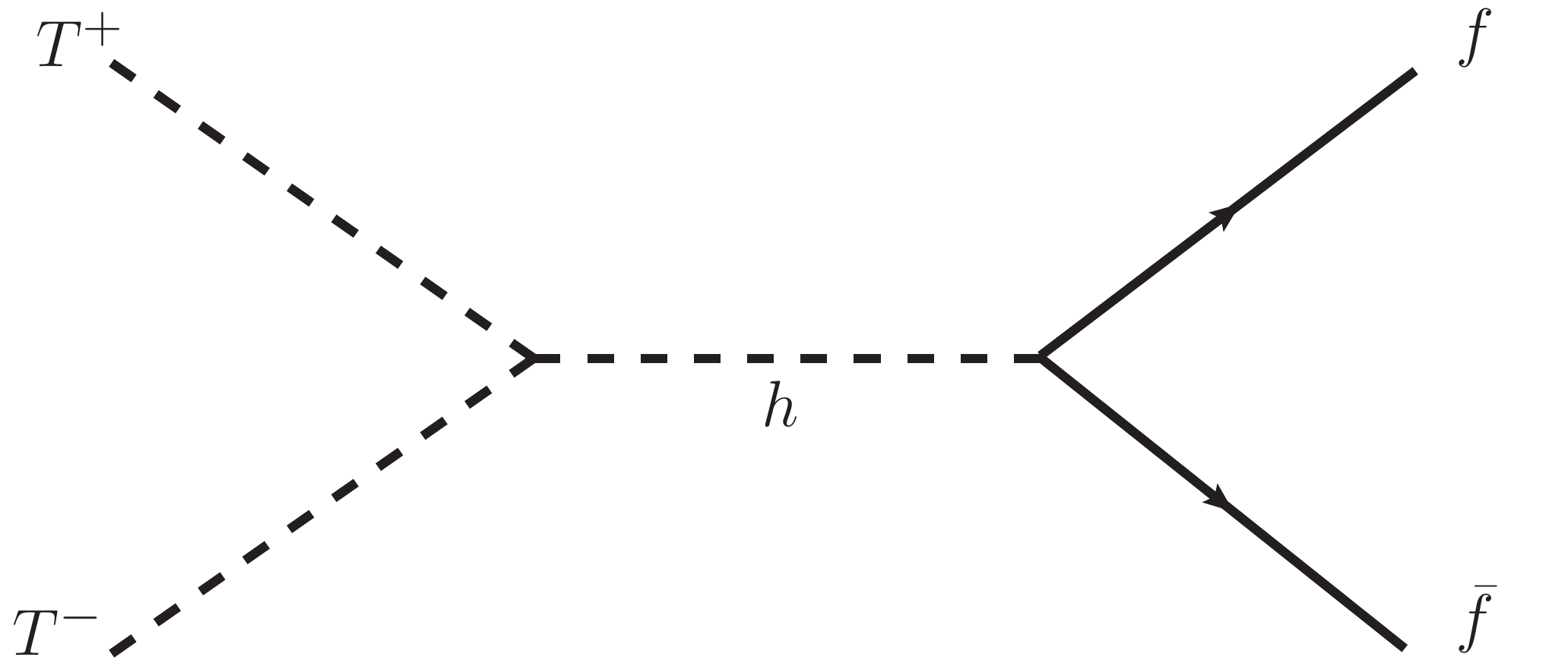}}
\subfigure[]{
\includegraphics[scale=0.22]{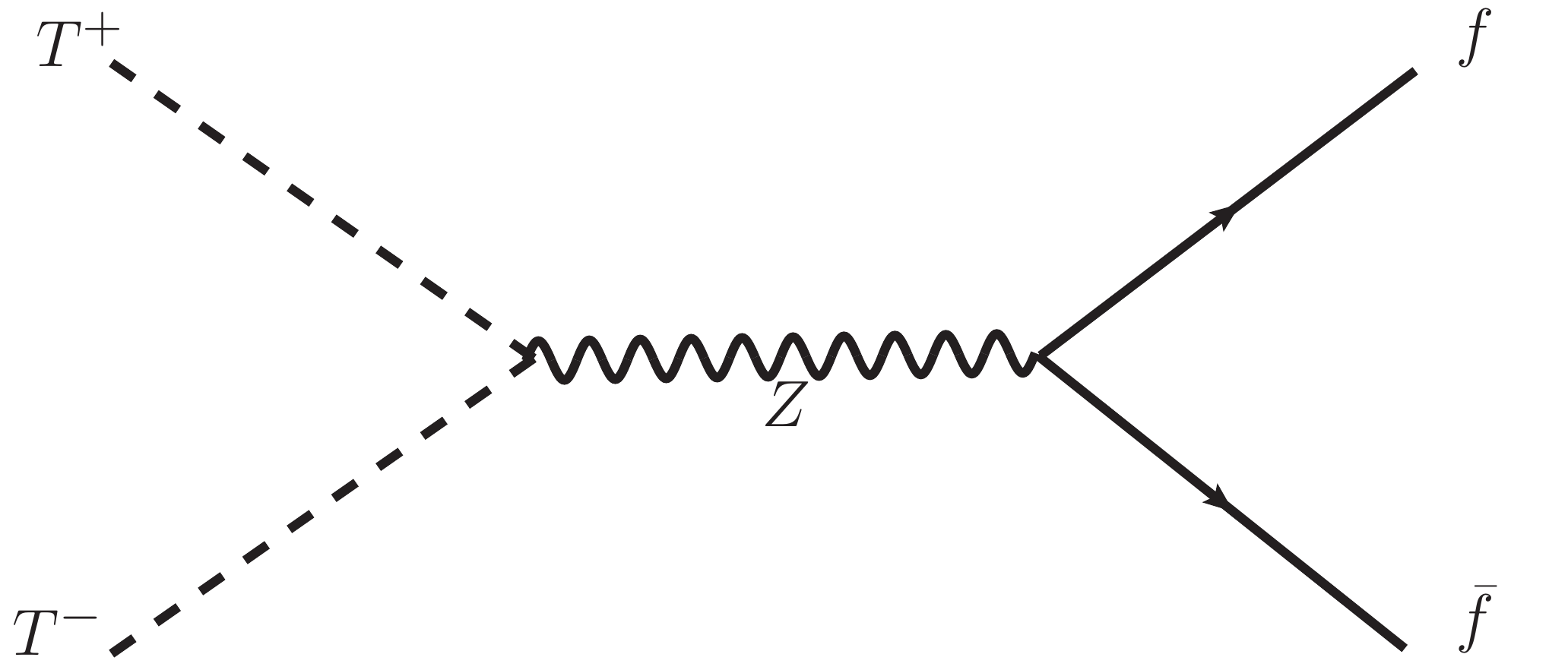}}
\subfigure[]{
\includegraphics[scale=0.22]{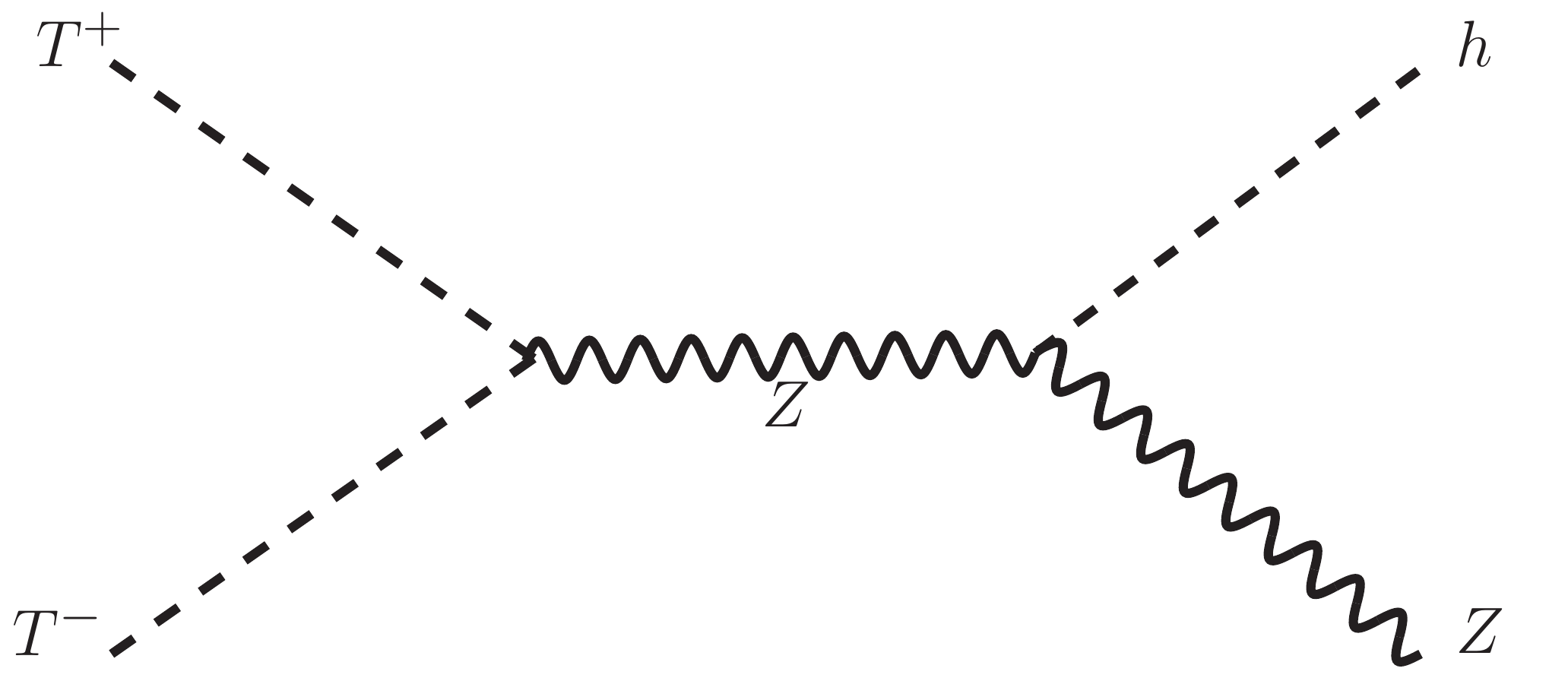}}
\subfigure[]{
\includegraphics[scale=0.22]{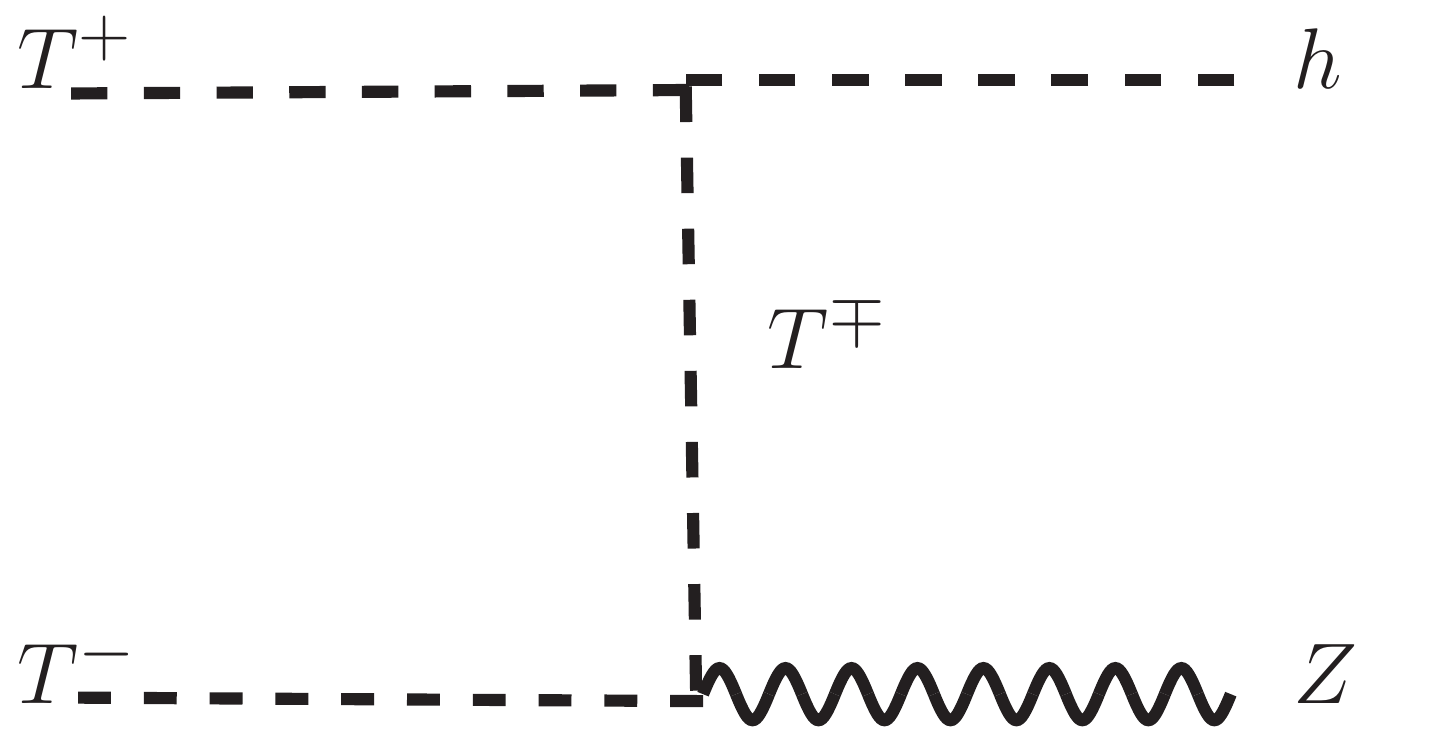}}
\subfigure[]{
\includegraphics[scale=0.22]{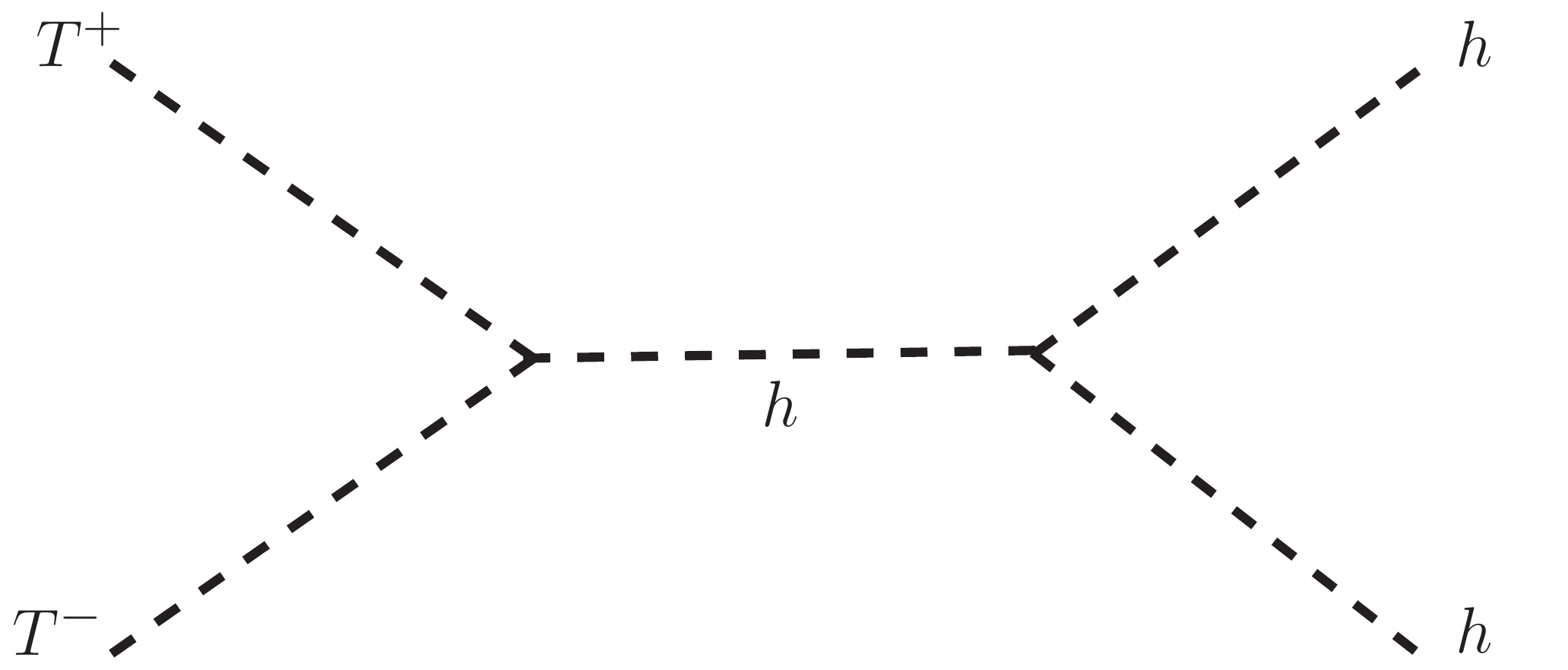}}
\subfigure[]{
\includegraphics[scale=0.22]{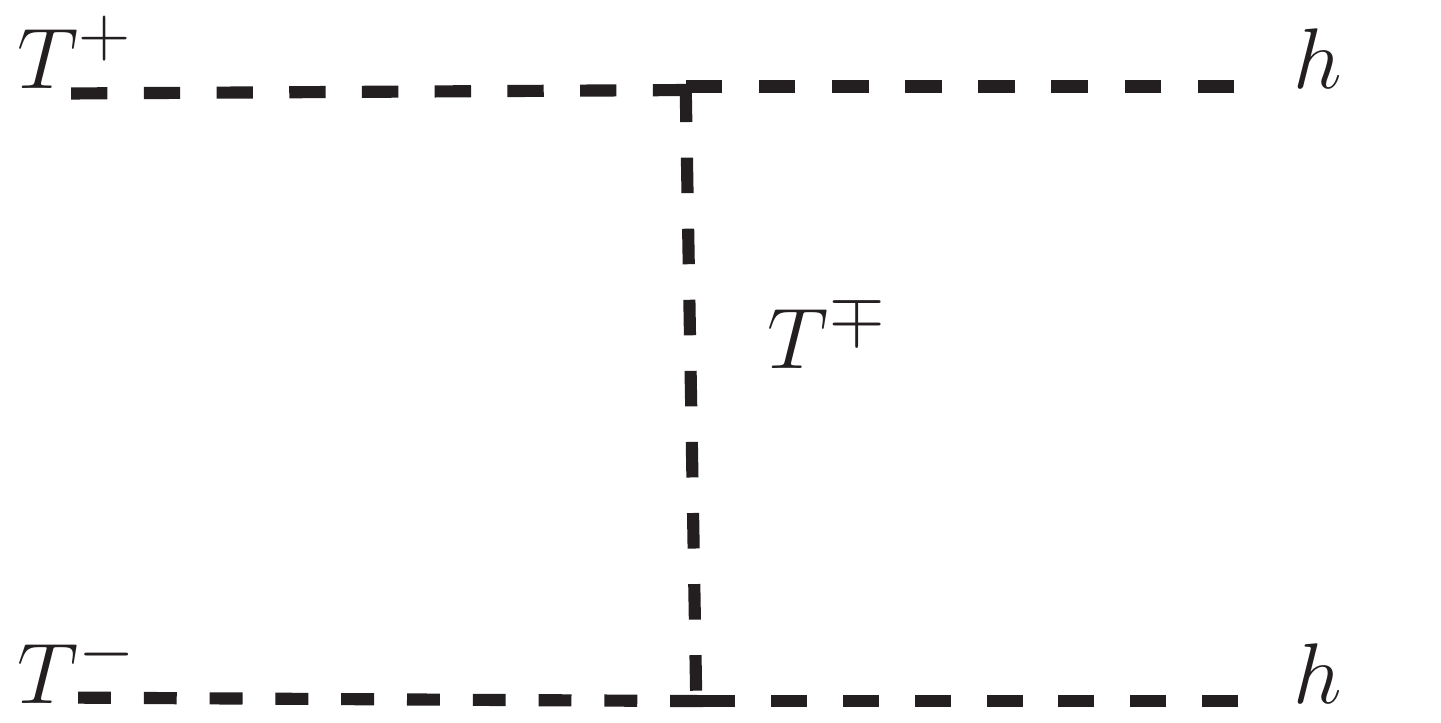}}
\subfigure[]{
\includegraphics[scale=0.22]{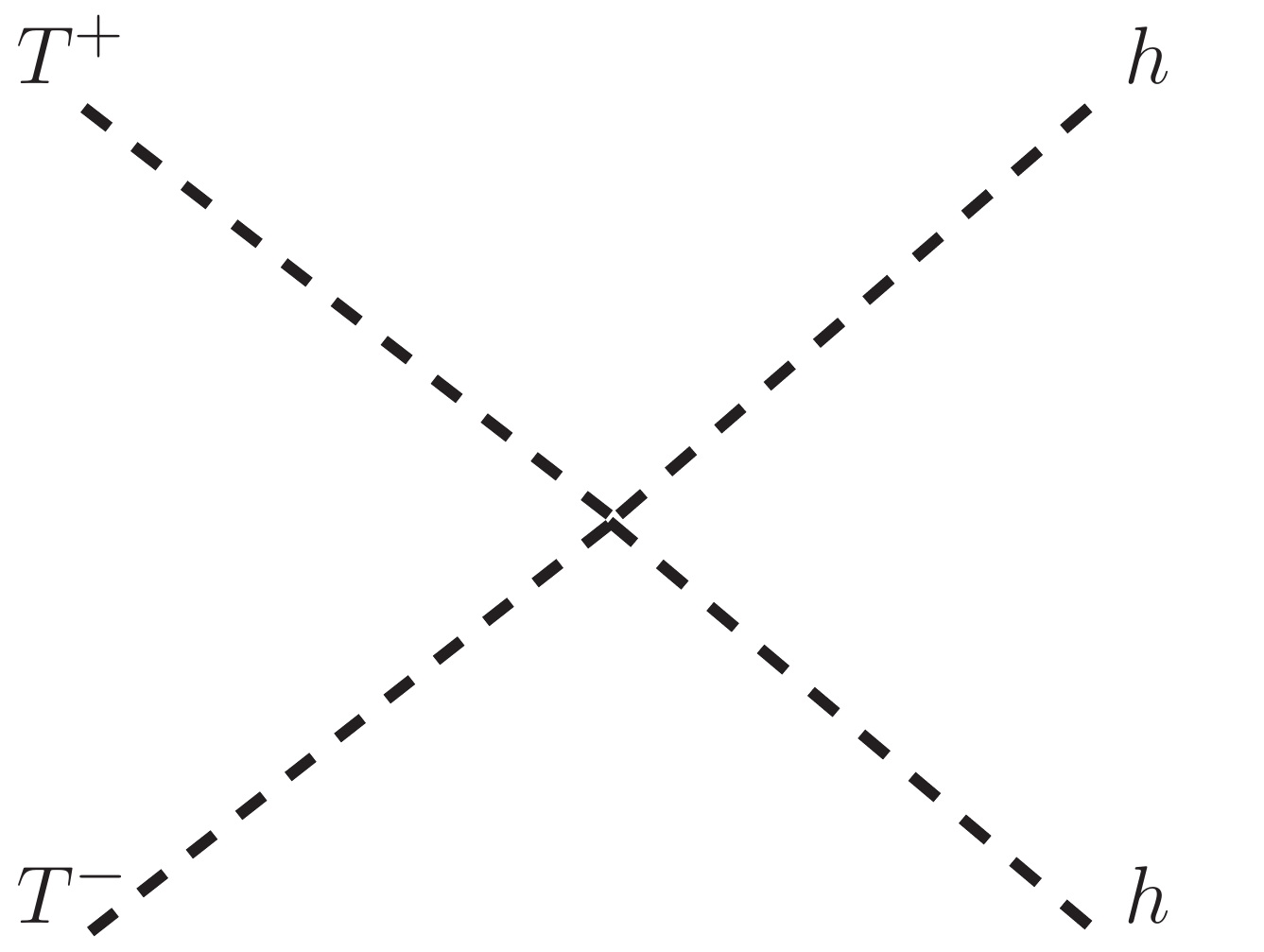}}
\subfigure[]{
\includegraphics[scale=0.22]{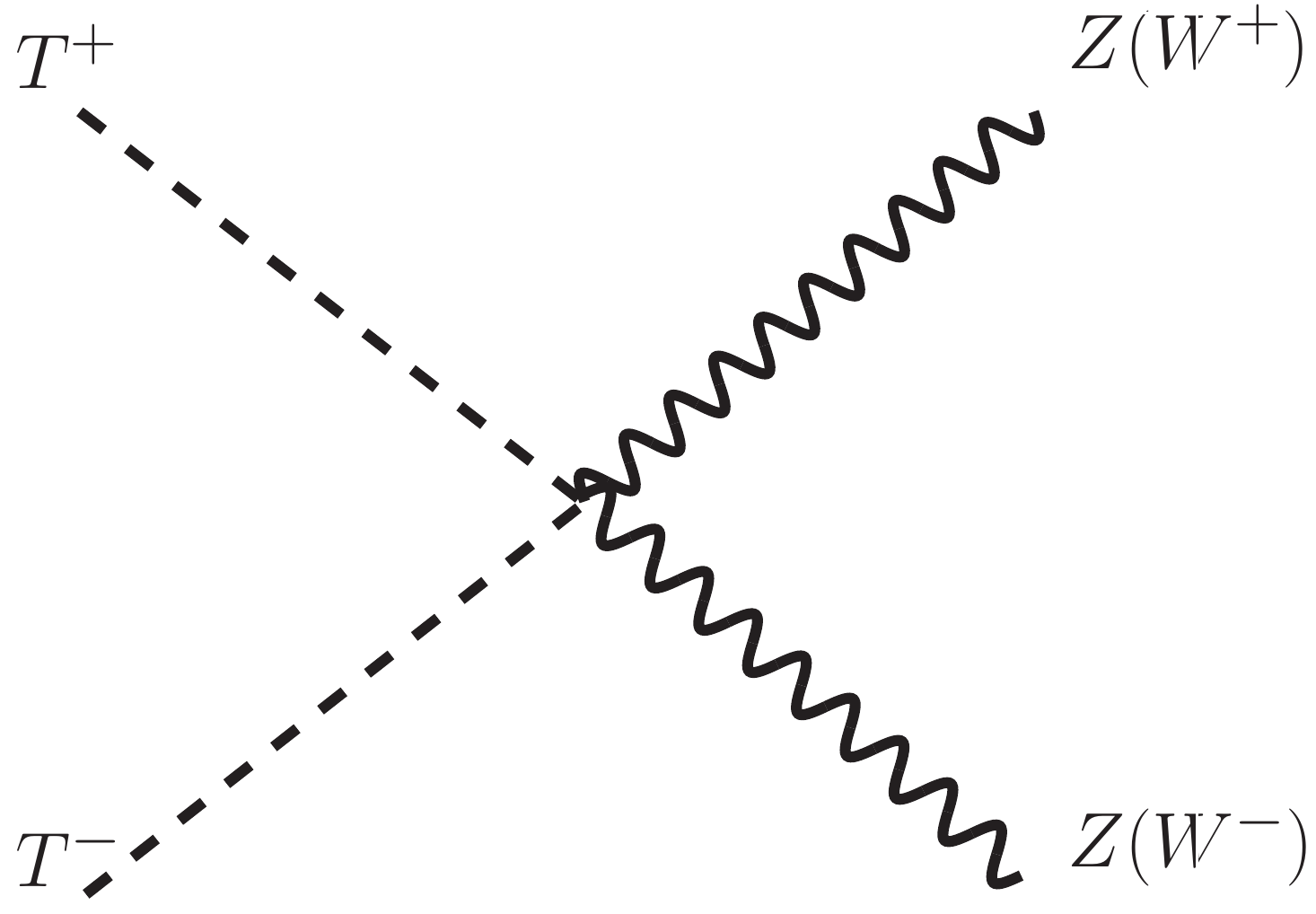}}
\subfigure[]{
\includegraphics[scale=0.22]{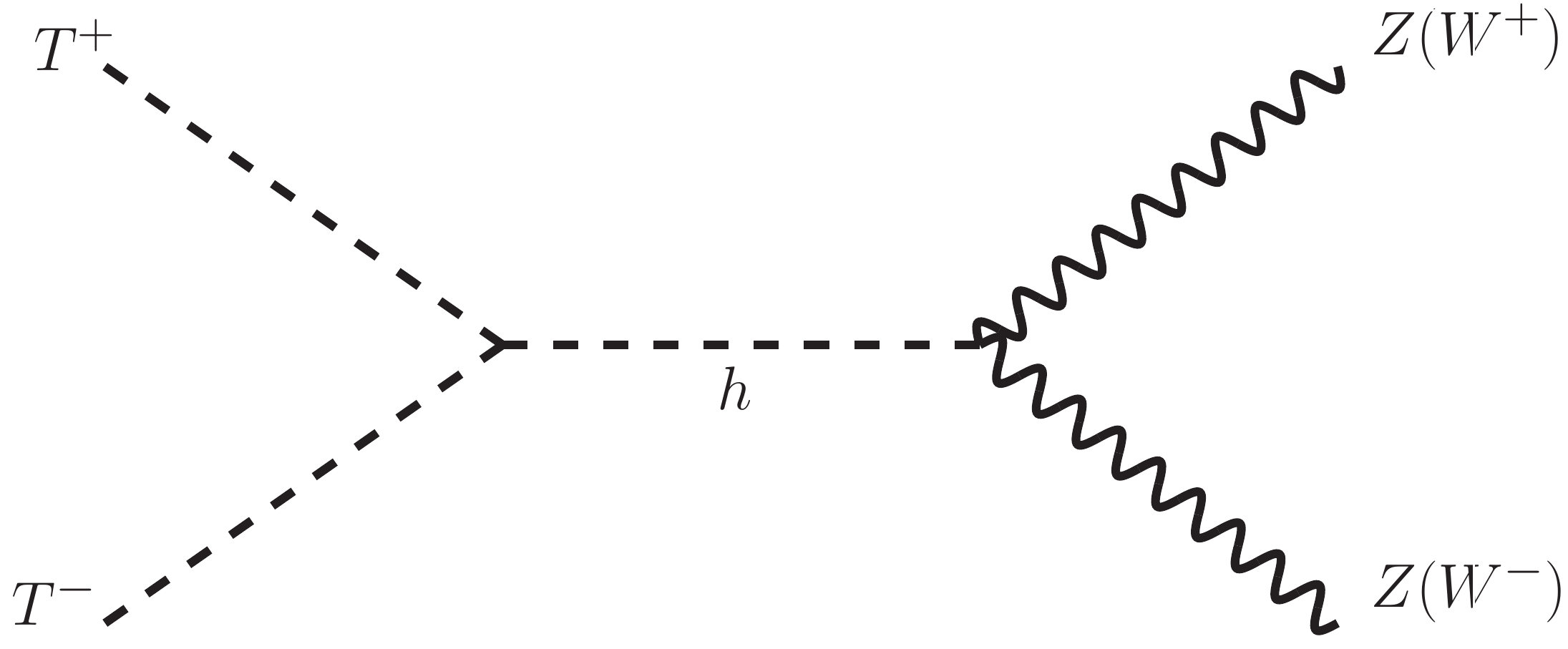}}
\subfigure[]{
\includegraphics[scale=0.22]{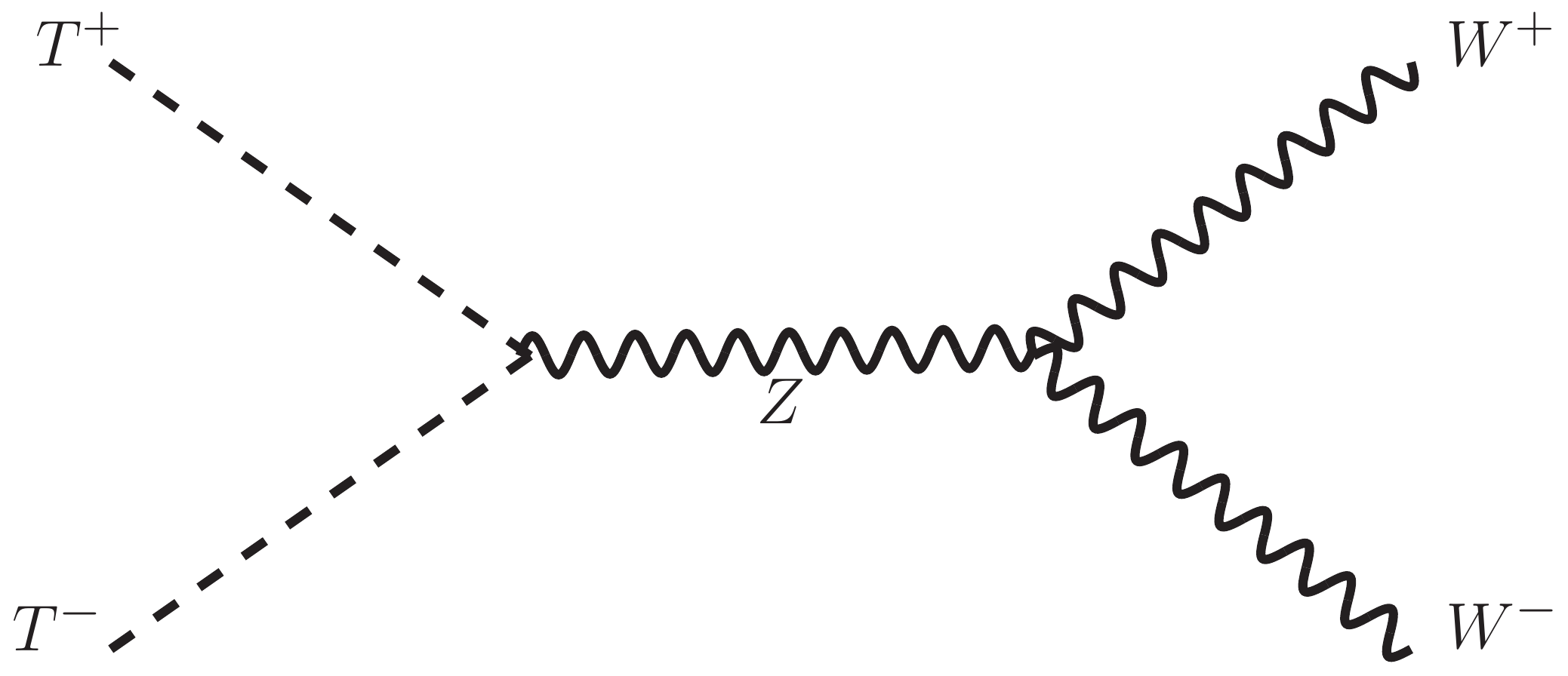}}
\subfigure[]{
\includegraphics[scale=0.22]{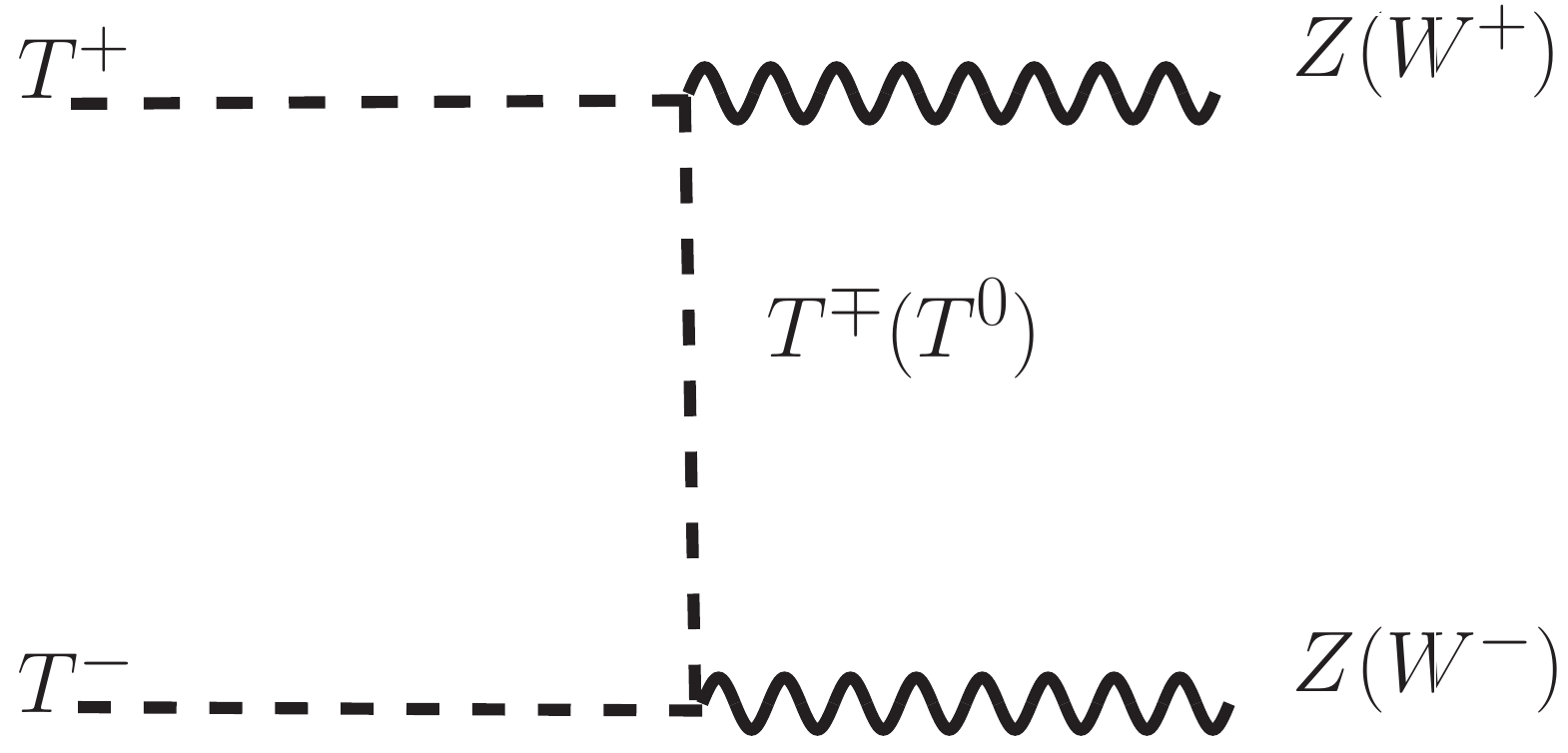}}
\caption{Co-annihilation (annihilation) channels for triplet scalar dark matter $T^0$ ($T^{\pm}$).}
\label{feynTco}
\end{figure}

\begin{figure}[]
\centering
\subfigure[]{
\includegraphics[scale=0.25]{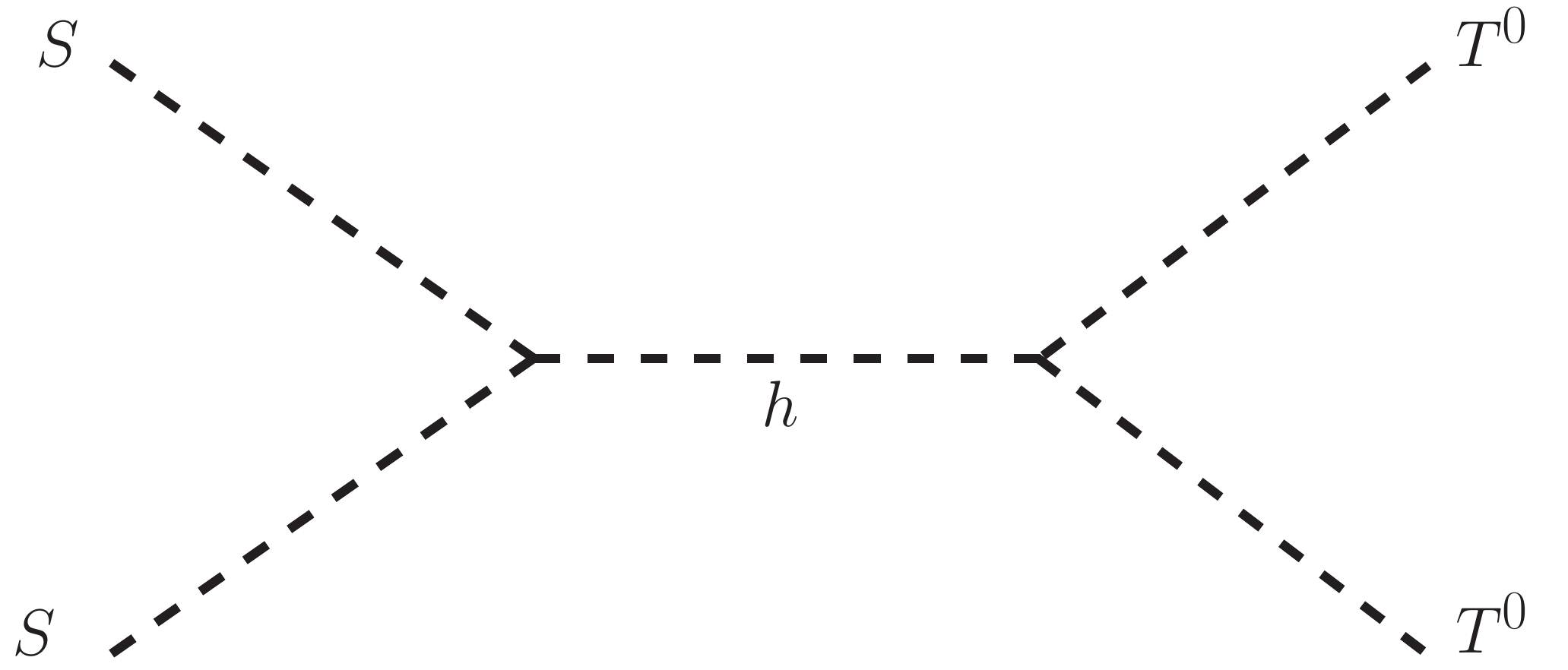}}
\subfigure[]{
\includegraphics[scale=0.25]{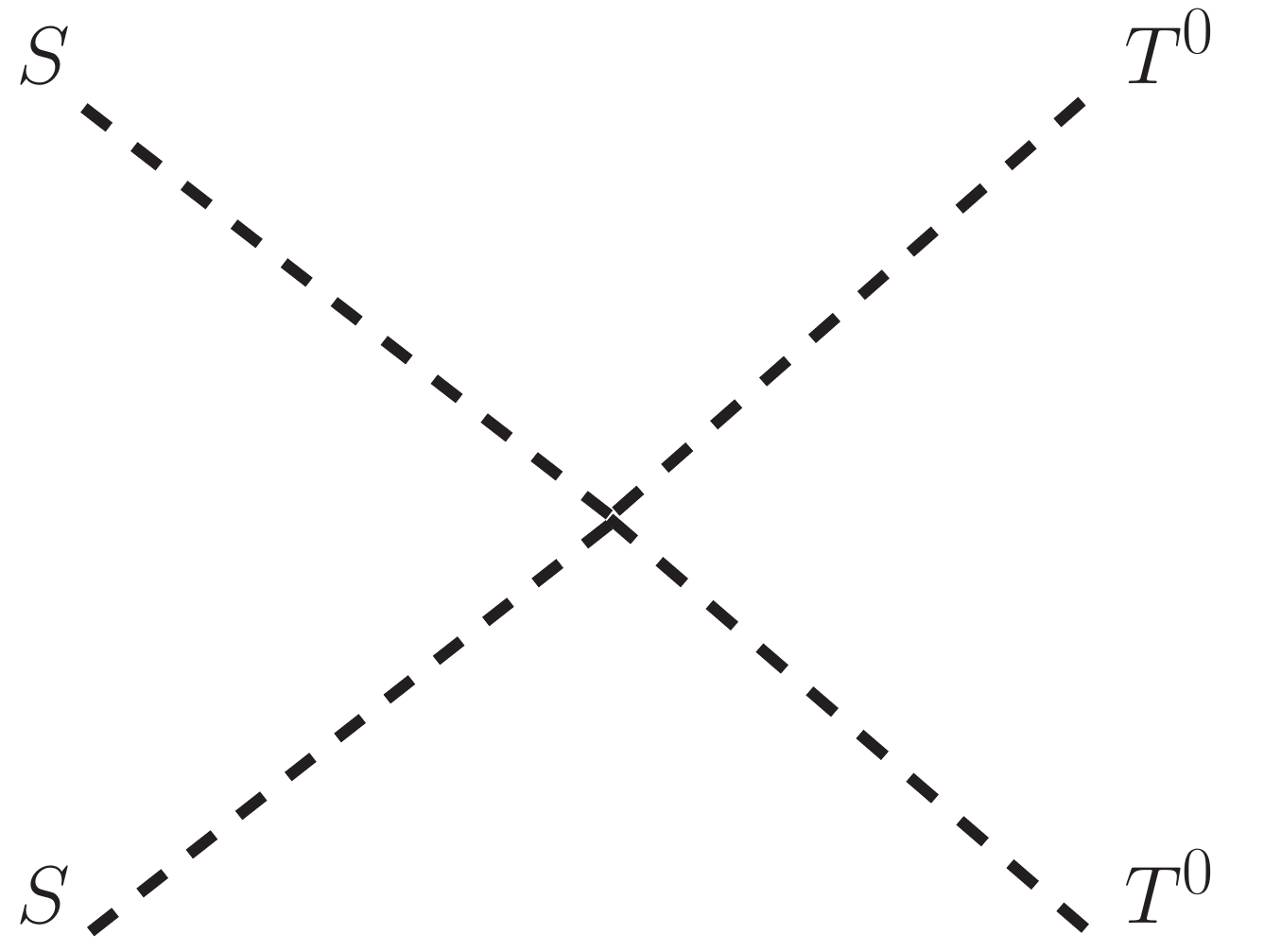}}
\subfigure[]{
\includegraphics[scale=0.25]{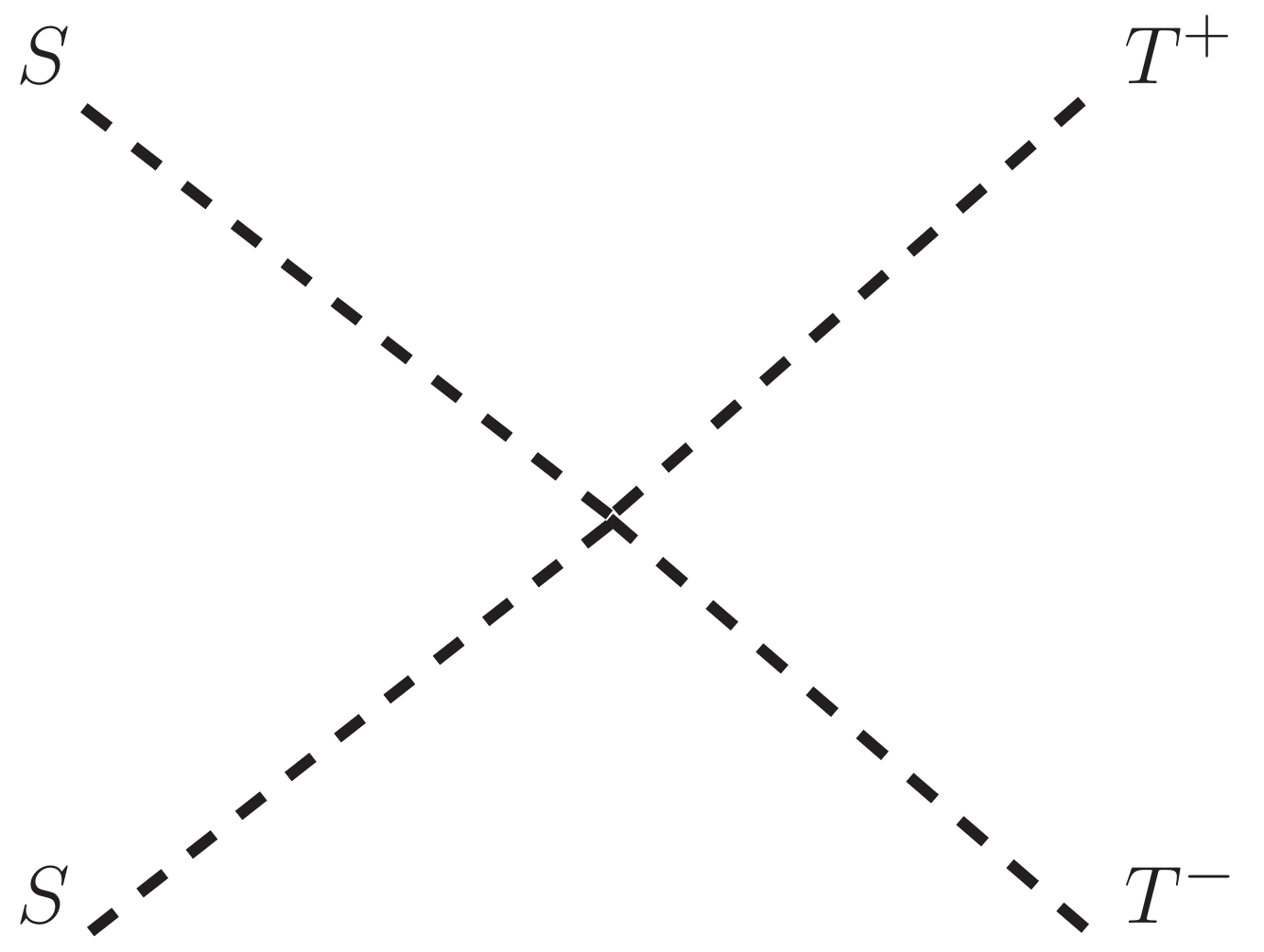}}
\subfigure[]{
\includegraphics[scale=0.25]{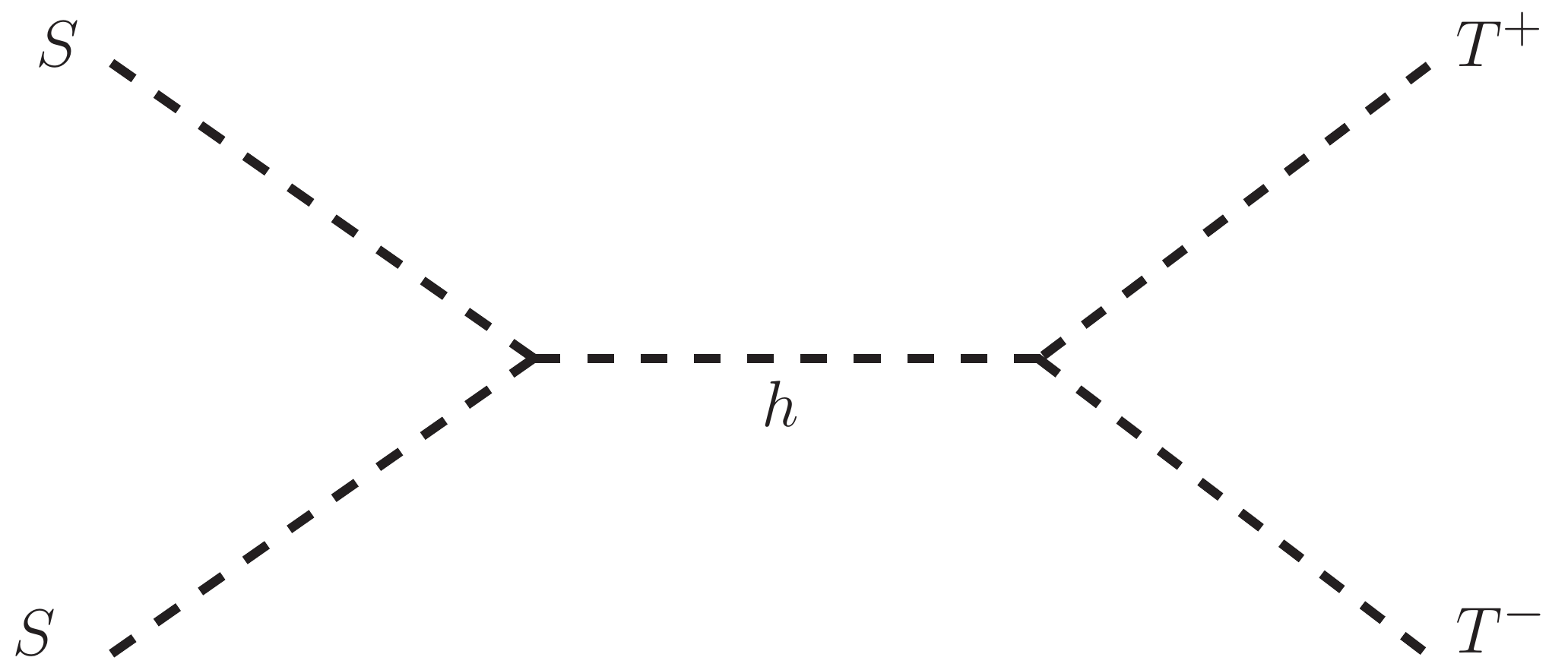}}
\caption{DM-DM conversion channels between Singlet scalar $S$ and triplet dark matter $T^0$, assuming $m_S>m_{T^0}$.}
\label{feynconv}
\end{figure}

The present setup contains two dark matter candidates $T^0$ and $S$, both are odd under different discrete symmetries
$Z_2$ and $Z_2^{\prime}$ which remain unbroken. To obtain the correct relic densities of the dark matter candidates one 
needs to solve the coupled Boltzmann equations. In order to do that we first identify all the relevant annihilation channels 
of both the dark matter candidates. In Fig.\ref{feynS} we show all the possible annihilation channels of $S$ which consists 
of $h$ mediated $s$-channels, $S$ mediated $t$-channel contribution as well as the four point interactions. Similarly in Fig.\ref{feynT}, all the relevant annihilation channels for $T^0$ are indicated. Co-annihilations of $T^0$ with heavier components 
of the triplet also contribute to the relic density which are shown in Fig.\ref{feynTco}. Finally in Fig.\ref{feynconv}, we show 
the channels through which one dark matter candidate (heavier one) can be converted to other one (lighter DM).  This DM-DM 
conversion turns out to be an important contribution in obtaining the final relic. 

\subsection{Relic Density}
To obtain the comoving relic densities corresponding to each dark matter candidates, we need to solve the coupled Boltzmann equations.  The involvement of two dark matter candidates leads to the modification in the definition of the $x$ parameter from $\frac{m_{DM}}{T}$ to $\frac{\mu_{dm}}{T}$, where $\mu_{dm}$ is the reduced mass expressed as:$~\mu_{dm}= \frac{m_{S}m_{T^0}}{m_{S}+m_{T^0}}$. One can write the coupled Boltzmann equations, in terms of newly defined parameter $x = \mu_{dm}/T$ and the 
co-moving number density $Y_{\rm DM} = n_{\rm DM}/s$ ($s$ being the entropy density), as follows\footnote{We use the notation from a recent article on two component DM  \cite{Bhattacharya:2018cgx}.},
\besub
\bea
\frac{dy_{S}}{dx} &=& \frac{-1}{x^2}\bigg{[}\langle \sigma v_{SS\rightarrow XX}\rangle \left(y_{S}^{2}-(y_{S}^{EQ})^2\right)~+~\langle \sigma v_{SS\rightarrow T^{0}T^{0}}\rangle \left ( y_{S}^{2}-\frac{(y_{S}^{EQ})^2}{(y_{T^{0}}^{EQ})^2} y_{T^{0}}^{2}\right)\Theta(m_{S}-m_{T^{0}}) \nonumber \\
&&
-~\langle\sigma v_{T^{0}T^{0} \rightarrow SS}\rangle \left( y_{T^{0}}^{2}-\frac{(y_{T^{0}}^{EQ})^2}{(y_{S}^{EQ})^2}y_{S}^{2}\right)~\Theta(m_{T^{0}}-m_{S})\bigg{]}  , \\
\frac{dy_{T^{0}}}{dx} &=& \frac{-1}{x^2}\bigg{[} \langle \sigma v_{T^{0}T^{0}\rightarrow XX} \rangle \left (y_{T^{0}}^{2}-(y_{T^{0}}^{EQ})^2\right )~+\langle \sigma v_{T^{0}T^{\pm}\rightarrow XX} \rangle 
\left (y_{T^{0}}y_{T^{\pm}}-y_{T^{0}}^{EQ}y_{T^{\pm}}^{EQ}\right )~+~\nonumber \\
&&
\langle \sigma v_{T^{0}T^{0}\rightarrow SS}\rangle \left (y_{T^{0}}^{2}-\frac{(y_{T^{0}}^{EQ})^2}{(y_{S}^{EQ})^2}y_{S}^{2}\right )\Theta(m_{T^{0}}-m_{S}) 
-~\langle \sigma v_{SS \rightarrow T^{0}T^{0}}\rangle \left (y_{S}^{2}-\frac{(y_{S}^{EQ})^2}{(y_{T^{0}}^{EQ})^2}y_{T^{0}}^{2}\right )\nonumber \\
&&
\Theta(m_{S}-m_{T^{0}})\bigg{]}.
\eea
\label{BE}
\eesub

\noindent Here one can relate $y_{i}$ ($i = S,T^{0}$) to $Y_i$ by  $y_i=0.264M_{Pl}\sqrt{g_*}\mu_{dm} Y_{i}$ whereas one can redefine 
$y_i^{EQ}= 0.264M_{Pl}\sqrt{g_*}\mu_{dm} Y_{i}^{EQ}$ in terms of equilibrium density $Y_{i}^{EQ}$, where the equilibrium distributions ($Y_{i}^{EQ}$) are now written in terms of $\mu_{dm}$ as
\bea
Y_{i}^{EQ}(x) = 0.145\frac{g}{g_*}x^{3/2}\bigg{(}\frac{m_{i}}{\mu_{dm}}\bigg{)}^{3/2}e^{-x\big{(}\frac{m_{i}}{\mu_{dm}}\big{)}}.
\eea 
Here $M_{\rm Pl} = 1.22\times10^{19} ~{\rm GeV}$,  $g_{*}=106.7$, $m_i=m_S,m_{T^0}$, $X$ represents all the SM particles and finally, the thermally averaged annihilation cross-section can be expressed as
\bea
\langle \sigma v\rangle = \frac{1}{8m^{4}_{i}T K_2^2(\frac{m_{i}}{T})}\int\limits^{\infty}_{4m_{i}^2}\sigma(s-4m_{i}^2)\sqrt{s}K_1\bigg{(}\frac{\sqrt{s}}{T}\bigg{)}ds
\label{eq:sigmav}
\eea
and is evaluated at $T_f$. The freeze-out temperature $T_f$ can be derived by equating the DM interaction rate $\Gamma = n_{\rm DM} \langle \sigma v \rangle$ with the  expansion rate of the universe $H(T) \simeq \sqrt{\frac{\pi^2 g_*}{90}}\frac{T^2}{M_{\rm Pl}}$. In  Eq.(\ref{eq:sigmav}), $K_{1,2}(x)$ represents the modified Bessel functions.

We use $\Theta$ function in Eq.(\ref{BE})  to explain the conversion process (corresponding to Fig.\ref{feynconv}) of one dark matter to another which strictly depends on the mass hierarchy of DM particles. These coupled equations can be solved numerically to find the asymptotic abundance of the DM particles, $y_{i} \left (\frac{\mu_{dm}}{m_{i}}x_{\infty} \right)$, which can be further used to calculate the relic:

\bea
\Omega_{i}h^2 &=& \frac{854.45\times 10^{-13}}{\sqrt{g_{*}}}\frac{m_{i}}{\mu_{dm}}y_{i}\left ( \frac{\mu_{dm}}{m_{i}}x_{\infty}\right ),
\eea

where $x_{\infty}$ indicates a very large value of $x$ after decoupling. Total DM relic abundance is then given as
\be
\Omega_{\rm Total} {h}^2 =\Omega_{T^0} {h}^2 + 
\Omega_{S} {h}^2\,\, .
\label{totalrelic}
\ee
It is to be noted that total relic abundance must satisfy the DM relic density
obtained from Planck \cite{Aghanim:2018eyx}
\be
\Omega_{\rm Total} {h}^2 =0.1199{\pm 0.0027}\, .
\label{totalrelic}
\ee

\subsection{Direct detection}
Direct detection (DD) experiments like LUX \cite{Akerib:2016vxi}, PandaX-II \cite{Tan:2016zwf,Cui:2017nnn} and Xenon1T \cite{Aprile:2017iyp,Aprile:2018dbl} look for the indication of the dark matter-nucleon scattering and provide bounds on the DM-nucleon scattering cross-section. In the present model, dark sector contains two dark matter particles. Therefore, both the dark matter can appear in direct search experiments.
However, one should take into account the fact that direct detection of both triplet and singlet DM are to be rescaled by factor $f_T^0$ ($f_S$) where
$f_j=\frac{\Omega_j}{\Omega_{\rm Total}}$ with $j=T^0,S$.  Therefore, the effective direct detection cross-section of triplet scalar DM $T^0$ is given as \cite{Ayazi:2015mva}
\bea
\sigma_{\rm {T,eff}}= f_{T^0}\frac{\lambda_{HT}^2}{4\pi}\frac{1}{m_h^4} f^2
\frac{m_N^4}{(m_{T^0}+m_N)^2},
\label{tripletdd}
\eea
and similarly the effective direct detection cross-section of scalar singlet is expressed as \cite{Athron:2017kgt}
\bea
\sigma_{\rm {S,eff}}=f_{S} \frac{\lambda_{HS}^2}{4\pi}\frac{1}{m_h^4} f^2
\frac{m_N^4}{(m_{S}+m_N)^2}\, .
\label{singletdd}
\eea 
where $m_N$ is the nucleon mass, $\l_{HT}$ and $\l_{HS}$ are the quartic couplings involved in the DM-Higgs interaction. A recent estimate of the Higgs-nucleon coupling $f$ gives $f=0.32$ \cite{Giedt:2009mr}. Below we provide the Feynman diagrams for the spin independent elastic scattering of  DM with nucleon.

\begin{figure}[H]
\centering
\subfigure[]{
\includegraphics[scale=0.450]{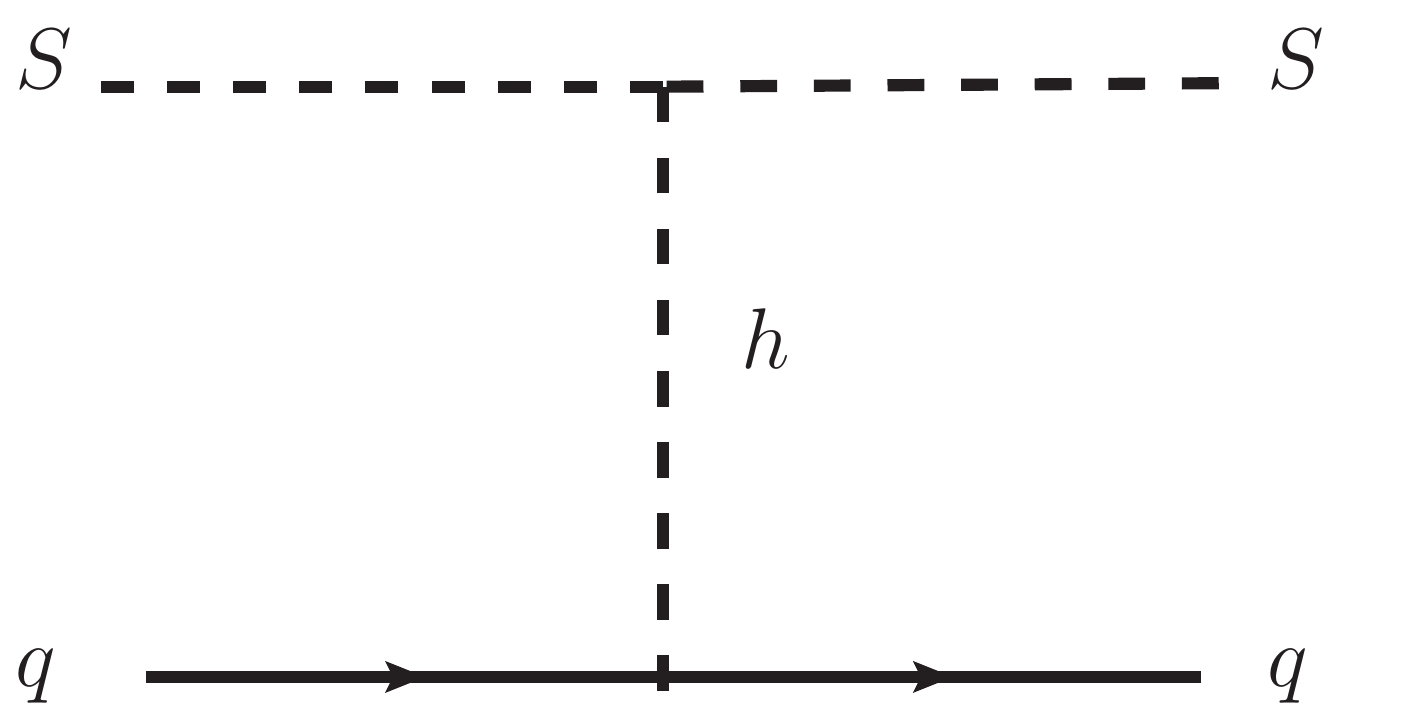}}
\subfigure[]{
\includegraphics[scale=0.450]{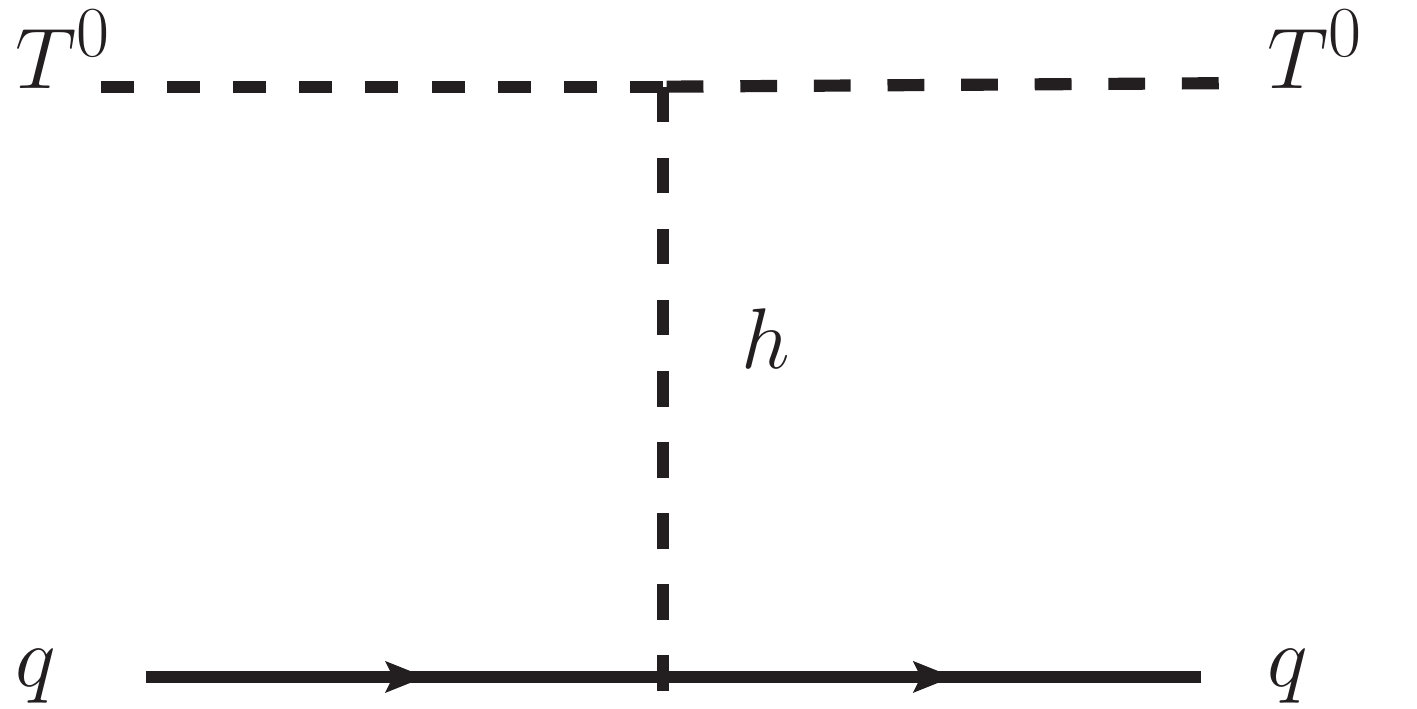}}
\caption{Spin independent elastic scattering of DM-nucleon.}
\label{DD}
\end{figure}

\subsection{Results}
\label{result}
To study the proposed two component DM scenario, we first write the model in LanHEP \cite{Semenov:2014rea} and 
then extract the model files to use in micrOMEGAs 4.3.5 \cite{Barducci:2016pcb}. In doing this analysis, all the relevant 
constraints as mentioned in section \ref{constraints} are considered. For the sake of better understanding, we divide 
our analysis in two parts: [A] $m_S > m_{T^0}$ and [B] $m_{T^0} > m_S$.
\begin{figure}[H]
\centering
\subfigure[]{
\includegraphics[scale=0.40]{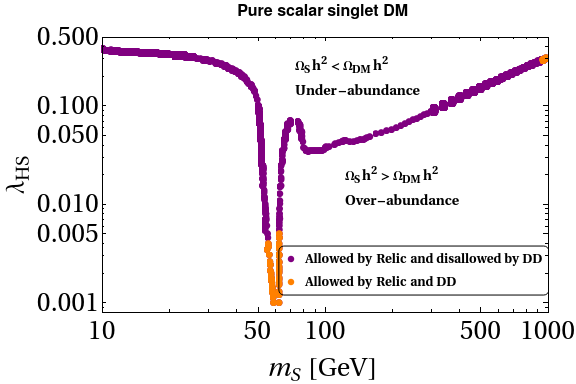}}
\subfigure[]{
\includegraphics[scale=0.38]{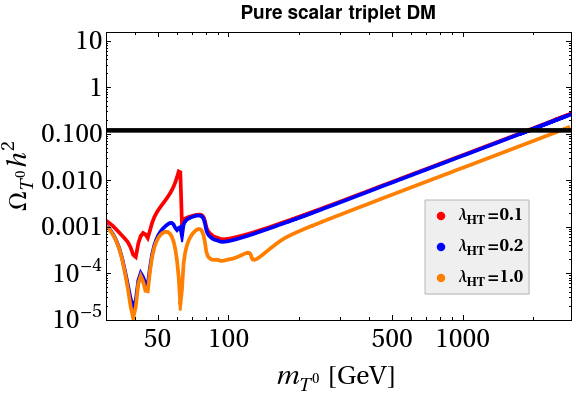}}
\caption{Single component DM: (a)Left panel:  relic contour for singlet scalar DM (b)Right panel: relic contribution from the 
triplet versus $m_{T^0}.$}
\label{relic-single-com-DM}
\end{figure}

It is known that for a single component DM scenario, both the singlet scalar as well as the triplet (with $Y=0$) DM are 
not allowed below TeV. To make it clear, we provide the relic contour for the singlet scalar in the $\l_{HS} - m_S$ plane 
in the left panel of Fig. \ref{relic-single-com-DM}, where except the resonance region, the entire range of $m_S$ up to 
$\sim$ TeV (the purple shaded region) is ruled out by the DD constraint. Similarly we also include the relic contribution 
from the triplet against its mass in the right panel of Fig. \ref{relic-single-com-DM} for different choices of the triplet-SM Higgs 
portal coupling $\l_{HT}$. It can clearly be seen that the relic (and DD too) can be satisfied for $m_{T^0}$ beyond 1.8 TeV. 
Changing the value of $\l_{HT}$ does not have much impact on this conclusion. This is because the effective annihilation 
cross-section is mostly dominated by the gauge bosons final states contributions, $i.e.,$ via Feynman diagrams shown in Fig.\ref{feynT} (annihilations) 
and Fig.\ref{feynTco} (co-annihilations). The presence of first three dips are due to the successive resonances mediated by the $ 
W^{\pm},Z$ and SM Higgs (as seen from the $s$-channel diagrams Fig.\ref{feynTco} and Fig.\ref{feynT}).  The later kinks around $80$ GeV and $125$ GeV 
are indicative of the openings of gauge and Higgs boson final states respectively.

Note that our aim is to have mass of both the DM candidates below TeV which is an interesting regime for experimental 
studies. Here we mostly rely on two facts to satisfy our goal: (i)  single component of DM does not require to produce the 
entire relic contribution and (ii) conversion involving two DMs is expected to contribute non-trivially. Below we proceed 
one after other cases. As we observe above that the triplet contribution to the relic is essentially under-abundant (irrespective 
of the choice of portal coupling $\l_{HT}$) in this region, we expect that the singlet scalar can make up the rest of relic 
while an important contribution to be contributed by the DM-DM conversion. As stated before, the relevant parameters that 
would control the study are $m_{T^0},~m_S,~\lambda_{HS},~\lambda_{HT}$, and $\kappa$ and we find below their 
importance. 

\subsubsection{Case I:  $m_S>m_{T^0}$ }
\begin{figure}[h]
\centering
\subfigure[]{
\includegraphics[scale=0.39]{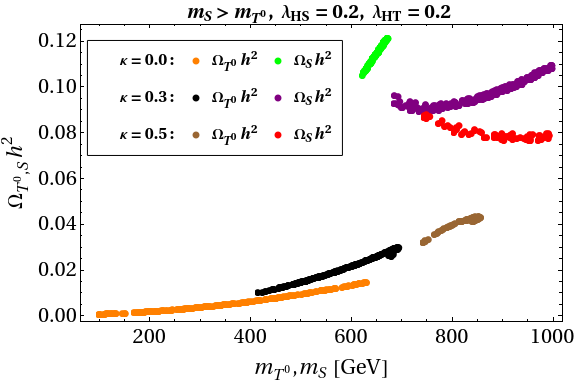}}
\subfigure[]{
\includegraphics[scale=0.4]{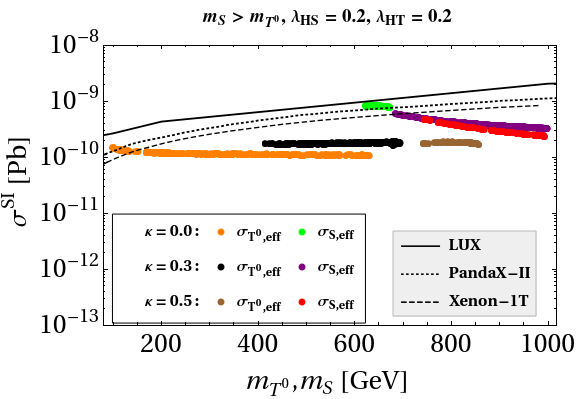}}
\caption{(a)Left panel: Points which satisfy the correct total DM relic abundance for different values of $\kappa$ while maintaining $m_S > m_{T^0}$ for $\lambda_{HS}= \lambda_{HT}=0.2$, (b)Right panel: Spin independent DM-nucleon scattering cross-section.  Limits from direct detection experiments are shown in black solid lines (LUX), dotted (PandaX-II), dashed (Xenon-1T). }
\label{relic1}
\end{figure}

In the left panel of Fig.\ref{relic1}, we show the variation of the individual contributions toward relic abundances from triplet ($\Omega_{T^0}h^2$) and singlet ($\Omega_Sh^2$) with their respective masses, $m_{T^0}$ and $m_S$ respectively, such that the total relic abundance $\Omega_{\rm  Total}h^2$ satisfies the Planck limit \cite{Aghanim:2018eyx}. In getting such plots, we chose different 
values of conversion coupling $\kappa=~0, 0.3$ and 0.5 and specifically consider the mass hierarchy as $m_S > m_{T_0}$. The respective variations of the relics versus their masses with different $\kappa$ 
are indicated by (i) orange ($T^0$ contribution) and green ($S$ contribution) patches with $\kappa = 0$, (ii) black ($T^0$) and purple ($S$) with $\kappa = 0.3$ and (iii) brown ($T^0$) and red ($S$) for $\kappa = 0.5$. Also for simplicity, we choose the Higgs portal couplings (with scalar singlet and triplet) to be same and a reference value is chosen as $\lambda_{HS} = \lambda_{HT}=0.2$. 
Note that such a value of the Higgs portal couplings of the singlet and the triplet scalar DMs is not allowed by the relic and DD constraints as seen from Fig. \ref{relic-single-com-DM}. Below we discuss implications of this plot in detail. 

In order to understand the importance of conversion coupling $\kappa$, we begin with $\kappa = 0$ 
case. It is to be noted that even when the conversion coupling $\kappa$ is set at 0, conversion between DM candidates ($SS \rightarrow T^0 T^0$) can take place via s-channel diagram as shown in Fig.\ref{feynconv}. With $\kappa=0$, we observe that 
the dominant contribution to the total relic comes from $S$ (the green patch on the top) whereas the contribution coming from 
$T^{0}$ is very small (orange patch near the bottom). To be more precise, a point in the leftmost side of the orange patch (say 
$m_{T^0} =100$ GeV having $\Omega_{T^0} h^2$ =0.0005) is correlated to a single point on the rightmost side of the green patch 
($m_S$= 667 GeV with $\Omega_S h^2$ =0.119 ). As stated earlier, since triplet annihilation channels are mainly gauge dominated, 
$\lambda_{HT}$ does not have a significant effect on the relic density. Hence $\Omega_{T^0}h^2$ has a limitation, it can't 
provide more than $\sim$ 10 percent contribution as seen from Fig. \ref{relic-single-com-DM}(b). However once the $\kappa$ has a 
sizeable magnitude, the $T^0$ contribution to the relic is enhanced to some extent due to the DM-DM conversion as can be 
seen from the black patch (paired with purple) for $\kappa = 0.3$ and brown patch (paired with red) for $\kappa = 0.5$.  
The numerical estimates of several parameters involved in the above discussion are tabulated in Table \ref{relic_case1} for two different choices of the conversion couplings $\kappa = 0.0 $ and 0.3.

\begin{table}[H]
\centering
\begin{tabular}{|c|c|c|c|c|}
\hline
$\kappa$ & $m_{S}$ [GeV] & $m_{T^0}$ [GeV] & $~\Omega_{S}h^2$ & $~\Omega_{T^0}h^2$  \\
\hline
0.0 & 667 & 100 & 0.119 & 0.0005  \\
    & 633 & 631 & 0.108 & 0.014    \\
\hline
0.3 & 999 & 416 & 0.108 & 0.009 \\ 
    & 748 & 695 & 0.092 & 0.029 \\
\hline
\end{tabular}
\caption{Table showing the variation of the relic densities of both the dark matters with thier respective masses for two different choices of $\kappa$. While generating the above results, we considered: $\l_{HS}=0.2$ and $\l_{HT}=0.2$ }
\label{relic_case1}
\end{table}

In Fig.\ref{relic1}(b), the evaluated DD cross-section corresponding to the respective pair of patches of left panel along 
with the upper limits on DM-nucleon scattering cross-section set by different direct search experiments are depicted. 
We already notice from the left panel of plots that with $\kappa = 0$, the dominant contribution to the relic comes from 
$S ~i.e.~\frac{\Omega_S}{\Omega_{\rm Total}} \sim 1$ and hence following Eq. (\ref{singletdd}), $\sigma_{S,eff}$ is quite 
large and turns out to be disallowed by the direct detection bounds. This shows that $\kappa~=~0$ is not an allowed 
possibility in this two-component framework. However, as we switch on the DM-DM conversion processes, $i.e.$ with 
$\kappa = 0.3,~0.5$ say, we notice that the intermediate mass range (below TeV) of DMs (which was otherwise disallowed 
in case of single component scenario for both triplet as well as singlet) becomes allowed from both the relic as well as 
the direct detection bounds. 

\begin{figure}[h]
\centering
\subfigure[]{
\includegraphics[scale=0.39]{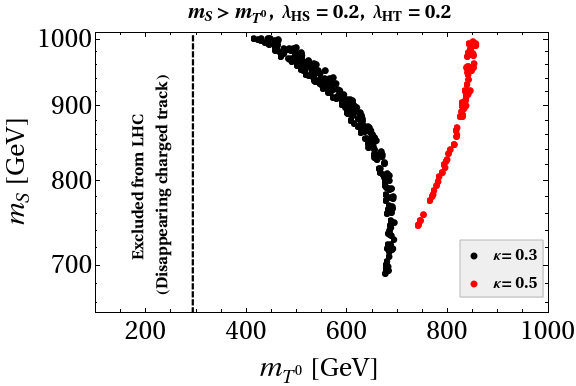}}
\subfigure[]{
\includegraphics[scale=0.4]{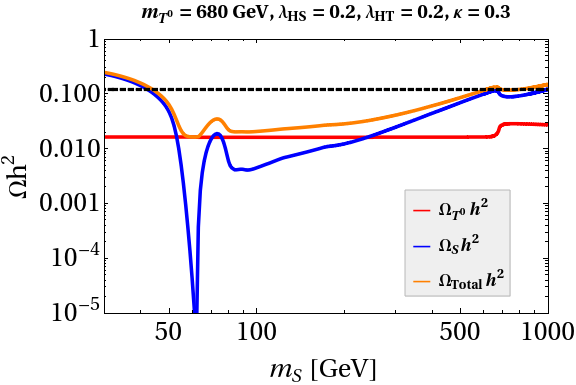}}
\caption{(a) Left panel shows all the points which satisfy the correct total DM relic abundance and are also allowed by direct detection for different values of $\kappa$ while maintaining $m_S > m_{T^0}$ for $\lambda_{HS}=0.2$ and $\lambda_{HT}=0.2$ in $m_S-m_{T^0}$ plane. (b) Right panel shows the variation of the relic density with $m_S$ for a fixed value $m_{T^0} = 680$ GeV while keeping $\kappa = 0.3$ and $\l_{HS}=\l_{HT}=0.2$.}
\label{msmt2}
\end{figure}

In Fig.\ref{msmt2}(a), we provide a relic contour plot in $m_S - m_{T^0}$ plane which is also in agreement with bounds from direct detection experiments. It clearly shows that in this two-component scenario 
allows both the DMs to have mass in the intermediate range or below 
TeV. It can be noticed that a parabolic pattern is prevalent for the relic contour. The reason of this would be clear if we look at the right panel
where in individual contributions to the relic ($\Omega_S h^2$ in blue and $\Omega_{T^0} h^2$ in red) are shown as a function of $m_S$. 
For this plot (b), the triplet DM mass is kept fixed at 680 GeV while $\kappa$ is considered to be 0.3 (one of the two benchmark values of Fig. \ref{msmt2}(a)). The total relic is shown here by the orange line. We observe that for the singlet scalar contribution, it exactly follows the 
pattern of its sole contribution (below 680 GeV) as shown in Fig. \ref{relic-single-com-DM} till it becomes heavier than $m_{T^0}$.
At this point (when $m_{T^0} < m_S$), the $SS \rightarrow T^0 T^0$ starts to take place. As a result, a mild dip is observed on the 
relic plot of $S$ field around this point and again it increases with the increase of $m_S$ value as usual. On the other hand, below 
$m_S = 680$ GeV, there exists a constant contribution (independent of $m_S$) from $T^0$ corresponding to fixed mass $m_{T^0}$ = 
680 GeV as expected. In this case also, when $m_{S}$ exceeds 680 GeV, we notice an increase in its relic which is reminiscent of 
the $SS \rightarrow T^0 T^0$ conversion process having $\kappa = 0.3$. The resultant relic plot (orange line) thereby touches the observed 
relic line ($\Omega_T h^2 = 0.12$) twice: first around $m_S =$ 690 GeV and then $\sim$ 801 GeV. The observation that for a fixed 
$m_{T^0}$, the total relic would be satisfied by two different values of $m_S$ explains the parabolic nature of black patch in the 
left panel figure. We also note that for the first pair, the two DM masses [(690, 680) GeV] are very close to each other while within 
the other pair, DM masses [(801, 680) GeV] are separated by a sizeable value. Once the $\kappa$ increases, the mass difference between 
the pair of DM masses (satisfying the relic and DD constraints for a fixed $m_{T^0}$) would also be increased. For this reason, though the similar observation (satisfaction of relic by two pair of points for a fixed $m_{T^0}$) is also present for the red patch (with $\kappa = 0.5$), 
due to the stipulated intermediate regime of DM mass ($i.e.$ below TeV) chosen here, the other (the one with heavier $m_S$) is not seen 
in the figure. 
\begin{figure}[h]
\centering
\subfigure[]{
\includegraphics[scale=0.39]{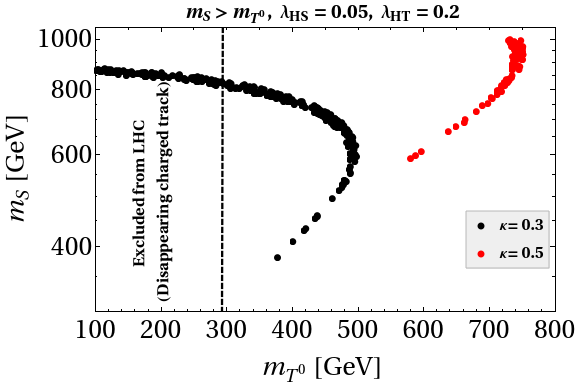}}
\subfigure[]{
\includegraphics[scale=0.4]{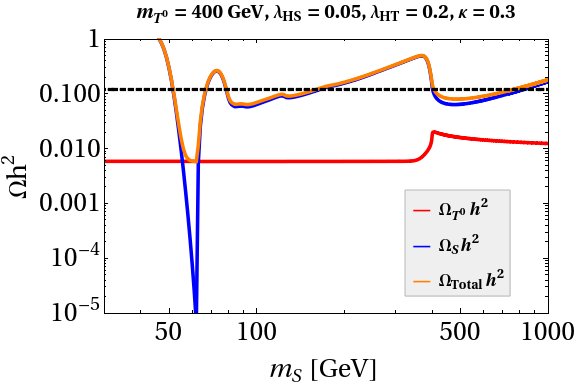}}
\caption{(a)Points which satisfy the correct total DM relic abundance and are also allowed by direct detection for different values of $\kappa$ while maintaining $m_S > m_{T^0}$ for $\lambda_{HS}=0.05$ and $\lambda_{HT}=0.2$ in $m_S-m_{T^0}$ plane. (b) Variation of $\Omega_{S}$ with $m_{S}$ for $m_{T^0}=400$ GeV and $\kappa=0.3$. }
\label{msmt1}
\end{figure}

In Fig.\ref{msmt1}, we repeat the plots of Fig.\ref{msmt2} for a smaller value of $\lambda_{HS}=0.05$, though keeping 
$\lambda_{HT}$ 
fixed at 0.2. As $\lambda_{HS}$ is decreased, the annihilation of $S$ into the SM particles is also decreased which in turn enhances the 
relic density of $S$ for a given mass. Hence, a relatively smaller contribution from $T^0$ (compared to Fig. \ref{msmt2}(a)) is required and 
as a result, lower mass of $m_{T^0}$ is allowed. In other words, a shift of the black patch (of parabolic nature) toward left ($i.e.$ shift 
toward lowered masses) is observed. For example, with the same value of $\kappa = 0.3$ is in Fig. \ref{msmt2} also, while a pair of DM 
masses $m_S, m_{T^0} = (690, 680)$ 
GeV satisfies the total relic in case with $\lambda_{HS} = 0.2$, a lower set of masses (407, 400) GeV can satisfy the relic in case with 
$\lambda_{HS} = 0.05.$ At this point, we can recall our finding from Fig. \ref{relic-single-com-DM} also. The relic contribution from 
$T^0$ is essentially governed by the $T^0 T^0$ annihilations to finals state gauge bosons, and being almost insensitive to $\lambda_{HT}$ value, the maximum contribution of $\Omega_{T^0} h^2$ incorporating a sizeable $\kappa$ can be around 30 percent of the total relic (provided we stick to the low mass regime of DMs, $i.e. \sim$ below TeV) with appropriate $\kappa$. Therefore the significant relic has to be obtained 
from $S$. Hence, the above conclusion that a smaller $\lambda_{HS}$ allows for a lighter DM pair remains valid for any choice of $\lambda_{HT}$. With a similar line of consideration as in Fig. \ref{msmt2}(a), here also we use the conservative bound on $m_{T^0}$ as 
$m_{T^0} > 287$ GeV. Finally in view of constraints on the mass of the triplet DM as stated in section \ref{constraints}, we put a vertical dashed line at $m_{T^0} = 287$ GeV such that the right side of it can be recognized as the allowed parameter space. As a result, some of the 
parameter space becomes disallowed for $\kappa = 0.3$. 

\subsubsection{Case II:  $m_S<m_{T^0}$}

We now study the DM phenomenology considering the mass hierarchy among DM components as $m_{S}<m_{T^0}$.
Note that in this case the DM-DM conversion can take place having the form: $T^0 T^0 \rightarrow SS$ and hence 
contribution from the singlet scalar would be more than that of the case-I. Following Fig. \ref{relic-single-com-DM}(b), 
we know that the maximum contribution of $\Omega_{T^0} h^2$ is less than 30 percent only provided we restrict $m_{T^0}$ 
to be in sub-TeV regime. Furthermore due to $T^0 T^0 \rightarrow SS$ conversion in this case, contribution to relic by 
$\Omega_{T^0} h^2$ would be even less. This particular case is therefore not very promising from the perspective of two 
component DM. Hence in this case, we extend the mass range of $T^0$ to be more than TeV (though less than 1.8 TeV) 
while $m_S$ is kept below 1 TeV. 

\begin{figure}[]
\centering
\subfigure[]{
\includegraphics[scale=0.37]{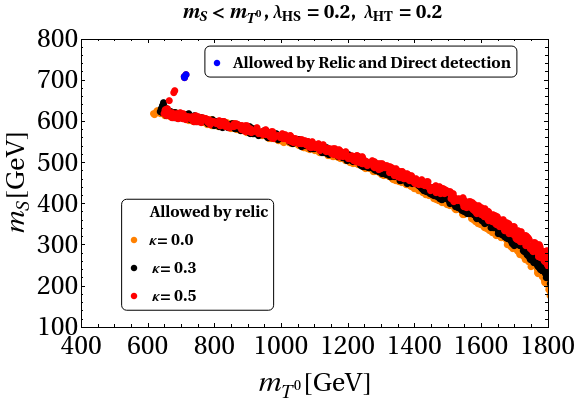}}
\subfigure[]{
\includegraphics[scale=0.38]{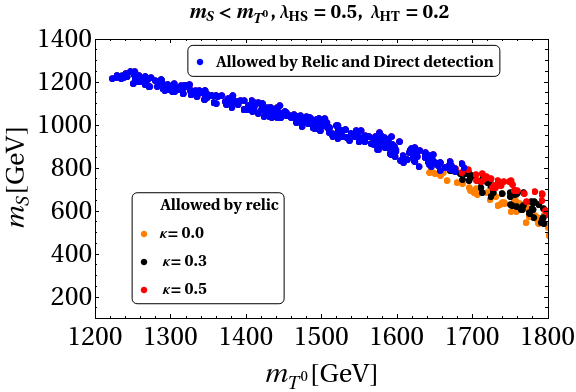}}
\caption{Points which satisfy the correct total DM relic abundance and are also allowed by direct detection (in blue) for different values of $\kappa$ while maintaining $m_S < m_{T^0}$ for $\lambda_{HT}=0.2$ and (a) $\lambda_{HS}=0.2$ (b) $\lambda_{HS}=0.5$ in $m_S-m_{T^0}$ plane. }
\label{relic-con-II}
\end{figure}

To analyse the case, we scan over the parameter space involving $m_S, m_{T^0}$ with different $\kappa$ values such that 
$\Omega_{\rm Total} h^2$ can satisfy the relic. Here initially the Higgs portal couplings are fixed at values, $\lambda_{HT} = 0.2$ and 
$\l_{HS} = 0.2$ while maintaining the mass hierarchy like $m_S<m_{T^0}$.  
Though it produces the expected pattern as shown in Fig.\ref{relic-con-II}(a), most of this parameter space are ruled out 
once DD constraints are applied. There exists only a very narrow regime corresponding to $\kappa = 0.5$, denoted by the 
blue shade having masses $m_{T^0} \sim 720$ GeV and $m_S \sim 710$ GeV, which satisfies both the relic and DD limits. 
Hence it is clear that a large DM-DM conversion is required. However, for the regime where relic satisfied but disallowed by 
DD, the conversion coupling $\kappa$ does not have much impact as they (orange points with $\kappa = 0$, black points 
with $\kappa = 0.3$ and red points with $\kappa = 0.5$) overlap each other. So one can come to a conclusion that this case 
$m_S<m_{T^0}$ is disfavored compared to the case-I for sub-TeV masses of both the DMs. Also it can be noted that in 
case-II, due to the parabolic nature of the plot, for a fixed $m_{T^0}$ there are two pairs of values of $m_{T^0}, m_S$ for 
which relic and DD constraints satisfaction can happen: one is where their masses are close enough and at another point 
where $m_S$ and $m_{T^0}$ are significantly apart. However such a possibility does not exist here as with a much lower 
mass of $m_S$ (around 600 GeV) compared to the blue shaded region, the DD constraint is more stringent.  

In the right panel, Fig. \ref{relic-con-II}(b), we consider a larger value for $\l_{HS} = 0.5$ and simultaneously open 
the above TeV (but below 1.8 TeV) regime for $T^0$ DM. Here the relic as well as DD satisfied points are denoted by blue patch 
and a sizable region of parameter space (compared to the left panel of the figure) becomes allowed and that too for all values 
of $\kappa$. Hence the scenario of two component DM works for a relatively heavier mass, above TeV, of $T^0$.

\section{Electroweak vacuum stability}
\label{VS}

In this section, we study the electroweak vacuum stability in this two-component DM framework. 
 As already mentioned in section 
\ref{constraints}, Eqs.(\ref{copos}) are to be fulfilled at any scale $\mu$ till $M_{Pl}$. Within SM itself, due to the presence 
of top quark Yukawa coupling $y_t \sim \mathcal{O}(1)$, the Higgs quartic coupling $\l_H$ becomes negative at a scale around 
$10^{10}$ GeV \cite{Buttazzo:2013uya,Degrassi:2012ry,Tang:2013bz,Ellis:2009tp,EliasMiro:2011aa}. However the present limits 
on the top quark mass suggests that the EW vacuum is a metastable one. It is well known that incorporating new scalars can 
modify the fate of the EW vacuum\cite{Haba:2013lga,Khan:2014kba,Khoze:2014xha,Gonderinger:2009jp,Gonderinger:2012rd,Chao:2012mx,Gabrielli:2013hma,DuttaBanik:2018emv,Ghosh:2017fmr,Borah:2020nsz}. In the present setup, presence of these new scalar fields $i.e. ~S$ and $T$ provides 
a positive contribution to the beta function of $\l_H$ through their Higgs portal interactions 
as  
\besub
\bea
\beta_{\l_H}  & =& \b_{\l_H}^{\rm{SM}}+\b_{\l_H}^{\rm{T}} +\b_{\l_H}^{\rm{S}}~= ~\b_{\l_H}^{\rm{SM}}+\frac{3}{2}\l^2_{HT}+\frac{1}{2}\l^2_{HS};\\
 \b_{\l_H}^{\rm{SM}} &=&  \frac{27}{200} g_{1}^{4} +\frac{9}{20} g_{1}^{2} g_{2}^{2} +\frac{9}{8} g_{2}^{4} -\frac{9}{5} g_{1}^{2} \lambda_H -9 g_{2}^{2} \lambda_H +24 \lambda_H^{2}+12 \lambda_H y_t^2 -6y_t^4.
\eea
\label{betaHSM}
\eesub
which (for details, see Appendix \ref{RGE}) helps in making the EW vacuum stable.

While $\l_H > 0$ till $M_{Pl}$ ensures the absolute stability of the EW vacuum, violation of this at a scale below $M_{Pl}$ 
could be problematic. In case $\l_{H}(\m)$ becomes 
negative at some scale (as happens for SM at $\Lambda_I$), there may exist another deeper minimum other than 
the EW one. Then the estimate of the tunneling probability $\mathcal{P}_T$ of the EW vacuum to the second 
minimum is essential to confirm the metastability of the Higgs vacuum. The Universe will be in a metastable state, 
provided the decay time of the EW vacuum is longer than the age of the Universe. The tunneling probability is given by \cite{Isidori:2001bm,Buttazzo:2013uya},
\bea
\mathcal{P}_T=T^4_U\m_B^4 e^{-\frac{8\pi^2}{3|\l_{H}(\m_B)|}},
\eea 
where $T_U$ is the age of the Universe, $\m_B$ is the scale at which the tunneling probability is maximized, determined from $\b_{\l_{H}}(\m_B)=0$. Solving the above equation, the metastability requires:
\bea
\l_{H}(\m_B)>\frac{-0.065}{1-\rm{ln}\bigg{(}\frac{v}{\m_B}\bigg{)}}.
\eea 
 
At high energies, the RG improved effective potential can be written as \cite{Degrassi:2012ry}
\bea
V^{\rm{eff}}_H&=&\frac{\l^{\rm{eff}}_H(\m)}{4}h^4, 
\label{effpot}   
\eea        
\noindent where $\l^{\rm{eff}}_H(\m)=\l^{\rm{SM,eff}}_H(\m)+\l^{S,\rm{eff}}_H(\m)+\l^{\rm{T,eff}}_H(\m)$. Here, $\l^{\rm{SM,eff}}_H(\m)$ is the contribution coming from the SM fields to $\l_H$ whereas $\l^{\rm{S,eff}}_H(\m)$ and $\l^{\rm{T,eff}}_H(\m)$ are contribution to the $\l_H$ coming from the additional fields $S$ and $T$ in the present setup. These new contributions can be expressed as :
\besub
\bea
\l^{\rm{S,eff}}_H(\m)&=& e^{4\G(h=\m)} \bigg[\frac{\l^2_{HS}}{64\pi^2}\bigg(\rm{ln}\frac{\l_{HS}}{2}-\frac{3}{2}\bigg)\bigg]\\
\l^{\rm{T,eff}}_H(\m)&=& e^{4\G(h=\m)} \bigg[\frac{3\l^2_{HT}}{64\pi^2}\bigg(\rm{ln}\frac{\l_{HT}}{2}-\frac{3}{2}\bigg)\bigg] .  
\eea 
\label{eff_lamH} 
\eesub
Here, $\G(h)=\int_{m_t}^{h}\g(\m)~\rm{d~ln(\m)}$ and $\g(\m)$ is the anomalous dimension of the Higgs field \cite{Buttazzo:2013uya}.

In a pure scalar singlet DM scenario, it is known that $m_S$ of the order of TeV is required to make the EW vacuum absolutely stable 
\cite{Bhattacharya:2019fgs}. On the other hand in a single component hypercharge-less scalar triplet scenario, it is shown that the EW vacuum becomes 
absolutely stable only if the mass of the scalar triplet particle is around 1.9 TeV. Following the 
analysis of section {\ref{result}} with two-component DM scenario made out of $S$ and $T^0$, we observe that both the DM can 
have sub-TeV masses along with relatively smaller values of Higgs portal couplings, $\l_{HS}$ and $\l_{HT}$. Therefore we would 
like to explore here whether the same parameter space can make the EW vacuum stable. 

\begin{table}[h]
\centering
\begin{tabular}{|c|c| c| c | c|c| c| c|c|}
\hline
Scale & $\l_{H} $  & ~$y_{t}$ &  ~$g_{1}$& ~$g_{2}$& ~$g_3$\\  
\hline
$\mu=m_t$ &$0.125932$ & $0.93610$ & $0.357606$ & $0.648216$ & $1.16655$  \\
 \hline
\end{tabular}
\caption{Values of the relevant SM couplings (top-quark Yukawa $y_t$ , gauge couplings $g_i (i = 1, 2, 3)$ and Higgs quartic
coupling $\l_H$ ) at energy scale $\mu= m_t= 173.2$ GeV with $m_h =125.09$ GeV and $\a_S(m_Z)= 0.1184$.}
\label{initial_conditions}
\end{table} 

For doing the analysis, the running of the SM couplings as well as all the other relevant BSM coupling involved in the present setup is done at two-loops from $\mu = m_t$ to $M_{Pl}$ energy scale\footnote{In Appendix \ref{RGE} we only provide the 1-loop $\b$ functions which were generated using the model implementation in SARAH \cite{Staub:2013tta}.}  while taking into account the two-loop boundary or matching conditions \cite{Coriano:2015sea}. In Table \ref{initial_conditions}, we provide the initial boundary values (at two-loops) for all SM couplings at an energy scale $\mu=m_t$ in 
line with \cite{Braathen:2017jvs}. We use these boundary values as evaluated in \cite{Buttazzo:2013uya} by taking various threshold corrections at $m_t$ and the mismatch between top pole mass and $\overline{MS}$ renormalized couplings, into account. Here, we consider
$m_h=125.09$ GeV, $m_t=173.2$ GeV, and $\a_S(m_Z)= 0.1184$.  A comment on the effect of additional fields (apart 
from the SM ones) in matching conditions can be pertinent here. In \cite{Braathen:2017jvs}, it has been shown that if the Higgs portal coupling(s) of the additional scalar singlet (one DM component here) remains reasonably small ($\mathcal{O}$(1)) while considering 
mass of the singlet (DM) $\sim$ TeV, the one loop correction observed in the Higgs quartic coupling turns out to be reasonably small 
in comparison to that of the pure SM. The same conclusion holds for the other DM component as well. As we have considered both 
the portal couplings as small along with not-so-heavy masses of them, we neglect such corrections in matching conditions while doing 
the present analysis. Even if those corrections are taken into account, we expect a very mild change in the analysis of vacuum stability.

\begin{figure}[H]
\centering
\subfigure[]{
\includegraphics[scale=0.39]{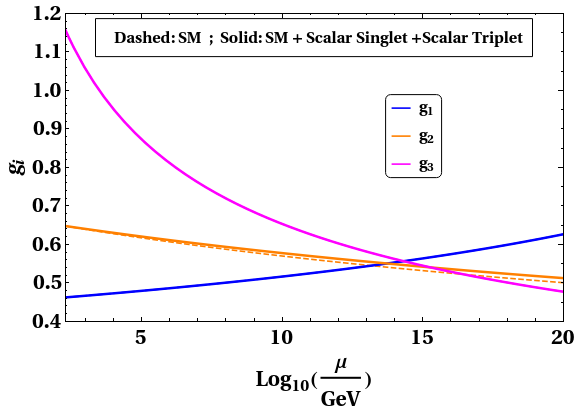}}
\subfigure[]{
\includegraphics[scale=0.4]{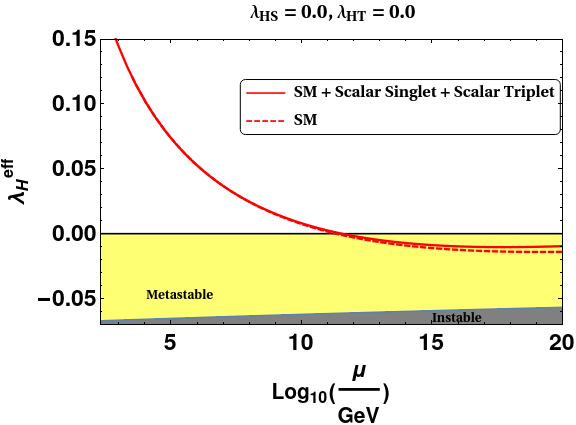}}
\caption{(a) Left panel compare the evolution of the gauge couplings in the present setup with that of the SM (b) Right panel shows evolution $\l^{\rm{eff}}_H$ and its comparision with the SM while keeping $\l_{HS}=\l_{HT}=0$. In both the panels we have kept  $\l_{HS} = 0.0$ and $\l_{HT} = 0.0$, while keeping $m_{T^0} = 680~\rm{GeV}$, $m_{S} = 690~\rm{GeV} $ and $\kappa=0.3$.  The colour codings are explained in the legends.}
\label{gauge}
\end{figure}
We first show the effect of the scalar triplet on all the gauge couplings $g_1,g_2$ and $g_3$ in Fig. \ref{gauge}(a). The newly introduced scalar triplet neither carry a colour charge nor owns hypercharge and hence no modification is observed 
in the evolution of $g_1$ (blue) and $g_3$ (magenta) when compared to the SM ones (solid lines overlaps with the dashed blue and magenta lines). However being charged under $SU(2)$, its inclusion in the present setup increases the number of particles carrying $SU(2)$ charges and hence a modification of the $\b$-function of $g_{2}$ is expected via Eq. \ref{RGE}(2) in appendix. 
This positive shift from the SM values (dashed orange line) is also depicted in the running of $g_2$  (orange line) in Fig. \ref{gauge}(a). The increase in the value of $g_2$ at high scales also impacts the of evolution of effective Higgs 
quartic coupling $\l^{\rm eff}_{H}$ to some extent which can be seen from the Fig. \ref{gauge}(b) where all the Higgs portal couplings are set to zero. This positive shift in $\l^{\rm eff}_{H}$ is observed due to the presence of term proportional to $g_2^4$ in Eq. (\ref{betaHSM} b). Even though a positive shift is observed in the evolution of $\l^{\rm eff}_{H}$ in Fig. \ref{gauge}(b), it is 
very moderate and hence fails to make the EW vacuum absolutely stable. The absolute stability of the EW vacuum can be obtained once the Higgs portal couplings are switched on.

\begin{table}[]
\centering
\begin{tabular}{|c|c| c| c | c|c| c| c|c|c|c|c|}
\hline
BP & $m_{T^0} ~\rm{[GeV]}$  & ~$m_{S} ~\rm{[GeV]}$ & ~$\lambda_{HS}$& ~$\lambda_{HT}$& ~$\kappa$ & $\Omega_{T^0}h^2$ &$\Omega_{S}h^2$&$\sigma_{T^0,eff}$ ($pb$)&$\sigma_{S,eff}$($pb$)&$\m_{\g\g}$\\  \hline
BP-I &$680$ & $690$  &  $0.2$ & $0.2$& $0.3$& $0.026$ &$0.094$ &$1.67\times10^{-10}$ &$5.77\times10^{-10}$&$0.998$\\
 \hline
BP-II &$1600$ & $850$  & $0.5$ & $0.2$& $0.3$& $0.083$ &$0.034$ &$9.75\times10^{-11}$ &$8.84\times10^{-10}$&$0.999$\\
 \hline
\end{tabular}
\caption{Benchmark points for which the total relic density satisfy the Planck limit and are also allowed by the direct detection experiments. The following BPs are aslo allowed by the constraints coming from ATLAS on the Higgs singnal strength $\m_{\g\g}$.
 }
\label{BP}
\end{table}

As stated above, both the couplings $\l_{HS}$ and $\l_{HT}$ (with different pre-factors) play a significant role in the running of effective Higgs 
quartic coupling $\l^{\rm eff}_{H}$. Presence of these Higgs portal couplings are therefore expected to make the EW vacuum stable. For the 
analysis purpose, 
we have chosen two benchmark points BP-I and BP-II as shown in Table \ref{BP}. Both these points satisfy the total relic density, 
the direct detection bounds and are also allowed by the constraints coming from ATLAS \cite{Aaboud:2018xdt} on the Higgs singnal strength $\m_{\g\g}$  ( as discussed in section \ref{constraints}). While choosing the benchmark points we have kept $\kappa$ fixed at 0.3 so that the conversion of 
the heavier dark matter to the lighter one remains effective. We here fix the scalar singlet DM mass $m_S=690~\rm{GeV}$ for BP-I and 850 GeV for BP-II along with the choices of $\l_{HS} = 0.2$ (BP-I) and $0.5$ (BP-II) respectively. Note that such choices of $m_S$ and $\l_{HS}$ neither allow $S$ to be 
a single component DM nor they make the effective Higgs quartic coupling $\l^{\rm{eff}}_{H}$ positive all the way till $M_{Pl}$. 

\begin{figure}[H]
\centering
\subfigure[]{
\includegraphics[scale=0.39]{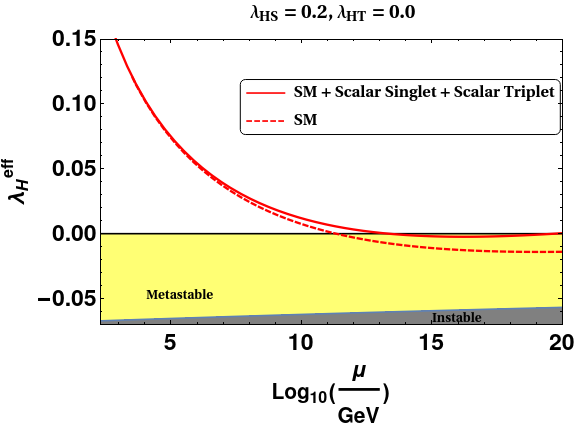}}
\subfigure[]{
\includegraphics[scale=0.39]{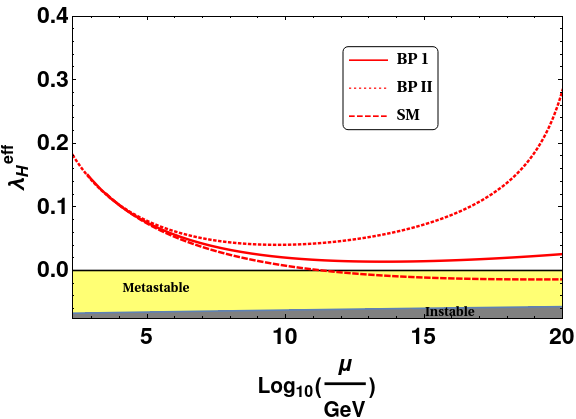}}
\caption{Evolution of effective Higgs quartic coupling $\l^{\rm{eff}}_H$ against the scale $\m$ for (a) $\l_{HS} = 0.2$ and $\l_{HT} = 0.0$, while keeping $m_{T^0} = 680~\rm{GeV}$, $m_{S} = 690~\rm{GeV} $ and $\kappa=0.3$ (b) two given benchmark points  BP-I (solid red lines) and BP-II (dotted red lines).}
\label{lamH}
\end{figure}


In Fig. \ref{lamH} (a) we first show the effect of the Higgs portal coupling $\l_{HS}$ on the running of $\l^{\rm{eff}}_{H}$ while keeping $\l_{HT}=0$. As discussed above, once the Higgs portal coupling $\l_{HS}$ is switched on, it tends to push the $\l^{\rm{eff}}_{H}$ towards the larger value. At this moment, one may recall that $\l_{HS} = 0.3$ or above is required to make 
the EW vacuum absolutely stable just by introducing the singlet scalar $S$ \cite{Garg:2017iva}. Here we will see that with $\l_{HS}\leq 0.2$, the EW vacuum can be absolutely stable, thanks to the other Higgs portal coupling $\l_{HT}$.
In Fig. \ref{lamH} (b) we show the running of the effective Higgs quartic coupling $\l^{\rm{eff}}_{H}$ in  
our model for the two benchmark points as mentioned in Table \ref{BP}, BP-I (solid red lines) as well as BP-II (dotted red lines) and compare it with that of the SM (dashed red lines). As expected, we observe in Fig.~\ref{lamH}(b) that  
due to the presence of both scalar couplings $\l_{HS}$ and $\l_{HT}$, the $\b_{\l_H}$ 
gets affected and hence make $\l^{\rm{eff}}_{H}$ positive till $M_{Pl}$. The conclusion remains valid for 
both the benchmark points, BP-I and BP-II. Increase in the value of  $\l^{\rm{eff}}_H$ for BP-II is due of the involvement of larger $\l_{HS} = 0.5$.

\begin{figure}[H]
\centering
\subfigure[]{
\includegraphics[scale=0.39]{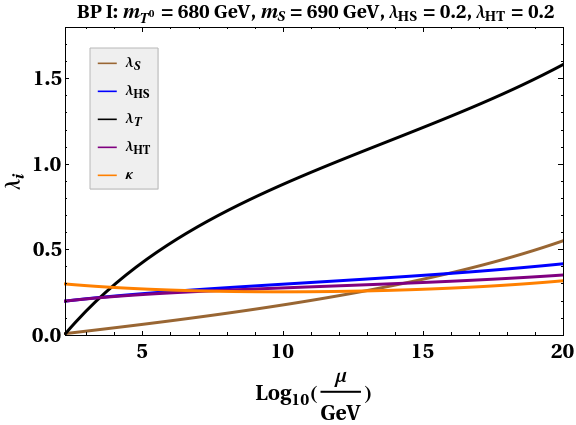}}
\subfigure[]{
\includegraphics[scale=0.39]{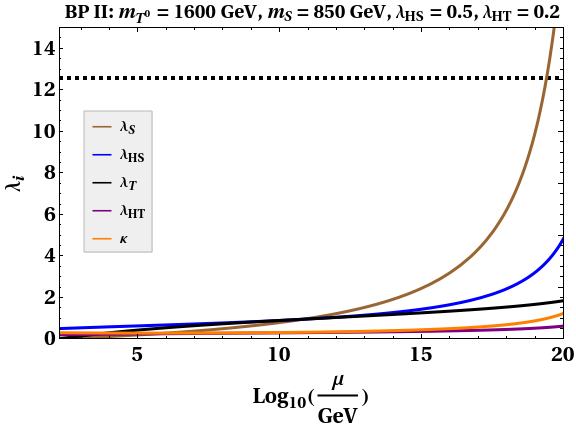}}
\caption{(a) Left panel shows evolution of the quartic couplings involved in the present setup for BP-I (b) Right panel shows evolution of the same quartic couplings for BP-II. The colour codings are explained in the legends.}
\label{quartic}
\end{figure}

In Fig. \ref{quartic}, we plot the running of all the scalar quartic couplings in our model for both the BPs.
We observe in  Fig. \ref{quartic} that all the couplings remain positive and perturbative till the 
Planck scale $M_{Pl}$ for both the BPs. 
Here we have used the central values of top mass and $\alpha_s$. It can be noted that if we allow a 
3$\sigma$ variation of $m_t$ and $m_s$, there would be some positive shift in the running of $\l^{\rm{eff}}_{H}$ even within 
the pure SM case corresponding to the smallest value of top mass and the largest value of $\a_s$ (in their $3\sigma$ allowed 
range). However we have found that that such a shift cannot be comparable to the ones obtained in our scenario for BP-I and 
II.
It is also interesting to note that the self quartic coupling of the scalar singlet $S$ in Fig. \ref{quartic} (b) shoots up, this happens because of the specific choice of $\l_{HS}=0.5$ made in BP-II as shown in Table \ref{BP}. This rapid increase in the evolution of $\l_S$ for the large value of $\l_{HS}$ is dictated by the presence of $12\l_{HS}^2$ term in the $\beta_{\l_S}$.

\section{Conclusions}
\label{conclusion}

In this work, we explore a two-component DM scenario made out of one singlet scalar and the neutral 
component of a hypercharge-less triplet scalar. As a single component dark matter, none of these 
candidates satisfies the relic density and the DD constraints having mass below TeV. While the singlet 
scalar starts to satisfy the relic and DD with its mass close to 1 TeV alone, the $Y=0$ triplet can do so 
with its mass close to 2 TeV. Hence we particularly focus in this sub-TeV region as this regime is 
otherwise an interesting one from the perspective of collider and dark matter experiments. 
We are able to show that the DM-DM conversions becomes helpful so as to realize our goal of 
restricting both the dark matters in sub-TeV regime for $m_S > m_{T^0}$. In case of reverse mass 
hierarchy, such a realization turns out to be not that favorable though. In this case where $m_S < 
m_{T^0}$, triplet mass beyond 1 TeV (but much less than 2 TeV) with $m_S$ below 1 TeV can do 
the job. 

In this entire analysis, the conversion coupling $\kappa$ plays a pivotal role. We observe that though it is 
mostly the scalar singlet contribution which contributes dominantly to the relic, the parameter space with 
$\kappa = 0$ is completely disallowed. This is due to the fact that the relic density then would be mainly 
followed from $S$ only and hence the effective cross-section in DD can not have adequate suppression 
which is otherwise expected via Eq.(\ref{singletdd}) with a sizeable $\kappa$. The parameter space that 
satisfy the relic and DD constraints is also consistent in making the electroweak vacuum absolutely stable. 
This is mainly achieved through the contributions of the Higgs portal couplings of the dark matters. The setup 
also bears an interesting discovery potential at LHC. Due to its multi-component nature, the present setup can 
accommodate smaller value of triplet scalar mass (below TeV) which provides a possibility of probing the charged scalar more 
proficiently at LHC via the disappearing charge track at the detector.  A detailed study in this direction remain 
an interesting possibility to explore in future. 

\acknowledgments
A.D.B and A.S acknowledge the support from DST, Government of India, under Grant No. PDF/2016/002148 
during the early phase of the work where A.D.B was supported by the SERB National Post-Doctoral 
fellowship under the same. A.D.B is also supported by the National Science Foundation of China (11422545,11947235).  
RR would like to thank Najimuddin Khan for various useful discussions during the course of this work. 
\appendix
\section{Tree level unitarity constraints}
\label{appenA}

In this section we discuss the perturbative unitarity limits on the quartic coupling present in our model. The scattering amplitude for any $2\rightarrow2$ process can be expressed in  terms of the Legendre polynomial as \cite{Lee:1977eg}

\bea
\mathcal{M}^{2\rightarrow2}&=& 16\pi \sum_{l=0}^{\infty}a_l(2l+1)P_l(\cos\theta)
\label{u1}
\eea

where $\theta$ is the scattering angle and $P_l(\cos\theta)$ is the Legendre polynomial of order $l$. In the high energy limit, only the s-wave ($l=0$) partial amplitude $a_0$ will determine the leading energy dependence of the scattering process. The unitarity constraint says 
\bea
\rm{Re}~|a_0|<\frac{1}{2}
\label{u2}
\eea
This constraint in Eq.(\ref{u2}) can be further converted to a bound on the scattering amplitude $\mathcal{M}$
\bea
|\mathcal{M}|<8\pi
\label{u3}
\eea

In our present setup, we have multiple possible $2\rightarrow2$ scattering process. Therefore, we need to construct a matrix ($\mathcal{M}_{i,j}^{2\rightarrow2}=\mathcal{M}_{i \rightarrow j}$) considering all the two particle states. Finally we need to calculate the eigenvalues of $\mathcal{M}$ and employ the bound as in Eq. (\ref{u3}). In the high-energy limit, we express the SM Higgs doublet as $H^T=(w^+ ~\frac{h+iz}{\sqrt{2}})$. Then the scalar potential in Eq.(p1) give rise to 13 neutral combinations of two particle states:

\bea
w^+w^-,~\frac{hh}{\sqrt{2}},~\frac{zz}{\sqrt{2}},~T^+T^-,~\frac{T^0T^0}{\sqrt{2}},~\frac{SS}{\sqrt{2}},~hT^0,~zT^0,~hS,~zS,~hz,~w^+T^-,~T^+w^-
\eea

and 8 singly charged two particle states:

\bea
w^+h,~w^+z,~w^+T^0,~w^+S,~T^+T^0,~T^+h,~T^+z,~T^+S.
\eea

Therefore, we can write the scattering amplitude matrix ($M$) in block diagonal form by decomposing it into a neutral (NS) and singly charged (CS) sector as

\bea
M_{21\times 21} = 
\begin{pmatrix}
(M^{NS})_{13\times 13} & 0  \\
0 & (M^{CS})_{8\times 8} \\
\end{pmatrix} 
. 
\label{u4} 
\eea

where the submatrices are given by

\bea
M_{13\times 13}^{NS} = 
\begin{pmatrix}
(M_1^{NS})_{6\times 6} & 0  \\
0 & (M_2^{NS})_{7\times 7} \\
\end{pmatrix} 
.  
\eea

with
\bea
M_{1}^{NS} = 
\begin{pmatrix}
4\l_H & \sqrt{2}\l_H & \sqrt{2}\l_H & \l_{HT} & \frac{\l_{HT}}{\sqrt{2}} & \frac{\l_{HS}}{\sqrt{2}}\\
 \sqrt{2}\l_H  & 3\l_H & \l_{H} & \frac{\l_{HT}}{\sqrt{2}} & \frac{\l_{HT}}{2} & \frac{\l_{HS}}{2}\\
  \sqrt{2}\l_H  & \l_H & 3\l_{H} & \frac{\l_{HT}}{\sqrt{2}} & \frac{\l_{HT}}{2} & \frac{\l_{HS}}{2}\\
\l_{HT} & \frac{\l_{HT}}{\sqrt{2}} & \frac{\l_{HT}}{\sqrt{2}} & \frac{2\l_{T}}{3} &\frac{\l_{T}}{3\sqrt{2}} & \frac{\kappa}{\sqrt{2}}\\
\frac{\l_{HT}}{\sqrt{2}} & \frac{\l_{HT}}{2} & \frac{\l_{HT}}{2} & \frac{\l_{T}}{3\sqrt{2}} &\frac{\l_{T}}{2} & \frac{\kappa}{2}\\
\frac{\l_{HS}}{\sqrt{2}} & \frac{\l_{HS}}{2} & \frac{\l_{HS}}{2} & \frac{\kappa}{\sqrt{2}} &\frac{\kappa}{2} & \frac{\l_S}{2}\\
\end{pmatrix} 
,  
\eea

\bea
M_{2}^{NS} = 
\begin{pmatrix}
\l_{HT} & 0 &0 &0 &0 &0 &0 \\
0 & \l_{HT} &0 &0 &0 &0 &0 \\
0 & 0 &\l_{HS} &0 &0 &0 &0 \\
0 & 0 &0 &\l_{HS} &0 &0 &0 \\
0 & 0 &0 &0 &2\l_{H} &0 &0 \\
0 & 0 &0 &0 &0 &\l_{HT} &0 \\
0 & 0 &0 &0 &0 &0 &\l_{HT} \\ 
\end{pmatrix} 
,  
\eea

and

\bea
M^{CS} = 
\begin{pmatrix}
2\l_{H} & 0 &0 &0 &0 &0 &0 &0 \\
0 & 2\l_{H} &0 &0 &0 &0 &0 &0 \\
0 & 0 &\l_{HT} &0 &0 &0 &0 &0 \\
0 & 0 &0 &\l_{HS} &0 &0 &0 &0 \\
0 & 0 &0 &0 &\frac{\l_{T}}{3} &0 &0 &0 \\
0 & 0 &0 &0 &0 &\l_{HT} &0 &0 \\
0 & 0 &0 &0 &0 &0 &\l_{HT} & 0 \\
0 & 0 &0 &0 &0 &0 &0 & \kappa \\ 
\end{pmatrix} 
.  
\eea

After determining the eigenvalues of Eq.(\ref{u4}) we conclude that the tree level unitarity constraints in this setup are the following:
\bea
|\l_H|<4\pi, ~\bigg{|}\frac{\l_{T}}{3}\bigg{|}< 8\pi ,\nonumber \\
|\l_{HT}|< 8\pi, |\l_{HS}|< 8\pi, |\kappa|< 8\pi, \nonumber \\
\rm{and}~ |x_{1,2,3}|< 16 \pi   
\eea
where $|x_{1,2,3}|$ are the roots of the following cubic equation:
\be
x^3+x^2(-36\l_H-3\l_S-5\l_T)+x(-27\kappa^2-36\l_{HS}^2-108\l_{HT}^2+108\l_H\l_S
+180\l_H\l_T\nonumber \\
+15\l_S\l_T)+972\kappa^2\l_H-648\kappa\l_{HS}\l_{HT}+324\l_{HT}^2+180\l_{HS}^2\l_T\l_S-540\l_H\l_T\l_S\nonumber \\
= 0.   
\ee

\section{1-loop $\b$-functions}
\label{RGE}

Below we provide the 1-loop $\b$-functions for all the couplings involved in the present setup. While generating the $\b-$functions we have considered one scalar singlet and one hypercharge-less scalar  triplet together with the SM particle spectrum. Since the new particles do not carry any colour charges and the Yukawa interactions of these particles are forbidden due to the symmetry assignment of the setup, no modification is observed in the $\b-$function of the strong coupling $g_3$ and the top Yukawa coupling $y_t$. The hypercharge being zero for both the BSM fields, the $\b-$function of gauge coupling $g_1$ remain same as  that of the $\b^{\rm{SM}}_{g_1}$ whereas $T$ being a $SU(2)$ triplet, a shift in the $\b-$function of $g_2$ can be observed in comparison to that of the $\b^{\rm{SM}}_{g_2}$. 
\subsubsection{SM Couplings}
{\allowdisplaybreaks  \begin{align} 
 \beta_{g_1}&= \b_{g_1}^{\rm{SM}} +\b_{g_1}^{\rm{T}}+\b_{g_1}^{\rm{S}}~ = ~ \b_{g_1}^{\rm{SM}}  \\  
\beta_{g_2}&= \b_{g_2}^{\rm{SM}} +\b_{g_2}^{\rm{T}}+\b_{g_2}^{\rm{S}}~ = ~\b_{g_2}^{\rm{SM}}+ \frac{g_2^3}{16 \pi^2}\bigg{(}\frac{1}{3}\bigg{)} \\
\beta_{g_3} & = \b_{g_3}^{\rm{SM}} +\b_{g_3}^{\rm{T}}+\b_{g_3}^{\rm{S}}~ = ~\b_{g_3}^{\rm{SM}}  \\   
\beta_{\l_H} & = \b_{\l_H}^{\rm{SM}}+\b_{\l_H}^{\rm{T}} +\b_{\l_H}^{\rm{S}}~ = ~\b_{\l_H}^{\rm{SM}}+\frac{3}{2}\l^2_{HT}+\frac{1}{2}\l^2_{HS} \\    
\beta_{y_t} &= \b_{y_t}^{\rm{SM}} +\b_{y_t}^{\rm{T}}+\b_{y_t}^{\rm{S}}~ = ~\b_{y_t}^{\rm{SM}}   
\end{align}} 
\subsubsection{BSM couplings}
{\allowdisplaybreaks  \begin{align} 
\beta_{\l_S}&  = 3 \Big(3 \kappa^{2}  + 4 \lambda_{HS}^{2}  + \lambda_{S}^{2}\Big)\\ 
\beta_{\l_{HS}}&  =3 \kappa \lambda_{HT} -\frac{9}{10} g_{1}^{2} \lambda_{HS} -\frac{9}{2} g_{2}^{2} \lambda_{HS} +12 \lambda_H \lambda_{HS} +\lambda_{S} \lambda_{HS} +4 \lambda_{HS}^{2}+6 \lambda_{HS}y_t^2 \\  
\beta_{\lambda_T}& =  
12 \lambda_{HT}^{2}  -24 g_{2}^{2} \lambda_T  + 3 \kappa^{2}  + 72 g_{2}^{4}  + \frac{11}{3} \lambda_{T}^{2} \\    
\beta_{\lambda_{HT}} & =  
6 g_{2}^{4} -\frac{9}{10} g_{1}^{2} \lambda_{HT} -\frac{33}{2} g_{2}^{2} \lambda_{HT} +12 \lambda_H \lambda_{HT} +4 \lambda_{HT}^{2} +\kappa \lambda_{HS} +\frac{5}{3} \lambda_{HT} \lambda_T +6 \lambda_{HT}y_t^2\\  
\beta_{\kappa} & =  
-12 g_{2}^{2} \kappa  + 4 \kappa^{2}  + 4 \lambda_{HT} \lambda_{HS}  + \kappa \Big(\frac{5}{3} \lambda_T  + \lambda_{S}\Big)
\end{align}}

\bibliographystyle{apsrev}
\bibliography{ref.bib}

\end{document}